\def\asec{$^{\prime\prime}$}
\def\Lya{Ly$\alpha$~}

\def\choneminchtwo{3.6$\mu$m - 4.5$\mu$m~}
\def\choneminchthree{3.6$\mu$m - 5.8$\mu$m~}
\documentclass[useAMS,usenatbib]{mn2e}

\usepackage{graphicx}
\usepackage{url}
\usepackage{times}
\usepackage{amsmath}
\usepackage{amssymb}
\usepackage{subfig}
\usepackage{caption}
\usepackage{verbatim}
\usepackage{color, soul}
\sethlcolor{yellow}

\title[Discovery of bright $z \simeq 7$ galaxies in the UltraVISTA survey]{Discovery of bright ${\bf z \simeq 7}$ galaxies in the UltraVISTA survey}
\author[R. A. A. Bowler et al.]{R. A. A. Bowler$^{1}$\thanks{E-mail:
raab@roe.ac.uk}, J. S. Dunlop$^{1}$, R. J. McLure$^{1}$, H. J. McCracken$^2$, 
B. Milvang-Jensen$^{3}$, \and H. Furusawa$^{4}$, J. P. U. Fynbo$^{3}$, O. Le F\`{e}vre$^{5}$, J. Holt$^{6}$, Y. Ideue$^{7}$, Y. Ihara$^{8}$, \and A. B. Rogers$^{1}$, Y. Taniguchi$^{7}$ \\
$^{1}$SUPA\thanks{Scottish Universities Physics Alliance}, Institute for Astronomy, University of Edinburgh, Royal Observatory, Edinburgh, EH9 3HJ \\
$^{2}$Institut d'Astrophysique de Paris, UMR 7095 CNRS, Universit\'{e} Pierre et Marie Curie, 98 bis Boulevard Arago, 75014 Paris, France \\
$^{3}$Dark Cosmology Centre, Niels Bohr Institute, University of Copenhagen, Juliane Maries Vej 30, 2100 Copenhagen, Denmark \\
$^{4}$Astronomical Data Center, National Astronomical Observatory of Japan, Mitaka, Tokyo 181-8588, Japan \\
$^{5}$Laboratoire d'Astrophysique de Marseille, CNRS and Aix-Marseille Universit\'{e}, 38 rue Fr\'{e}d\'{e}ric Joliot-Curie, 13388 Marselle Cedex 13, France \\
$^{6}$Leiden Observatory, Leiden University, P. O. Box 9513, NL-2300 RA Leiden, The Netherlands \\
$^{7}$Research Institute for Space and Cosmic Evolution, Ehime University, 2-5 Bunkyo-cho, Matsuyama 790-8577, Japan \\
$^{8}$TOYOTA InfoTechnology Center CO., LTD, 6-6-20 Akasaka, Minato-ku, Tokyo, 107-0052, Japan
}

\begin{document}

\date{}

\pagerange{\pageref{firstpage}--\pageref{lastpage}} \pubyear{2012}

\maketitle

\label{firstpage}

\begin{abstract}
We have exploited the new, deep, near-infrared UltraVISTA imaging of the COSMOS field, in tandem with deep optical and mid-infrared 
imaging, to conduct a new search for luminous galaxies at redshifts $z \simeq 7$. 
The year-one UltraVISTA data provide contiguous $Y,J,H,K_{s}$ imaging over 1.5\,deg$^2$, reaching a 5$\sigma$ detection limit of $Y$+\,$J \simeq 25$ (AB mag, 2-arcsec diameter aperture).
The central $\simeq 1$\,deg$^2$ of this imaging coincides with the final deep optical ($u^*,g,r,i$) data provided by the Canada France Hawaii Telescope (CFHT) 
Legacy Survey and new deep Subaru Suprime-Cam $z'$-band imaging obtained specifically 
to enable full exploitation of UltraVISTA. It also lies within the {\it Hubble Space Telescope} 
(HST) $I_{814}$-band and {\it Spitzer} IRAC imaging 
obtained as part of the COSMOS survey. We have utilised 
this unique multi-wavelength dataset to select galaxy candidates at redshifts $z > 6.5$ by searching first for 
$Y$+\,$J$-detected objects 
which are undetected in the CFHT and HST optical data. 
This sample was then refined using a photometric redshift fitting code, enabling the rejection of lower-redshift galaxy 
contaminants and cool galactic M, L, T dwarf stars.
The final result of this process is a small sample of (at most) ten 
credible galaxy candidates at $z > 6.5$ (from over 200,000 galaxies detected in the year-one UltraVISTA data) which we present 
in this paper. The first four of these appear to be 
robust galaxies at $z > 6.5$, and fitting to their stacked spectral energy distribution yields $z_{\rm phot} = 6.98 \pm 0.05$ with a stellar mass $M_*\simeq 5\times10^9\,{\rm M_{\odot}}$ 
and rest-frame UV spectral slope $\beta \simeq -2.0 \pm 0.2$ (where $f_{\lambda} \propto \lambda^{\beta}$). 
The next three are also good candidates for $z > 6.5$ galaxies, 
but the possibility that they are dwarf stars cannot be completely excluded.
Our final subset of three additional candidates is afflicted not only by potential 
dwarf-star contamination, but also contains objects likely to lie at redshifts just below $z = 6.5$.
We show that the three even-brighter $z \gtrsim 7$ galaxy candidates reported in the COSMOS field by~\citet{Capak2011} are in fact all lower-redshift galaxies at $z \simeq 1.5-3.5$. Consequently 
the new $z \simeq 7$ galaxies reported here are the first credible $z \simeq 7$ Lyman-break galaxies discovered in the COSMOS field and, as the most UV-luminous discovered to date at these redshifts, are 
prime targets for deep follow-up spectroscopy. 
We explore their physical properties, and briefly consider the implications of their inferred number density for the form of the galaxy luminosity function 
at $z \simeq 7$.

\end{abstract}


\section{Introduction}

The advent of deep near-infrared imaging on the {\it Hubble Space Telescope} (HST) ushered in a new era in the discovery of galaxies at redshifts
$z \simeq 7 - 8$. Installed in 2009, the infrared channel of Wide Field Camera 3 (WFC3/IR) has already been used to obtain near-infrared imaging to detection limits
$Y,J,H \simeq 29$ (AB mag). This ultra-deep near-infrared imaging, when combined with existing deep optical imaging in fields such as the 
Hubble Ultra Deep Field (HUDF) and GOODS-South, has enabled the first detection of significant numbers of Lyman-break galaxies (LBGs) at $z > 6.5$, and hence
the first meaningful studies of the rest-frame ultraviolet (UV; $\lambda_{\rm rest} \simeq 1500$\,\AA) galaxy 
luminosity function (LF) at $z \simeq 7$ (e.g. \citealp{McLure2010, Oesch2010, Bouwens2010, Finkelstein2010, McLure2011, Bouwens2011a}).  

The depth of this HST imaging, coupled with the (relatively) small field-of-view of WFC3/IR has meant that, to date, these new studies of the $z \simeq 7$
LF have been largely focussed on the faint end, with the WFC3/IR samples dominated by sub-$L^*$ galaxies ($M_{1500} \simeq -20 \rightarrow -18$\,mag). This 
has of course been of enormous value, as it has revealed that the faint-end slope of the $z \simeq 7$ LF is steep (e.g. $\alpha = -1.7$,~\citealp{McLure2010}; 
$\alpha = -2.0$,~\citealp{Bouwens2011a}), implying that it is the large population of fainter galaxies which is likely responsible for the re-ionisation
of the Universe (e.g.~\citealp{Robertson2010}). Indeed, to further explore the faint end of the $z \simeq 7$ LF, and extend such studies out to $z \simeq 9$,
even deeper WFC3/IR imaging of the HUDF will be completed before the end of 2012 (HST Cycle-19 program GO12498).

However, at the same time the importance of better determining the number density of brighter galaxies at this crucial epoch has not been over-looked.
In particular, the 3-year, 900-orbit Cosmic Assembly Near-infrared Deep Extragalactic Survey (CANDELS) HST 
Treasury Program (\citealp{Grogin2011, Koekemoer2011}),
reaching detection limits of $J,H \simeq 27$\,mag, 
is already providing better statistics on the number density of $z \simeq 7$ galaxies at luminosities around the (apparent) break luminosity 
of the LF (i.e. at $M^*_{1500} \simeq -20.1$;~\citealt{Ouchi2009, McLure2010,Bouwens2011a}). 
Meanwhile, CANDELS and parallel HST WFC3/IR imaging programs such as Brightest of the Reionizing Galaxies survey (BoRG,~\citealp{Trenti2011}) are also now yielding
the first significant numbers of brighter galaxies at $z \simeq 8$ (e.g.~\citealp{Oesch2012}, \citealp{Bradley2012}).   

However even CANDELS, when complete, will cover an area of 
only $\simeq 800$\,arcmin$^2$. There thus remains a 
key role for wider-area, albeit shallower, 
ground-based near-infrared imaging to better constrain the number density of rarer more luminous galaxies at $z > 6.5$, and hence properly 
determining the bright end of the galaxy LF at these epochs. The value of degree-scale imaging surveys of the high-redshift Universe (sampling 
comoving volumes $\simeq 100 \times 100 \times 100$\,Mpc$^3$) was demonstrated by~\citet{McLure2006, McLure2009} and~\citet{CurtisLake2012}, who utilised the Subaru Suprime-Cam 
and UKIRT WFCAM imaging of the SXDS field to search for rare bright galaxies at $z \simeq 5$ and $z \simeq 6$, complementing 
deeper smaller-area imaging work with HST ACS (e.g.~\citealp{Bouwens2007}), and ultimately yielding determinations of the $z \simeq 5$ and $
z \simeq 6$ LF spanning a dynamic range of over five magnitudes. 

An attempt to push ground-based studies of LBGs out to $z \simeq 7$ has been made by~\citet{Ouchi2009} and~\citet{Castellano2010a,Castellano2010b}. However the latter 
study (with Hawk-I on ESO's Very Large Telescope; VLT) 
has covered an area smaller than the final CANDELS area, while the former was based on pushing the silicon-based CCDs in Subaru Suprime-CAM to their red limit, and 
lacked the longer-wavelength ($\lambda > 1$\,$\mu$m) near-infrared imaging required to confirm a blue spectral slope longward 
of the putative Lyman break (resulting in highly-contaminated galaxy samples).

Now, however, degree-scale near-infrared imaging reaching the depth required to uncover credible galaxy candidates at $z > 6.5$ has finally 
arrived with the UltraVISTA survey. This 5-year public survey on the COSMOS field (\citealp{Scoville2007}) with the near-infrared camera VIRCAM~\citep{Dalton2006} 
on the new VISTA survey telescope at Paranal in Chile~\citep{Emerson2010} commenced in 2010, and the first public data release of the 
fully-reduced year-1 data was made through ESO in Feb 2012\footnote{\url{http://www.eso.org/sci/observing/phase3/data_releases.html}}(\citealp{McCracken2012}).  
This new near-infrared imaging covers 1.5\,deg$^2$ to depths of $Y=24.7, J = 24.5, H = 24.0, K_s = 23.8$ (5$\sigma$; 2-arcsec aperture diameter). 
At the longer wavelengths this is not yet as deep as the 0.8\,deg$^2$ imaging being delivered in the SXDS field by the ongoing UKIDSS 
Ultra Deep Survey (UDS;~\citealp{Lawrence2007}).  However, as well as covering a larger area, UltraVISTA provides 
the crucial new ingredient of the first large-area $Y$-band imaging reaching $\simeq 25$\,mag 
(due, in part, to the vastly superior short-wavelength sensitivity of the VISTA Raytheon detectors).
As discussed further below, this deep $Y$-band imaging is of vital 
importance, both for the effective selection of credible LBGs at $z > 6.5$ and for the robust 
rejection of cool brown-dwarf star contaminants. Also crucial is the new availability of very deep Subaru $z'$-band imaging over the central 
1\,deg$^2$ of the UltraVISTA field, obtained over the last three years with the refurbished 
Suprime-Cam (equipped
with its new red-sensitive Hamamatsu CCDs); these data were obtained specifically to
provide red optical imaging of the depth necessary to exploit UltraVISTA in the search for extreme-redshift LBGs.

\begin{figure} 
\includegraphics[width = 0.5\textwidth,  trim = 0cm 1.5cm 0.5cm 2cm, clip = true]{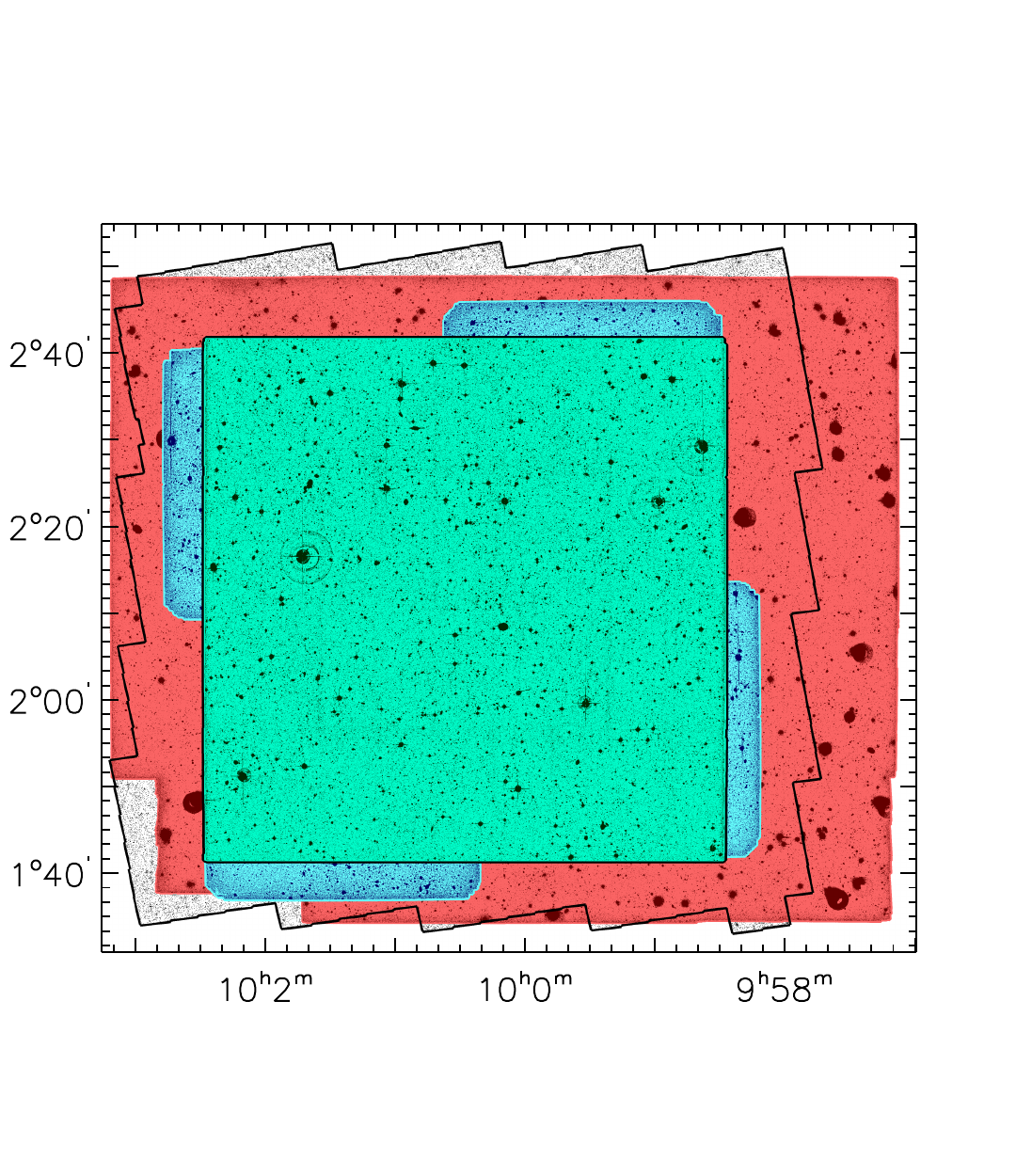}

\caption{The multi-band coverage map of the UltraVISTA/COSMOS field utilised in this study.  Working from the outside in, the year-one
1.5\,deg$^2$ UltraVISTA imaging is shown as the large red rectangle, with the HST/ACS $I_{814}$-band coverage indicated by the jagged outline.
The blue irregular shape within the UltraVISTA data is the Subaru $z'$-band mosaic, formed from four individual Suprime-Cam pointings.
Finally, the central green outlined square is the CFHTLS D2 optical data. This final square area, covering $\simeq 1$\,deg$^2$, is
the area utilised in this study, as it contains all the required overlapping multi-wavelength data.}
\label{fig:fieldmap}
\end{figure}

\begin{table}
\caption{The 5$\sigma$ limiting magnitudes for the relevant optical and near-infrared data used in this study, obtained from the 
rms of the flux from apertures placed in blank regions of each image 
(see Section~\ref{sect:initialdet}).  All magnitudes were calculated within a 2-arcsec
diameter aperture, apart from the ACS $I_{814} $ values, which used a 0.6-arcsec diameter aperture and the IRAC $3.6\,\mu$m and $4.5\,\mu$m values, which were calculated in a 2.8-arcsec diameter aperture.}

\begin{tabular}{ l | c l }

     \hline
 Filter  & m$_{\rm AB}$(5$\sigma$)  & Source \\
  \hline
  $u^*$ & 26.9  & CFHT/MegaCam \\
  $g$ & 27.0 & CFHT/MegaCam \\
  $r$ & 26.6 & CFHT/MegaCam \\
  $i$ & 26.4 & CFHT/MegaCam \\
  $I_{814}$ & 26.7 & HST/ACS \\
  $z$ & 25.2 & CFHT/MegaCam\\
  $z'$ & 26.3 & Subaru/SuprimeCam \\
  $Y$ & 24.7  & UltraVISTA\\
  $J$ & 24.5 & UltraVISTA \\
  $Y+J $& 24.9 & UltraVISTA\\
  $H$ & 24.0 & UltraVISTA \\
  $K_{s}$ & 23.8  & UltraVISTA\\
  $3.6\mu$m & 24.2  & Spitzer/IRAC \\
  $4.5\mu$m & 23.8 & Spitzer/IRAC \\

  \hline
\end{tabular}
\label{table:depths}
\end{table}

We note that UltraVISTA does not represent the first near-infrared imaging of the COSMOS field. Indeed,~\citet{Capak2011} exploited the somewhat 
shallower $H, K_s$ imaging obtained with WIRCam on the Canada-France-Hawaii Telescope (CFHT) and $J$ imaging from WFCAM on UKIRT to report three potential $z > 7$ galaxies in COSMOS 
at surprisingly bright magnitudes ($J < 23.3$), some with proposed spectroscopic confirmation. 
However, the power of the new UltraVISTA imaging is well demonstrated by the ease with which we can now show that 
all of these galaxies in fact lie at much lower redshift (see below).
The main aim of this paper is, however, not only to check previous claims of $z > 6.5$ LBGs in the COSMOS field but more importantly to 
demonstrate that UltraVISTA, even in its first few months of data taking, has delivered images of the depth and quality necessary 
to produce a small but robust sample of luminous $z \simeq 7$ galaxy candidates, and to present this sample to the community for spectroscopic follow-up.

It is important to revisit why the discovery and study of rare luminous galaxies at these early epochs is of interest. First, while the study of faint 
galaxies may appear of more importance for understanding reionisation, the errors on the faint-end slope ($\alpha$) of the Schechter function parameterisation ($\phi(L) = \phi^*(L/L^*)^{\alpha}e^{-L/L^*}$) can be substantially 
reduced by a better determination of the form of the brighter end of the galaxy LF, due to the resulting improved  constraints 
on the other key parameters (i.e. density normalisation $\phi^*$, and characteristic luminosity $L^*$). 
Second, it remains possible that, at some epoch, the exponential cutoff at the bright end of of the LF (still apparently present at $z \simeq 5$ and, arguably, at 
$z \simeq 6$;~\citealp{McLure2009,Willott2012}) may disappear and, for example, be replaced by a shallower power-law 
due to early inefficiency (or absence) of the physical processes (e.g. AGN feedback) purported to be responsible for the relative inefficiency of star-formation 
in higher-mass dark-matter halos (e.g.~\citealp{Finlator2011}, but see also~\citealp{Trenti2010} and~\citealp{Jaacks2012}). Finally, quite apart from the issue of the form of the LF, the discovery of
luminous $z \simeq 7$ galaxies is of intrinsic interest, as it is often the most luminous galaxies which place the most stringent demands on models of galaxy
formation (e.g. Benson et al. 2003) and $z \simeq 7$ galaxies as bright as $Y \simeq 25$\,mag are clearly very attractive targets for follow-up near-infrared 
spectroscopy.

This paper is structured as follows. In Section~\ref{sect:data},
we present the new UltraVISTA survey near-infrared data, and summarise the crucial supporting datasets: these comprise deep optical
imaging (including the final CFHT Legacy Survey data, the long-public HST ACS $i$-band imaging, new ultra-deep Subaru $z'$-band imaging) 
and the mid-infrared (3.6\,$\mu$m and 4.5\,$\mu$m) {\it Spitzer} IRAC imaging obtained via the S-COSMOS survey~\citep{Sanders2007}. In Section~\ref{sect:candsel} we then describe the 
creation and subsequent progressive refinement of our galaxy sample based primarily on spectral energy distribution (SED) fitting,   
and also show colour-colour plots to clarify the validity of our selection method and help expose the possible contaminant populations.
Our final ten candidate $z > 6.5$ galaxies are presented in Section~\ref{sect:cand}, where we also discuss potential contamination, with special emphasis 
on an extremely careful elimination of cool brown-dwarf (L, M, T) stars; as discussed by many authors (e.g.~\citealp{Stanway2008}), contamination by cool dwarf stars is a much more serious 
issue for ground-based searches for $z \simeq 6 - 7$ galaxies than for ultra-deep HST WFC3/IR imaging surveys, 
both because extreme-redshift galaxies 
are generally unresolved in even good-seeing ground-based imaging, and because the relative number of brown-dwarf 
stars to genuine high-redshift galaxies is much larger
at brighter magnitudes (e.g. with a typical absolute magnitude $J \simeq 19$, a T-dwarf star with an apparent magnitude of $J=24$
still lies well within the galactic disc; see~\citealt{Dunlopbook2012}).  
Next, to place our results in context, in Section~\ref{sect:previous} we provide 
a discussion/re-analysis of the previous claims of bright $z > 7$ galaxies recently advanced  
by~\citet{Capak2011} and~\citet{Hsieh2012}, as well as the claimed highest redshift X-ray source candidate presented by~\citet{Salvato2011}. Then, in Section~\ref{sect:discussion},
we analyse the properties displayed by the stacked photometry of our top four galaxy candidates, and provide a very brief 
discussion of the implications of our results for the form of the LF at $z = 7$ (a detailed 
reanalysis of the $z \simeq 7$ luminosity function is deferred to a future paper, where we aim to also incorporate data from 
the UKIDSS UDS survey). We conclude with a summary of our main results in Section 7.

All magnitudes quoted are in the AB system~\citep{Oke1983}, including data points from the dwarf-star literature, where Vega magnitudes and AB magnitudes
are, unfortunately, frequently mixed. For the calculation of physical quantities we 
assume a $\Lambda$CDM cosmology with $H_0 = 70\, {\rm kms}^{-1}{\rm Mpc}^{-1}$, $\Omega_{\rm m}  = 0.30$ and $\Omega_{\Lambda} = 0.70$.

 \begin{figure*}
\subfloat{\includegraphics[width = 0.5\textwidth, trim = 0.9cm 0cm 0cm 0.5cm, clip = true]{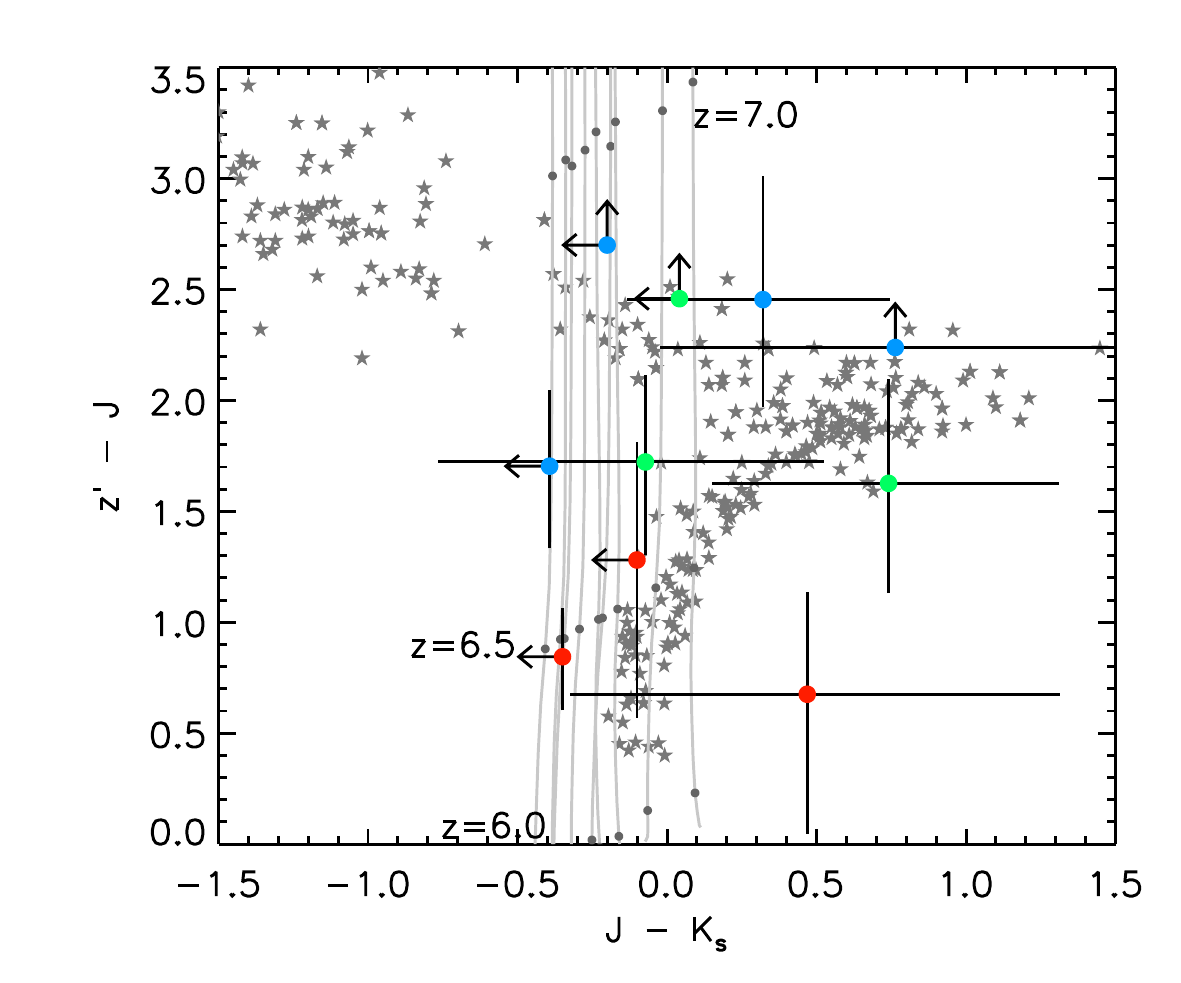}}
\subfloat{\includegraphics[width = 0.5\textwidth, trim = 0.9cm 0cm 0cm 0.5cm, clip = true]{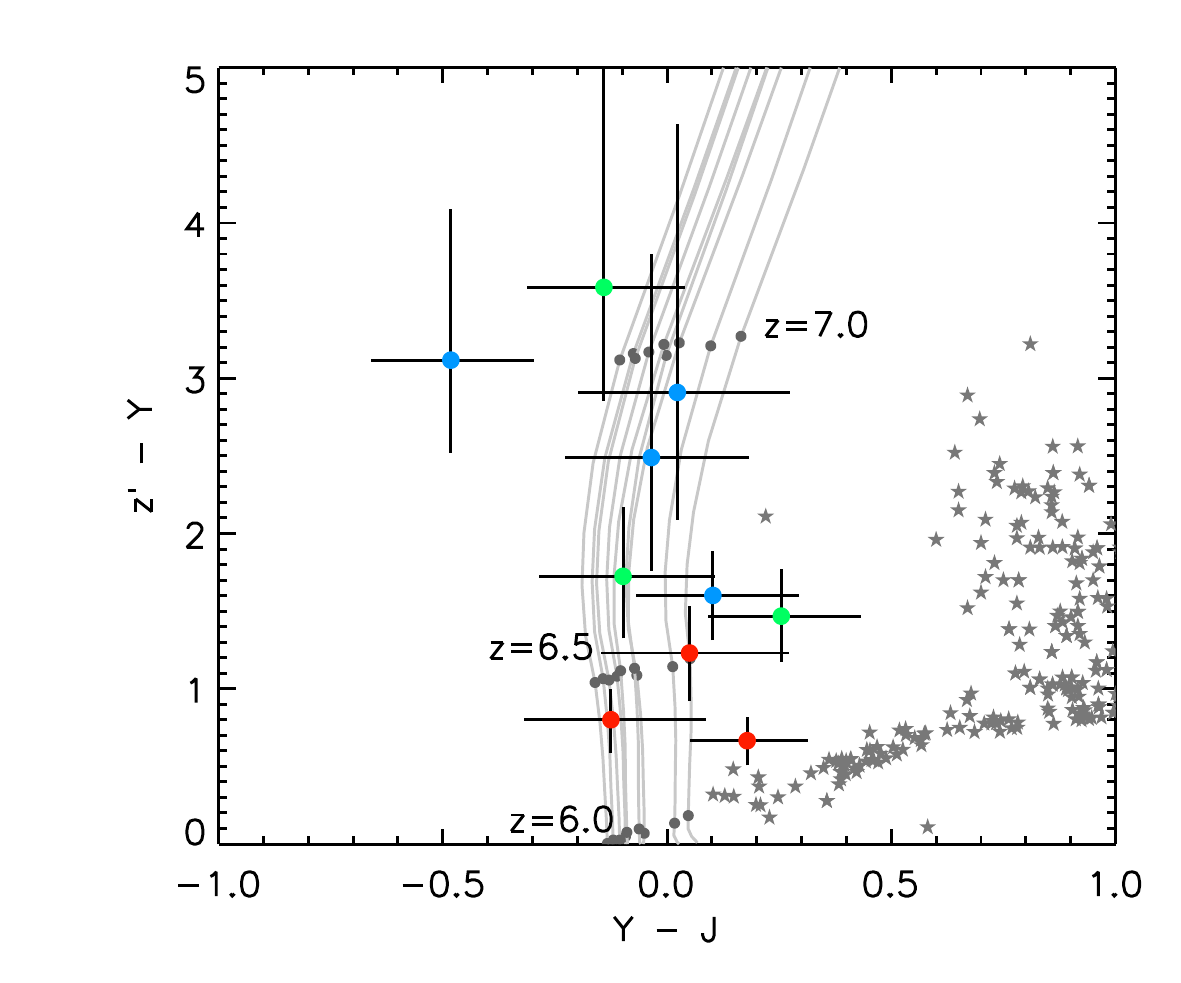}}
\caption{Colour-colour plots of our final ten candidates selected by SED fitting.  
The most secure high-redshift candidates are shown in blue. Candidates that
are likely to be at high redshift, but where we cannot completely 
exclude the possibility they are dwarf stars or low-redshift galaxy contaminants, are shown in green.
Finally, our our three least secure candidates are shown in red.  The large uncertainty in the $J-K_s$ colour
is a consequence of the relative shallowness of the $K_s$ imaging as compared to the $Y$+\,$J$ detection band (which also
results in a few of our candidates being formally undetected in the $K_s$ image).
The colours of M, L and T-dwarf stars are shown with a star symbol, and were obtained from photometric
surveys~\citep{Knapp2004, Burningham2010} and by calculating synthetic magnitudes
from spectra obtained from several libraries as described in~\citet{Findlay2012}.
The grey tracks indicate the evolution of colours with redshift for high-redshift Lyman-break galaxies
as synthesised from the theoretical SEDs of~\citet{Bruzual2003}, assuming a constant star-formation history, and a range of other
parameter values ($Z = 0.2~{\rm Z}_{\sun}$, $A_V = 0.0-0.5$, ${\rm age} = 50-500$\,Myr).  Appropriate redshift steps are illustrated with grey circles situated along the tracks at $z = 6, 6.5$ and $7$.  Absorption by the Intergalactic Medium 
is applied with the~\citet{Madau1995} prescription, with additional suppression of the flux blueward 
of $\lambda_{\rm rest} = 1216~$\AA~at the very highest redshifts (parameterised as $e^{-2((1+z)/6)^4}$ from~\citealp{Patel2010}).}
\label{fig:JKz}
\end{figure*}

\section{Data}\label{sect:data}

The analysis presented in this paper is based on the first year of near-infrared imaging obtained by the UltraVISTA survey, 
in combination with the deep multi-band optical imaging obtained as part of the CFHT Legacy Survey (CFHTLS), new deep $z'$-band imaging obtained
with the refurbished Suprime-Cam on the Subaru telescope, and other publicly-available HST and {\it Spitzer} data obtained as part of the 
Cosmological Evolution Survey (COSMOS)~\citep{Scoville2007}. The production of multi-wavelength catalogues required the use of images on a 
common pixel grid and so the region of overlapping data defined by the CFHTLS imaging (described below in Section~\ref{sect:CFHT}) 
was used to define the search area as shown in Fig.~\ref{fig:fieldmap}.  In Table~\ref{table:depths} we give the 5$\sigma$ depths 
which we have calculated for each of the relevant imaging datasets (see caption for details). All images were matched to the astrometric grid of the UltraVISTA 
$Y$-band image using the {\sc iraf} package {\sc ccmap}.  Finally, the images were resampled to match the pixel scale and image size of the CFHTLS D2 field using the {\sc iraf} package {\sc sregister}.

\subsection{UltraVISTA near-infrared imaging}\label{uvistadata}

Over the next $\sim 5$ years it is planned that the UltraVISTA survey will provide deep near-infrared imaging covering the central region of the COSMOS field 
in four broad-band near-infrared filters ($Y, J, H, K_{s}$) and one narrow-band filter (NB118) to unprecedented depths~\citep{McCracken2012}\footnote{\url{http://ultravista.org}} . The survey was designed to commence 
with a ``deep'' programme (212 hours) providing fully-sampled imaging over a contiguous 1.5\,deg$^2$ field, with the majority (1408 hours) of 
the observing time being subsequently devoted to the ``ultradeep'' programme, comprising substantially deeper imaging over four strips (0.73\,deg$^2$) 
within the field. The first phase of ``deep'' imaging was completed by July 2011, and released to the public in fully-reduced form via ESO in Feb 2012. It is 
this year-one imaging that is utilised here, and the relevant coverage map and photometric depths are summarised in Fig. 1. The UltraVISTA 
images use the COSMOS CFHT $i^*$-band image from 2003-2004 as the astrometric reference~\citep{McCracken2012,Capak2007}.  For the present purpose of searching for high-redshift Lyman-break galaxies, 
an inverse-variance weighted stack of the $Y$-band and $J$-band UltraVISTA images was created 
with the aim of increasing the sensitivity of the survey for objects with a near-flat (in $f_{\nu}$) near-infrared spectral slope.

\subsection{CFHT optical imaging}\label{sect:CFHT}

Deep optical data are essential for the identification of genuine high-redshift galaxies, as it is vital to confirm 
that no flux is detected at wavelengths shortward of the putative Lyman-break. The COSMOS field benefits from extensive multi-band 
optical imaging obtained with both the CFHT and the Subaru telescope.  In this study we use the deepest available data from the CFHT 
Legacy Survey, which provides imaging over a 1\,deg$^2$ subsection of the COSMOS field centred on 
RA ${\rm 10^h00^m28.00^s}$, Dec ${\rm +2^{\circ}12^{\prime}30^{\prime \prime}}$; this defines the area of the survey appropriate for 
a consistent high-quality search for high-redshift optical drop-out galaxies.  
The CFHTLS T0006 release\footnote{\url{http://terapix.iap.fr/cplt/T0006-doc.pdf}}, in the deep field D2, provides data in the $u^{*}, g, r, i_{1}, i_{2}(``y")$ and $z$ 
optical filters from the MegaCam instrument, with data in the $u^*$-band obtained as part of the COSMOS survey~\citep{Capak2007}.  
The $i_{1}$ and $i_{2}$ (or $y$-band) image distinction is a consequence of a new $i$-band filter installed since October 2007. However, 
since the filter transmission curves are similar and probe a region of the spectrum of $z > 6.5$ galaxies where we impose a 
non-detection condition, and the images are of comparable depth (to within 0.1 mag), 
we combined these images to form a deeper inverse-variance weighted stack, hereafter simply termed the $i$-band image.   
The astrometry of the CFHTLS T0006 data release was based on the 2MASS catalogue rather 
than the standard COSMOS CFHT $i^*$-band catalogue.  Hence, the CFHTLS images were 
mapped onto the astrometric solution of the $Y$-band UltraVISTA imaging 
using the technique described in Section~\ref{sect:data}, with a correction of typically 0.2\asec, although the pixel scale was retained 
and used as a base for all other multi-wavelength images.

\subsection{HST/ACS $\bmath{I_{814}}$-band imaging}

The COSMOS field has been imaged to single-orbit depth by HST/ACS in the $I_{814}$-band through a 
HST Treasury Programme~\citep{Scoville2007a, Koekemoer2007, Massey2010}.  Due to the large size of the 1.8\,deg$^2$ high-resolution mosaic (0.03\asec/pix), 
individual postage stamps of galaxy candidates were retrieved from the NASA/IPAC 
archive\footnote{\url{http://irsa.ipac.caltech.edu/data/COSMOS/index_cutouts.html}} 
and visually inspected (after smoothing) as described in Section~\ref{sect:candsel}.  
The $I_{814}$-band image uses the same base astrometric reference as the COSMOS CFHT $i^*$-band 
image used as the standard reference for subsequent imaging within the COSMOS field and the UltraVISTA data as described above, hence no further transformations were applied.

\subsection{Subaru Suprime-Cam $z'$-band imaging}

Deep $z'$-band imaging is crucial for the selection of galaxies at $z>6.5$. 
Since 2009 we have exploited the new red sensitivity provided by the refurbished Suprime-Cam instrument on Subaru to obtain 
very deep $z'$-band imaging over the central 1\,deg$^2$ of the COSMOS field (i.e. matching, as near as possible, the CFHT imaging).
The imaging consists of four Suprime-Cam pointings, each with $\gtrsim 15$ hours of exposure time, rotated by 90\,deg with respect to each other. This yields
a final $z'$-band mosaic reaching a minimum 5$\sigma$ depth of $z' = 26.3$ (in a 2-arcsec diameter aperture), 
with the deepest panel reaching 0.3\,mag deeper.  
To create the mosaic, the astrometry of each of the four pointings was matched to that of the $Y$-band UltraVISTA imaging 
using the {\sc iraf} package {\sc ccmap}, background-subtracted using {\sc sextractor}~\citep{Bertin1996}, and 
the zero points  were equalised to take into account the different exposure time of each pointing.  A mosaic in the native pixel size (0.202\asec/pix) 
was produced using {\sc swarp}~\citep{Bertin2002}, where overlapping sections were combined with the 
{\sc weighted} keyword using weights taken from an rms map created by {\sc sextractor}.  A science image for use with the UltraVISTA datasets, 
overlapping the CFHT Legacy Survey image region with a pixel scale of 0.186\asec/pix, was created using the {\sc iraf} package {\sc sregister}.  
The full Subaru mosaic overlaps the CFHTLS field that forms the search area in this paper, with some sections extending beyond as can be seen in 
Fig.~\ref{fig:fieldmap}.

\subsection{Spitzer IRAC mid-infrared imaging}\label{sect:irac}

The full 2\,deg$^2$ COSMOS field is covered by publicly-available mid-infrared data from {\it Spitzer} IRAC and MIPS 
obtained as part of the S-COSMOS 
survey~\citep{Sanders2007}.  All channels were imaged with an integration time of 1200 seconds.  The IRAC 
$5.8\mu$m and $8.0\mu$m data are too shallow ($m_{\rm AB} = 21.3$ and $21.0$ respectively,~\citealt{Sanders2007}) 
for effective use in the present study, beyond visual inspection of our final candidate sample (see Section~\ref{sect:candsel}). 
However, the IRAC $3.6\,\mu$m and $4.5\,\mu$m images reach 5$\sigma$ depths in a 2.8\asec diameter aperture of 24.2 and 23.8\,mag 
respectively, are sufficiently well-matched to the UltraVISTA $H$ and $K_{s}$ depths to be of potential use in the 
galaxy selection process. In particular, low-redshift galaxy contaminants are predicted to be particularly 
bright at these wavelengths, in contrast to the much flatter near-infrared slope anticipated for genuine 
high-redshift galaxies.  The {\it Spitzer} IRAC photometry was obtained in a 2.8\asec diameter aperture, from 
images that had been background subtracted with a large filter length using the {\sc global} keyword in 
{\sc sextractor} and then matched to the common pixel grid and area defined above.  
In acknowledgement of the limitations of photometry 
from the confused IRAC imaging, the 3.6\,$\mu$m and 4.5\,$\mu$m photometric errors used in the photometric redshift fitting 
described below were set to a minimum level of 20\% of the measured flux density.

\section{Candidate Selection}\label{sect:candsel}

\subsection{Initial detection, photometry and depth analysis}\label{sect:initialdet}

The primary catalogue was created using {\sc sextractor} v2.8.6~\citep{Bertin1996} on the $Y$+\,$J$ stacked image, 
with photometry in the additional filters collected in the dual-image mode.  
Objects were also added to the catalogue from $Y$- and $J$-band selected catalogues, 
to ensure no galaxies were missed. However, all objects retained in the final sample were in fact selected 
from the $Y$+\,$J$ catalogue.  All optical and near-infrared photometry was measured in a 2\asec diameter circular aperture, 
which corresponds to 70\% enclosed flux (for a point source) in the $Y$-band imaging.  

The global limiting magnitude for each image was calculated by finding the sigma-clipped standard deviation of the flux enclosed 
within 2\asec (or 1.2\asec where relevant) diameter circular apertures placed on the image in a grid.  
Apertures that contain flux from an object were removed using the segmentation-map output produced by {\sc SExtractor}, 
where aggressive parameter values were chosen to ensure detection of the majority of low-significance 
objects (exceeding 2$\sigma$ significance in the detection procedure).  
For the final sample of objects presented in this paper, we undertook a local 
estimate of the photometric errors by finding the standard deviation of the 50 
closest blank apertures placed around the object, with 2.5$\sigma$ clipping.

The 5$\sigma$ depth of the $Y$+\,$J$ image is 24.9\,mag, and so, for simplicity, we cut our catalogue at $Y$+\,$J < 25.0$, 
resulting in an initial near-infrared selected sample of 175075 sources.  
To confine our search to potential objects at redshifts $z>6.0$, we then applied a non-detection ($< 2\,\sigma$) criterion in all 
optical filters blueward of and including the $i$-band.  
Removing artefacts around bright stars and the small region of missing UltraVISTA data in the corner of our central survey area, resulted in a catalogue of 146 objects with $Y$+\,$J < 25.0$ and non-detections in all optical bands except for the CFHT 
$z$-band and/or Subaru $z'$-band imaging.

\subsection{Sample refinement via SED fitting}\label{sect:refine}


\begin{figure}
\includegraphics[width = 0.5\textwidth, trim = 0.9cm 0cm 0cm 0.5cm, clip = true]{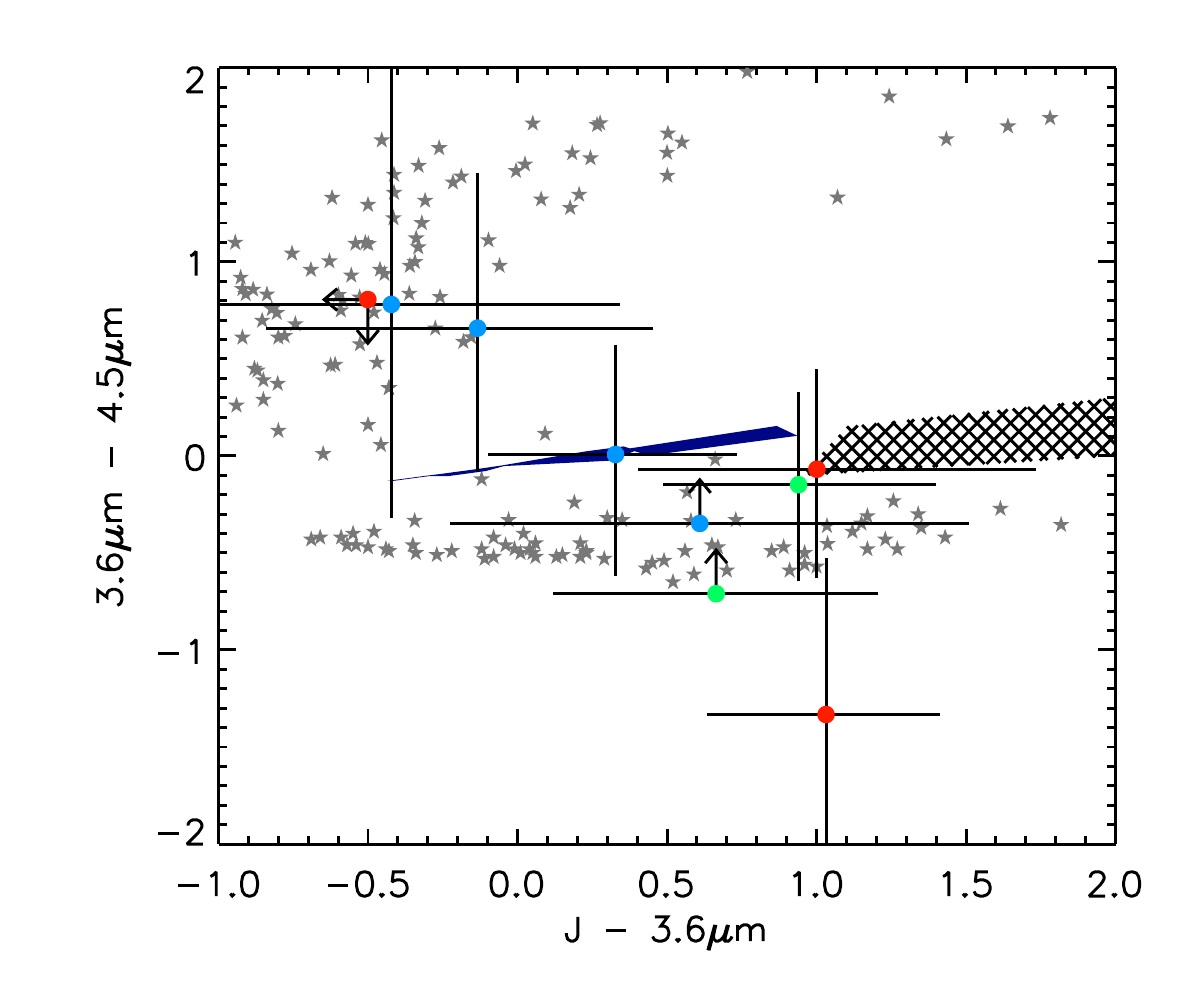}
\caption{The colours of our high-redshift galaxy candidates (shown as the coloured circles) using the Spitzer IRAC information, where the most secure candidates are shown in blue, the robust/contaminant category are shown in green and the insecure category are shown in red.
The colours of M, L and T-dwarf stars are shown with grey star symbols from~\citet{DavyKirkpatrick2011} and~\citet{Patten2006},
where the error bars are all below 0.05\,mag.  Predicted colours of low-redshift contaminant galaxies ($1.3 < z < 1.9$)
are shown by the hatched area, where reddening increases to redder $J-3.6\mu$m colours (from $A_{V}$ = 0 to $\sim$ 2 at the right-hand edge
of the plot).
The thin navy wedge represents the predicted colours of high-redshift galaxies using the models and parameter ranges as in Fig.~\ref{fig:JKz}.
Flux densities below the 2$\sigma$ level were set to the 2$\sigma$ depth and the resulting colour limits plotted as an arrow.
Note that T-dwarf stars have the reddest 3.6$\mu$m - 4.5$\mu$m colours,
whereas L and M types all have slightly bluer colours and occupy the lower horizontal band of stars.}
\label{fig:Jch1ch2}
\end{figure}

In contrast to other studies which have endeavoured to select bright $z \simeq 7$ galaxies from ground-based data, 
we do not attempt to define areas in colour-colour space to isolate high-redshift galaxies.  Fig.~\ref{fig:JKz} 
clearly illustrates the difficulties in separating high-redshift galaxies from dwarf-star contaminants 
based on $z', J$ and $K_s$ photometry alone.  
The addition of $Y$-band data is crucial (as shown in Fig. 2b), but again the use of only two colours 
does not make optimal use of the available multi-wavelength data. Therefore, as in our previous studies (e.g.~McLure et al. 2009; 2010; 2011) 
we have used a full SED-fitting analysis 
to derive the redshift-probability distribution for each galaxy candidate.  We employed the~\citet{Bruzual2003} stellar evolution models, considering models with metallicities ranging from solar ($Z_{\sun}$) to 1/50th solar (0.02$Z_{\sun}$).  The star-formation histories considered were instantaneous bursts, constant and exponentially declining with  characteristic timescales in the range 50 Myr $< \tau < $ 10 Gyr.  The ages of the stellar populations models were allowed to range from 10 Myr to 13.7 Gyr, but were required to be less than the age of the Universe at each redshift.    Dust reddening was described by the~\citet{Calzetti2000} attenuation law, and allowed to vary within the range 0.0 $\leq A_{\rm V} \leq$ 4.0.  Inter-galactic medium absorption short-ward of \Lya was described by the~\citet{Madau1995} prescription, and a~\citet{Chabrier2003} IMF was assumed in all cases.  Further details can be found in~\citet{McLure2011}. Specifically for this study we have included detailed fitting of the spectra of L, M and T dwarf stars, 
where the reference spectra from the speX library\footnote{\url{http://pono.ucsd.edu/~adam/browndwarfs/spexprism/}} 
were used for each spectral type from M4 through to T8.  Only for the final sample of galaxies presented in 
Section~\ref{sect:cand} did we introduce the equivalent width ($EW$) of the \Lya-line in the 
SED-fitting analysis as an extra free parameter, with the rest-frame equivalent width within the range $0 $~\AA\, $ <   EW_0 < 240~$\AA.

SED fitting was performed using all the multi-wavelength data, including the CFHTLS $z$-band data (although it is 
significantly shallower than the Subaru $z'$-band image, it has a slightly bluer effective wavelength and thus 
adds some extra spectral resolution to the fitting process). The IRAC photometry was only included in the 
final stages of candidate selection due to the large uncertainties involved and the high number of objects where the photometry is confused.  All near- and far-infrared photometry was corrected to the 84\% enclosed flux level of the $z'$-band imaging for the SED fitting, where the enclosed flux within a 2\asec diameter aperture 
was calculated from the point spread function obtained by taking the median of unsaturated stars extracted from each image.

The output redshift-$\chi^2$ distribution from the SED fitting (without the IRAC photometry) 
was used to determine which of the 146 candidates were consistent with being at high redshift.  
We applied the criteria that the object must have an acceptable solution ($ \chi^2 < 11.3$, which corresponds to $2\sigma$ significance given 5 degrees of freedom) at $z > 6.0$, and that 
the $z > 6.0$ template must be preferred ($ \Delta \chi^2 > 0.0$) over the alternative low-redshift galaxy solution (usually produced at a redshift 
where the putative Lyman-break can be interpreted as a Balmer or 4000\AA\ break).  
These conditions resulted in a reduced sample of 68 objects consistent with being at $ z > 6.0$.  
Further inspection of the $i$-band image, in combination with the 
HST/ACS $I_{814}$ image and an inverse-variance weighted stack of the optical bands up 
to and including $i$, resulted in the removal of a further 21 objects with weak detections in these bands implying $z < 6.5$.

As a result of the SED fitting and manual optical checks, the sample of potential $z > 6.5$ galaxies was reduced to 47 objects.
However many of these still had formally-acceptable solutions at much lower redshift. At this point the IRAC $3.6\,\mu$m 
and $4.5\,\mu$m measurements were incorporated into the SED-fitting analysis, in effect to remove the red dusty low-redshift galaxy interlopers
which display a much redder $J - 3.6\mu$m colour than genuine high-redshift galaxies. In practice the IRAC photometry was simply 
incorporated into the full SED fitting, but the process is illustrated in colour-colour space in Fig.~\ref{fig:Jch1ch2} which shows the
(hashed) area at $J-3.6\,\mu$m $ > 1$ where low-redshift dusty interlopers are generally found. At this stage all the observed 
optical--to--near-infrared SEDs were also fitted 
to the SEDs of L, M, T dwarf stars and objects were rejected if the dwarf-star solution 
was formally preferred over the high-redshift galaxy solution. 
The lack of mid-infrared spectroscopy of dwarf stars prevented the simple incorporation of the IRAC measurements
into the dwarf-star SED-fitting process, but we were able to confirm a number of ambiguous dwarf-star solutions
on the basis of the position of the object on the colour-colour plots shown in Fig.~\ref{fig:JKz} and Fig.~\ref{fig:Jch1ch2}. 
The final result of this IRAC-based cull was a remaining sample of 17 potential $z > 6.5$ galaxies.

The final stage of sample refinement involved detailed re-investigation of every single galaxy image, and
repetition of the dwarf-star fitting using photometry taken in a 
smaller 1.2\asec diameter circular aperture (with the expectation that, in the case of a real star, the fit should improve with the 
smaller aperture photometry). Galaxy colours were also calculated from smaller-aperture photometry for comparison.  
As a consequence of these final checks, three objects were excluded because of low-level flux detected in the optical 
stack, in combination with insecure photometry that resulted in the objects drifting substantially around the 
colour-colour diagrams when smaller apertures were used.  A further two objects were only detected in the 
$Y$-band image, and closer inspection revealed them to be part of a faint diffraction halo around a 
bright star. Finally two objects were confirmed as secure $z > 6$ galaxies, but their redshift-probability distributions 
indicated that they could not lie above $z = 6.5$.

The final outcome is a sample of 10 objects which have survived as credible candidate Lyman-break galaxies at $z > 6.5$, 
and which we deem worthy of presentation in this paper.  The magnitude errors and SED fits for these objects presented in subsequent sections 
are all based on the local error estimates (although these are all formally consistent with the global error estimates).  
We visually inspected the $5.8\,\mu$m and $8.0\,\mu$m \emph{Spitzer}/IRAC imaging for our final 10 candidates, 
finding no evidence for even low-level flux (as expected given the magnitudes predicted by the SED fits, and 
the depths of the S-COSMOS IRAC imaging at these longer wavelengths; see Section~\ref{sect:irac}).

\section{Candidate ${\bf \lowercase{z} > 6.5}$ Galaxies}\label{sect:cand}

Table~\ref{table:phot} details the photometry of the final ten $z > 6.5$ galaxy candidates, where  
we have grouped them into three categories depending on the statistical security of their $z > 6.5$ solutions.
The final galaxy and dwarf-star SED fits for these ten candidates are presented in Fig.~\ref{fig:SEDfits}. Their 
photometric redshifts and derived physical properties such as stellar masses and star-formation rates are presented in Table~\ref{table:properties}.

\begin{figure*}
\includegraphics[width = 1.0\textwidth]{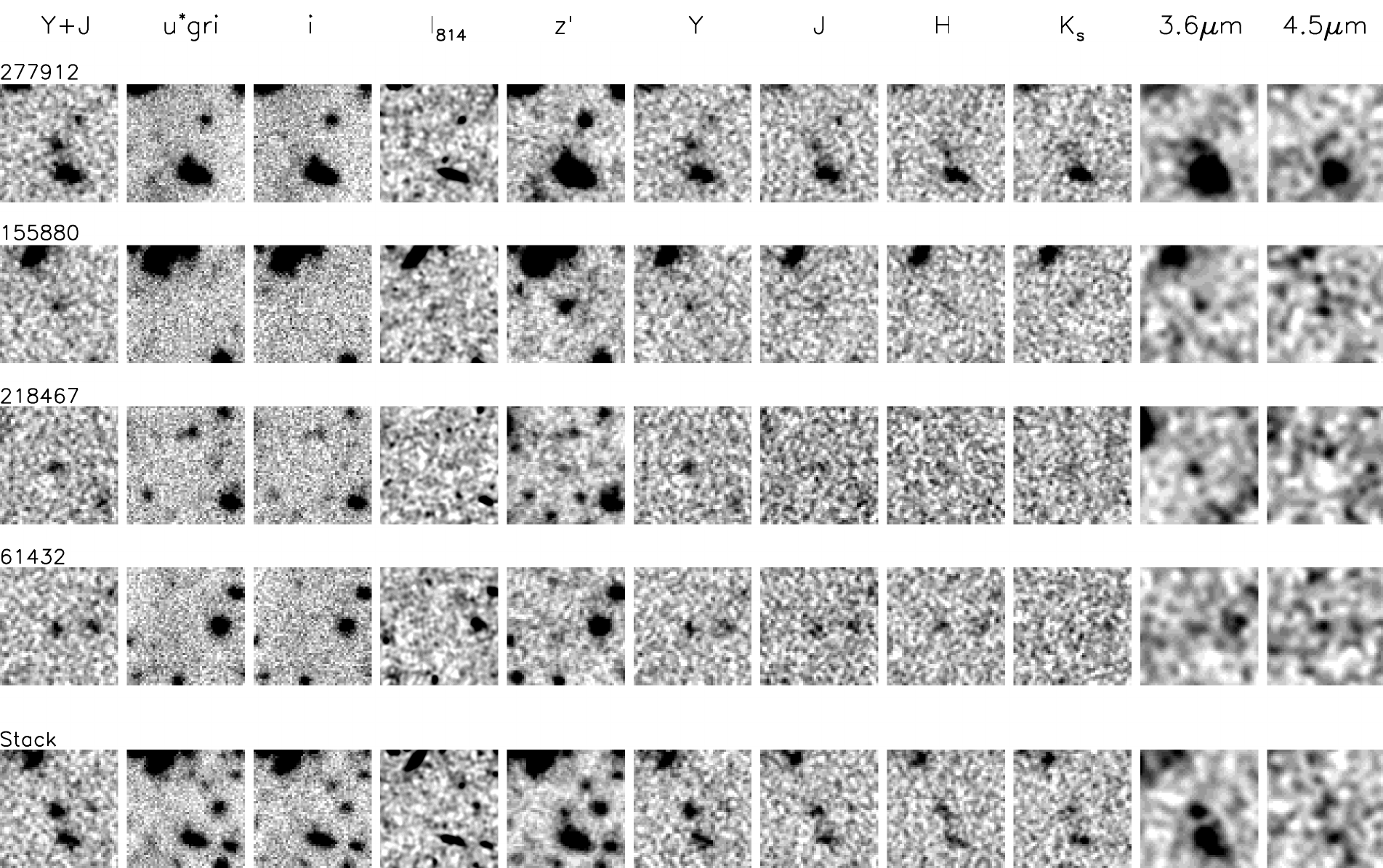}
\caption{Multi-band postage-stamp images of the four most secure $z \simeq 7$ galaxies presented in this paper, 
with the detection image ($Y + J$) shown on the left.  A stack of these four objects is presented in the bottom row.  
The stamps are 10\asec $\times$ 10\asec, with North up, and East to the left.  Each filter stamp has a linear grayscale, 
with the lower level (white) set to 2$\sigma$ below the background level and objects (black) saturated at 4.5$\sigma$ above the background level.  To aid the visual detection of low-level flux indicative of a $z < 6.5$ contaminant object, the $I_{814}$ stamp has been smoothed using the {\sc iraf gauss} package, with a standard deviation of 5 pixels.}
\label{fig:stamps}
\end{figure*}

\begin{table*}
 \centering
  \caption{Photometry of the ten $z > 6.5$ galaxy candidates, grouped into the three categories described in the main text.  
From top to bottom the top four galaxies are classed as ``Robust" (followed by the photometry for a stack of this subset of galaxies), the next three as ``Robust/Contaminant", and the final three can be regarded 
as ``Insecure".  The optical and near-infrared 
magnitudes were measured within a 2\asec diameter circular aperture and corrected to 84\% enclosed flux, and the IRAC magnitudes were 
measured within a 2.8\asec diameter circular aperture and corrected to 84\% enclosed flux.  As a consequence of the correction up to 84\% enclosed flux, 
the $Y$- and $J$-band magnitudes presented here appear typically $\sim 0.2$ brighter than the quoted 2-arcsec diameter $5\sigma$ limit (see Table~\ref{table:depths}).  
Wherever the measured flux density lies 
below the 2$\sigma$ limit in that band, the magnitude has been replaced here by the 2$\sigma$ limiting magnitude depth of the appropriate image, and is given 
as a lower limit on the apparent magnitude. }
 \begin{tabular}{l| c| c| c|c|c|c|c|c|c}

\hline


ID & RA & DEC & $z'$ & $Y$ & $J$ & $H$ & $K_s$ & $3.6\mu$m & $4.5\mu$m  \\
  
  \hline
  
    277912  &  10:00:43.38  &  +02:37:51.8 &  $26.7_{- 0.3}^{+ 0.4}  $ &  $24.2_{- 0.1}^{+ 0.2}  $ &  $24.3_{- 0.2}^{+ 0.2}  $ &  $24.0_{- 0.2}^{+ 0.3}  $ &  $23.9_{- 0.2}^{+ 0.3}  $ &  $23.9_{- 0.2}^{+ 0.3}  $ &  $23.9_{- 0.3}^{+ 0.4}  $ \\
      155880  &  10:02:06.49  &  +02:13:24.1 &  $26.1_{- 0.1}^{+ 0.1}  $ &  $24.5_{- 0.2}^{+ 0.2}  $ &  $24.4_{- 0.2}^{+ 0.2}  $ &  $24.4_{- 0.3}^{+ 0.4}  $   &  $>24.8$ &  $24.5_{- 0.3}^{+ 0.5}  $ &  $23.9_{- 0.3}^{+ 0.4}  $ \\
      218467  &  10:01:52.31  &  +02:25:42.3   &  $>27.3$ &  $24.6_{- 0.2}^{+ 0.2}  $ &  $25.1_{- 0.4}^{+ 0.6}  $   &  $>25.0$ &  $24.3_{- 0.3}^{+ 0.4}  $ &  $24.4_{- 0.3}^{+ 0.5}  $   &  $>24.8$ \\
       61432  &  10:01:40.70  &  +01:54:52.5   &  $>27.3$ &  $24.6_{- 0.2}^{+ 0.2}  $ &  $24.6_{- 0.2}^{+ 0.3}  $ &  $24.4_{- 0.3}^{+ 0.4}  $   &  $>24.8$ &  $25.0_{- 0.5}^{+ 1.0}  $ &  $24.2_{- 0.4}^{+ 0.6}  $ \\
     
     \hline
    Stack  &    &   &  $27.0_{- 0.2}^{+ 0.2}  $ &  $24.5_{- 0.1}^{+ 0.1}  $ &  $24.6_{- 0.1}^{+ 0.1}  $ &  $24.4_{- 0.2}^{+ 0.2}  $ &  $24.6_{- 0.2}^{+ 0.3}  $ &  $24.4_{- 0.2}^{+ 0.2}  $ &  $24.6_{- 0.3}^{+ 0.4}  $ \\

       \hline
       
      277880  &  10:01:36.86  &  +02:37:49.3 &  $26.4_{- 0.2}^{+ 0.2}  $ &  $24.7_{- 0.2}^{+ 0.2}  $ &  $24.7_{- 0.2}^{+ 0.3}  $ &  $24.9_{- 0.5}^{+ 1.0}  $ &  $24.0_{- 0.3}^{+ 0.4}  $ &  $24.1_{- 0.2}^{+ 0.3}  $   &  $>24.8$ \\
      268511  &  10:00:02.36  &  +02:35:52.2   &  $>27.3$ &  $24.7_{- 0.2}^{+ 0.3}  $ &  $24.8_{- 0.3}^{+ 0.4}  $ &  $24.8_{- 0.3}^{+ 0.5}  $   &  $>24.8$   &  $>25.2$   &  $>24.8$ \\
      271105  &  09:59:07.60  &  +02:36:24.4 &  $26.1_{- 0.2}^{+ 0.2}  $ &  $24.7_{- 0.2}^{+ 0.2}  $ &  $24.4_{- 0.2}^{+ 0.3}  $ &  $24.2_{- 0.2}^{+ 0.2}  $ &  $24.5_{- 0.3}^{+ 0.5}  $ &  $23.5_{- 0.2}^{+ 0.2}  $ &  $23.6_{- 0.2}^{+ 0.3}  $ \\
      
      \hline
      
       95661  &  10:01:20.70  &  +02:01:44.0 &  $25.4_{- 0.1}^{+ 0.1}  $ &  $24.6_{- 0.3}^{+ 0.5}  $ &  $24.7_{- 0.4}^{+ 0.5}  $ &  $25.0_{- 0.5}^{+ 1.0}  $ &  $24.3_{- 0.3}^{+ 0.4}  $ &  $23.7_{- 0.2}^{+ 0.2}  $ &  $23.8_{- 0.3}^{+ 0.4}  $ \\
       28400  &  10:00:58.01  &  +01:48:15.6 &  $25.3_{- 0.1}^{+ 0.1}  $ &  $24.6_{- 0.1}^{+ 0.2}  $ &  $24.4_{- 0.2}^{+ 0.2}  $ &  $24.5_{- 0.3}^{+ 0.4}  $   &  $>24.8$ &  $23.4_{- 0.2}^{+ 0.2}  $ &  $24.7_{- 0.6}^{+ 1.2}  $ \\
        2233  &  10:01:43.16  &  +01:42:53.5 &  $26.0_{- 0.1}^{+ 0.2}  $ &  $24.7_{- 0.3}^{+ 0.4}  $ &  $24.7_{- 0.4}^{+ 0.6}  $   &  $>25.0$   &  $>24.8$   &  $>25.2$ &  $24.4_{- 0.4}^{+ 0.7}  $ \\

  \hline
  
     \end{tabular}
 \label{table:phot}
\end{table*}

\subsection{Category 1 - Robust}\label{sect:robust}

Our four most secure $z > 6.5$ galaxy candidates have completely 
unacceptable alternative fits for either a low-redshift dusty galaxy or a late-type dwarf star from the SED fitting analysis, 
and are all very well described by a high-redshift galaxy template at $z > 6.5$, as can been seen in Fig.~\ref{fig:SEDfits}.

Object 277912 has $z'-Y = 2.5$ and a flat near-infrared SED through to the IRAC bands, with a best-fitting spectral template 
corresponding to a galaxy with $z_{\rm phot} = 6.97^{+0.06}_{-0.07}$. As can be seen from the postage stamps presented in Fig.~\ref{fig:stamps}, 
a low-redshift galaxy is present $\sim$2.5\asec\,away from the position of our high-redshift candidate in the $Y$-band imaging.  
The companion has a photometric redshift of $z_{\rm phot} = 0.94 \pm 0.10$ from the COSMOS Photometric Redshift Catalogue~\citep{Ilbert2008} 
and $z_{\rm phot} = 1.18 \pm 0.10$ from our own fitting of the 2\asec diameter aperture photometry.  
The $\chi^2$ distribution for object 277912 indicates that the low-redshift solution is not only extremely 
unlikely, but would also lie at $z \sim 1.7$. Hence it seems clear that this object is at high redshift and 
not associated with the low-redshift foreground galaxy.  
Unfortunately the presence of the foreground galaxy means that the IRAC photometry, 
where the FWHM of the PSF is $\sim$ 1.7\asec, is confused, and therefore the stellar mass estimate is uncertain.
Allowance for a possible contribution from Lyman-$\alpha$ emission in the SED fitting does not change the photometric redshift,
with zero Lyman-$\alpha$ emission remaining the preferred option.

Object 155880 also has a blue spectral slope in the near-infrared bands 
and is clearly detected in the deep Subaru $z'$-band imaging giving a $z_{\rm phot} = 6.78^{+0.06}_{-0.09}$.  
We stress that the $z'$-band detection ($z' = 26.1\pm0.1$) is achieved because the Subaru Suprime-CAM $z'$-band imaging is so deep
and the filter extends to $\sim 1\mu$m; the measured spectral break is still large $z'-Y > 1.5$, 
although not large enough to warrant inclusion as a high-redshift candidate in some traditional colour-colour cut 
selection methods (e.g. $z' - Y/J \gtrsim 2$).  
Fitting with the addition of a \Lya line results in a slightly higher-redshift fit with $z_{\rm phot} = 6.98$ (with a Ly$\alpha$-line $EW_0 = 40~$\AA).  
The best-fitting stellar template of an earlier type T-dwarf star (T3) can produce the $z'-Y$ colour, but not the other near-infrared colours,
with the result that the high-redshift galaxy solution is the only acceptable SED fit.  
The IRAC $3.6\mu$m - $4.5\mu$m = 0.6 colour is redder than would be predicted directly from constant SFR models, but 
there is evidence (Curtis-Lake et al., in preparation) that a multiple-component star-formation history or a significant contribution from 
nebular emission can reproduce this spectral shape.  
Inspection of the smoothed ACS $I_{814}$-band imaging shown in Fig.~\ref{fig:stamps} shows a low-significance ($2.3\sigma$)
detection 0.5\asec\,from our object. However, given that our astrometric accuracy is better than 0.2\asec,
we conclude that this detection is unrelated to our candidate.

Object 218467 has the bluest $Y-J$ colour of all our candidates as shown in the colour-colour plot in Fig.~\ref{fig:JKz}, 
which results in unacceptable SED fits for both the low-redshift and stellar templates, 
as neither can reproduce both the magnitude drop between $z'$ and $Y$ and the blue spectral slope at longer wavelengths.  
The $J$-band detection is below the 5$\sigma$ limit of the data and hence there is a large uncertainty in the extremely blue $Y-J = -0.5$ 
colour (bluer than predicted by our range of models; see Fig.~\ref{fig:JKz}).  
The candidate is formally undetected in the current $K_s$ and $H$-band imaging which leads to a large uncertainty on these data points in the 
SED fitting. However, when a smaller 1.2\asec diameter aperture is used, marginal detections produce a flat spectral shape as 
seen in the dwarf-star SED fit figure.  
A faint detection in the $z'$-band is consistent with the $z_{\rm phot} = 7.04^{+0.10}_{-0.08}$ 
within the errors, although the detection here is best fitted with inclusion of a strong \Lya line of $EW_0 = 110$\,\AA\, at the very red end of the 
Subaru $z'$-band filter at $z_{\rm phot} = 7.20$.  
Fig.~\ref{fig:stamps} shows a detection in the IRAC 3.6\,$\mu$m image 
consistent with a flat SED extending from the near-infrared data, 
but the 4.5\,$\mu$m image appears to have a negative hole at the position of our object.  
The implied stellar mass of 218467 is the lowest of the four robust candidates, 
indicating that it may be underestimated as a result of the potentially spurious lack of 4.5\,$\mu$m flux.  
Alternatively, the implied blue 3.6$\mu$m - 4.5$\mu$m slope could indicate the presence of strong nebular emission lines.

\begin{figure*}
\subfloat{\includegraphics[width=0.38\textwidth, trim = 1cm 8cm 10cm 4cm, clip = true]{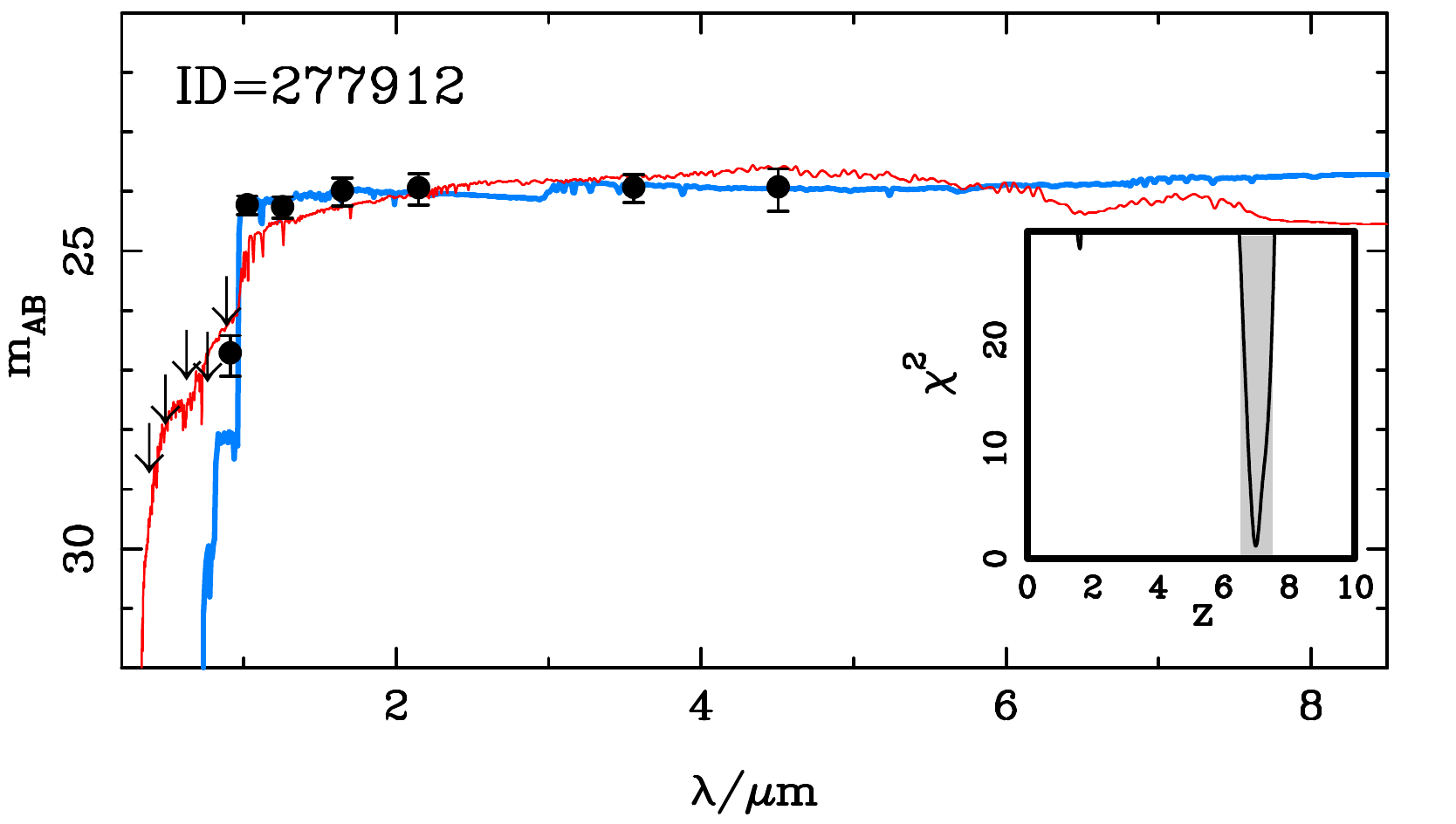}}\hspace{1cm}
\subfloat{\includegraphics[width=0.38\textwidth, trim = 1cm 8cm 10cm 4cm, clip = true]{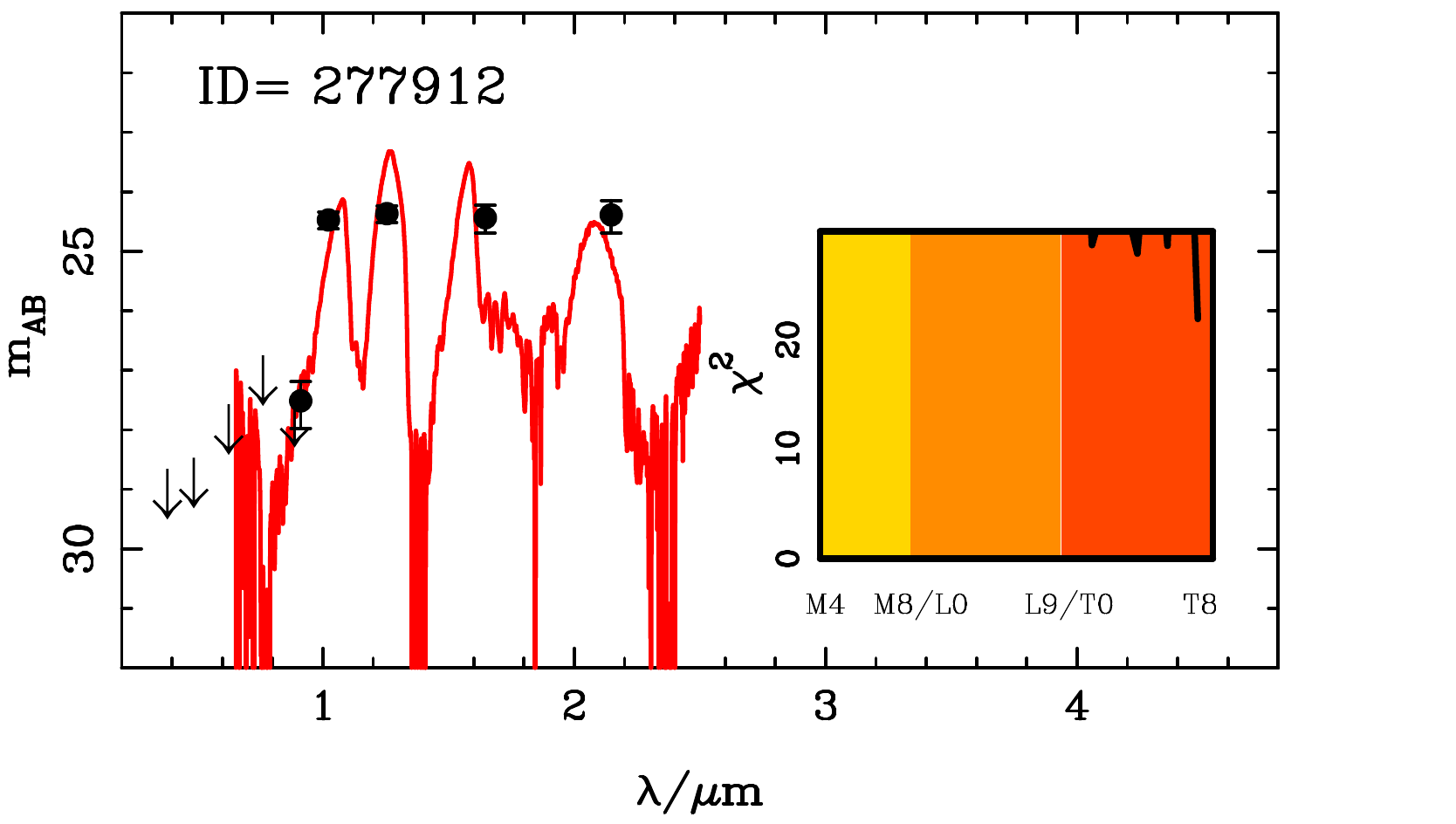}}\\
\subfloat{\includegraphics[width=0.38\textwidth, trim = 1cm 8cm 10cm 4cm, clip = true]{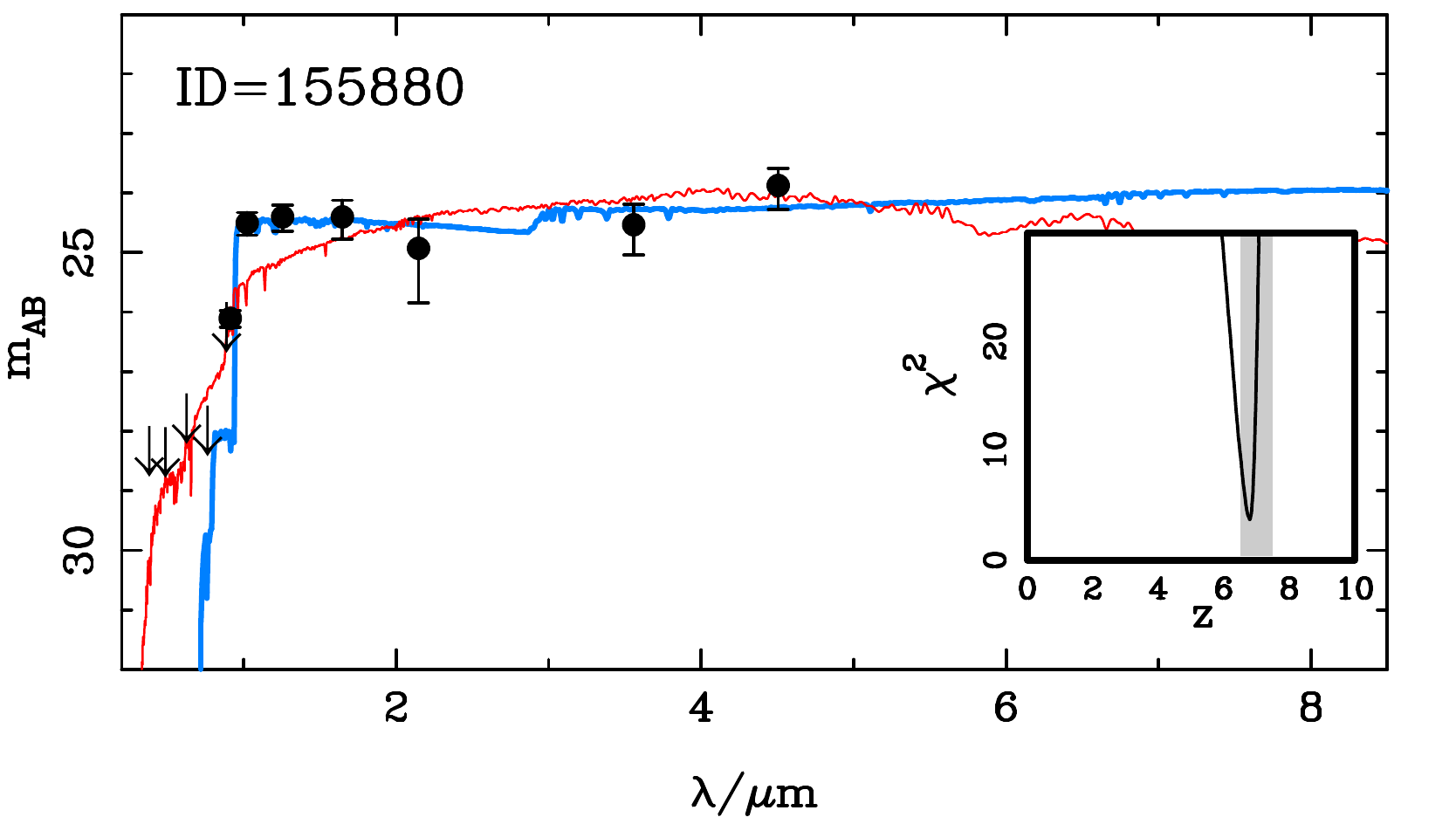}}\hspace{1cm}
\subfloat{\includegraphics[width=0.38\textwidth, trim = 1cm 8cm 10cm 4cm, clip = true]{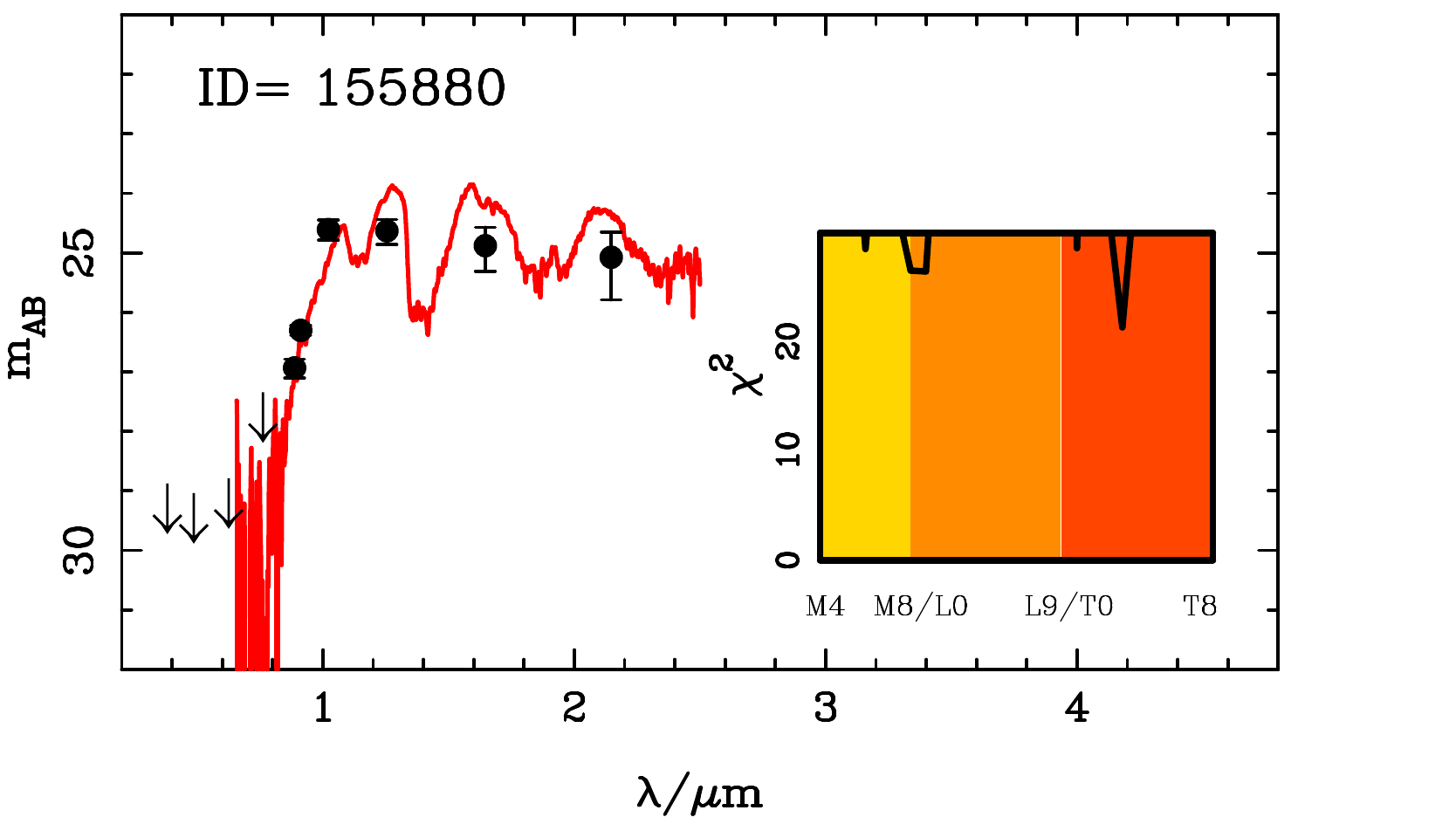}}\\
\subfloat{\includegraphics[width=0.38\textwidth, trim = 1cm 8cm 10cm 4cm, clip = true]{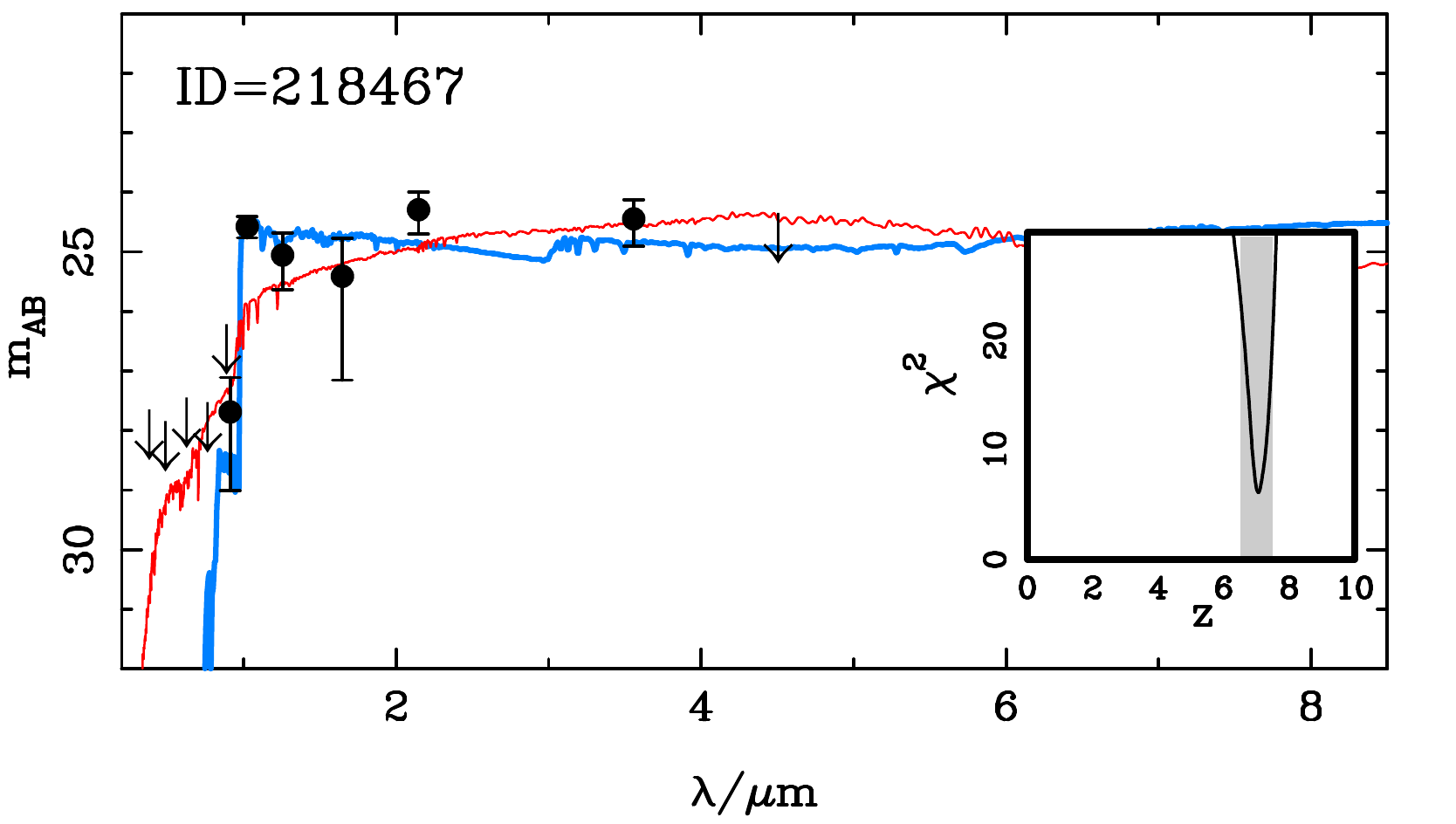}}\hspace{1cm}
\subfloat{\includegraphics[width=0.38\textwidth, trim = 1cm 8cm 10cm 4cm, clip = true]{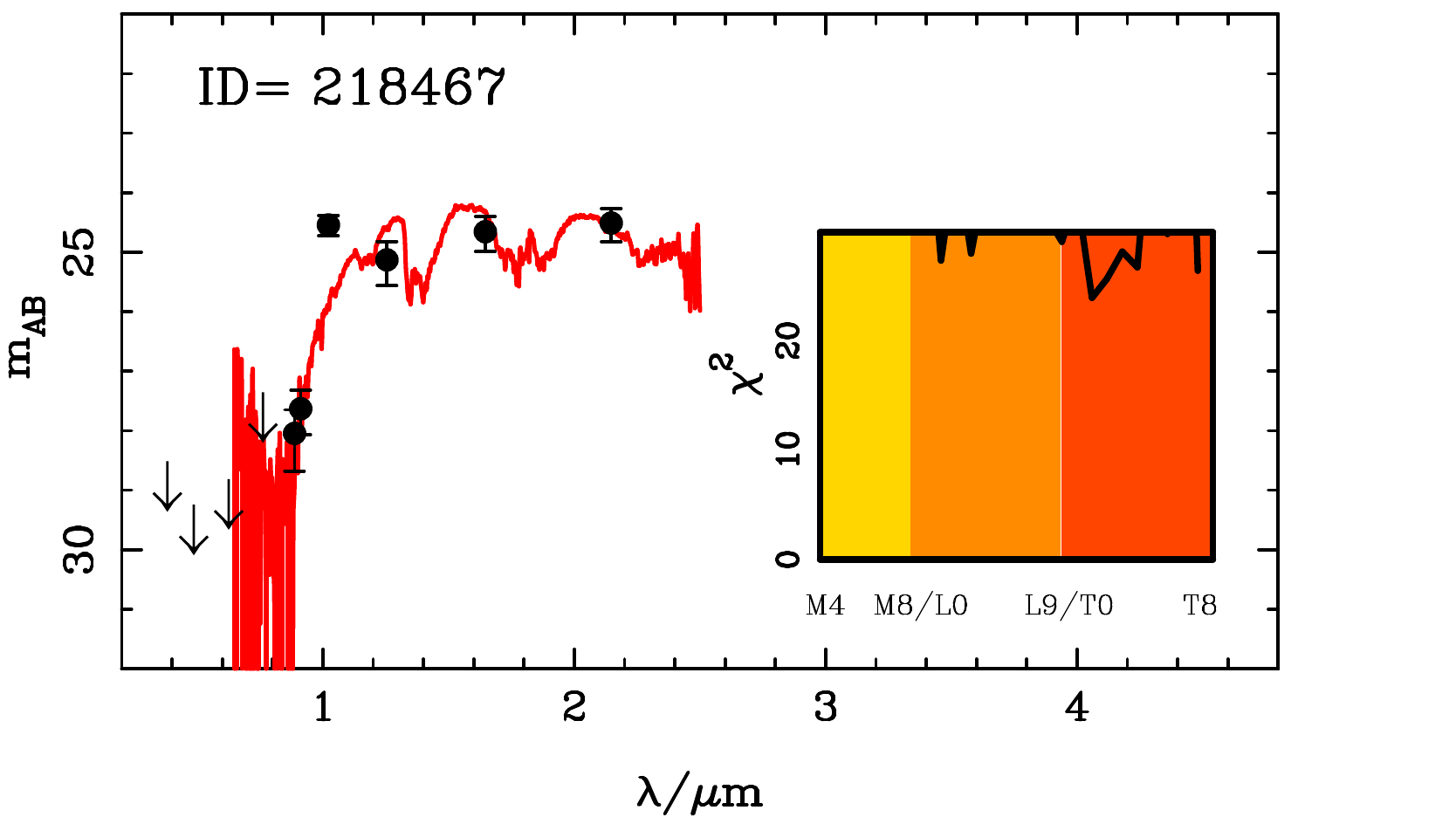}}\\
\subfloat{\includegraphics[width=0.38\textwidth, trim = 1cm 8cm 10cm 4cm, clip = true]{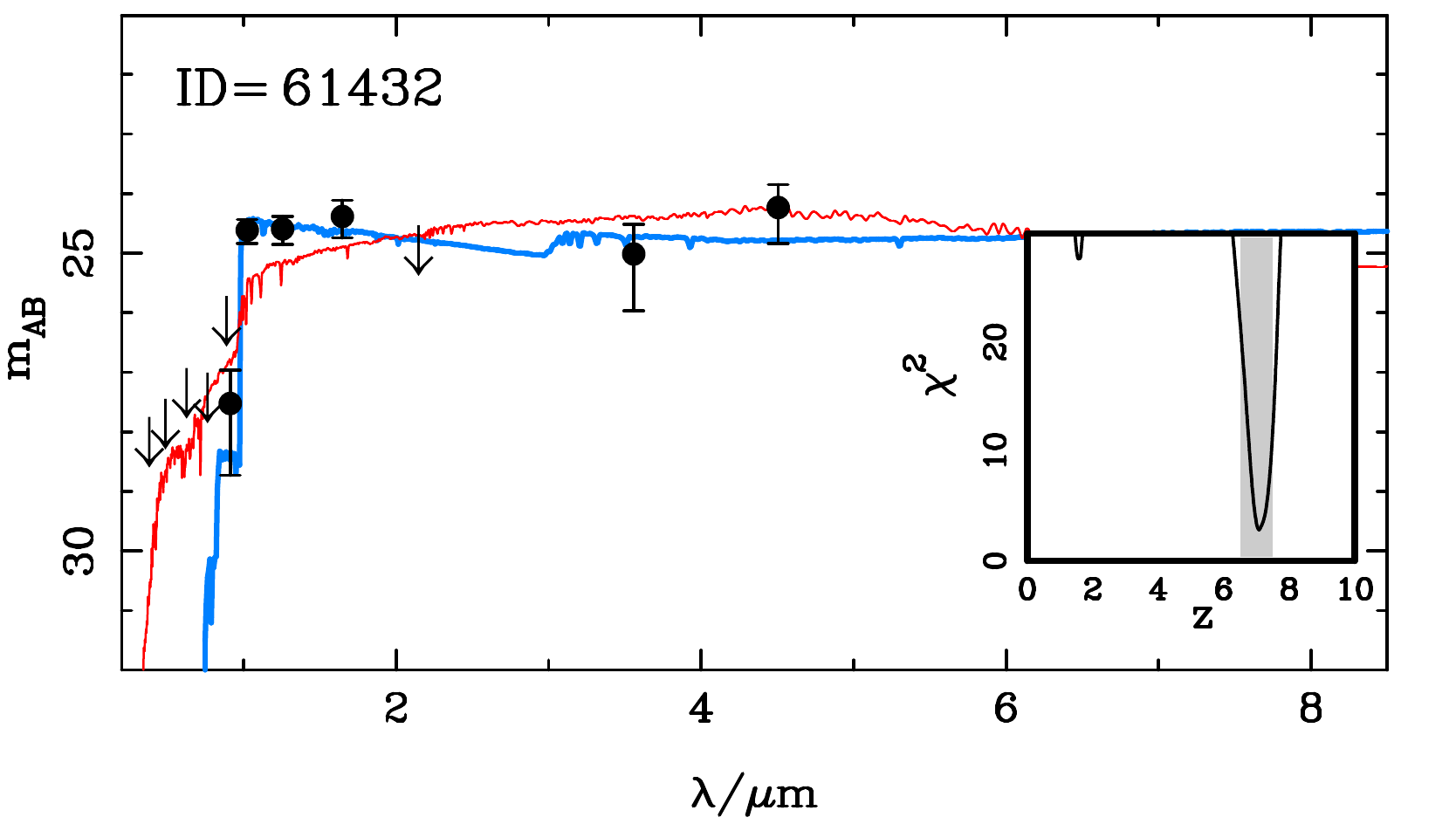}}\hspace{1cm}
\subfloat{\includegraphics[width=0.38\textwidth, trim = 1cm 8cm 10cm 4cm, clip = true]{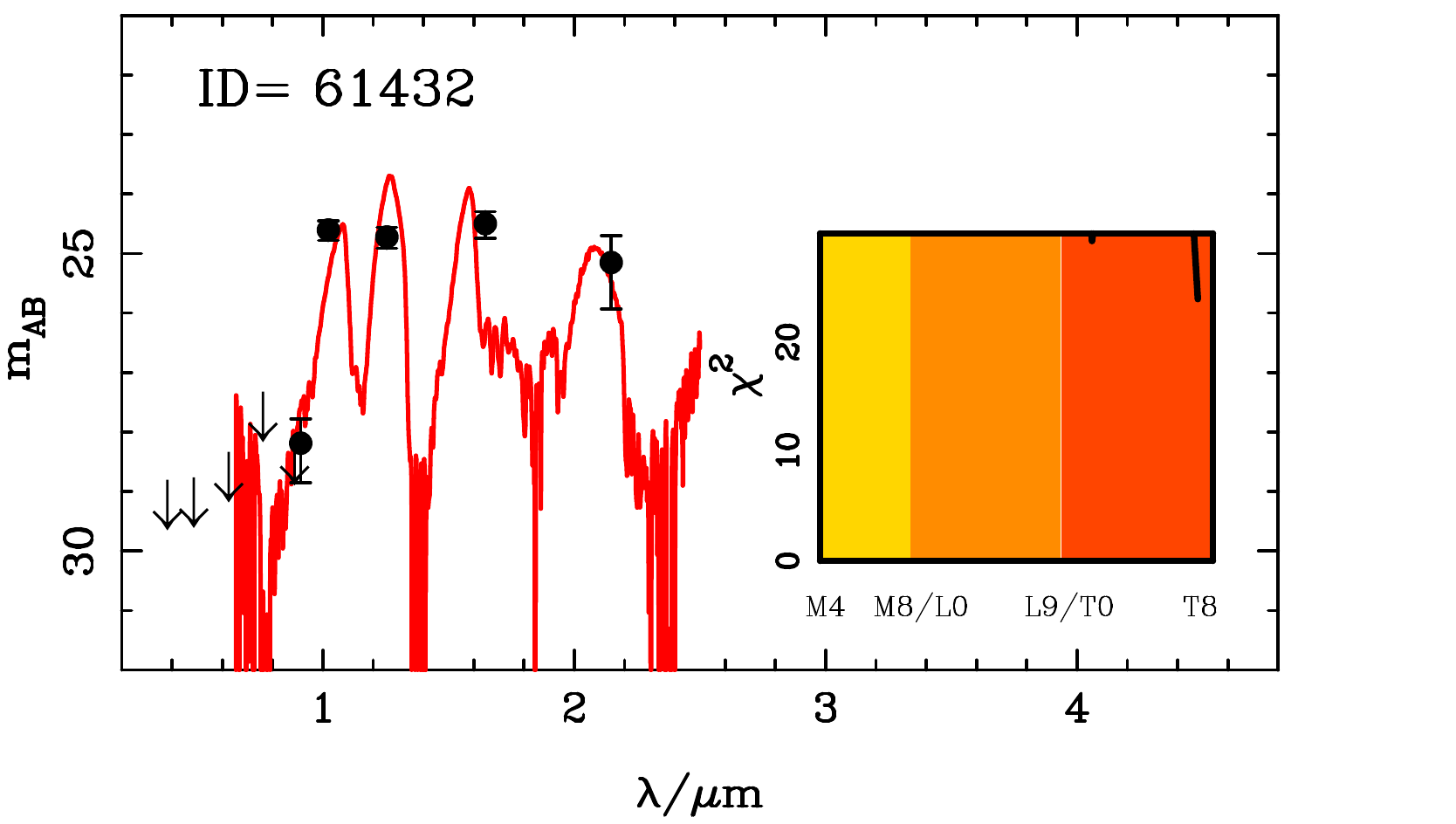}}\\
\subfloat{\includegraphics[width=0.38\textwidth, trim = 1cm 8cm 10cm 4cm, clip = true]{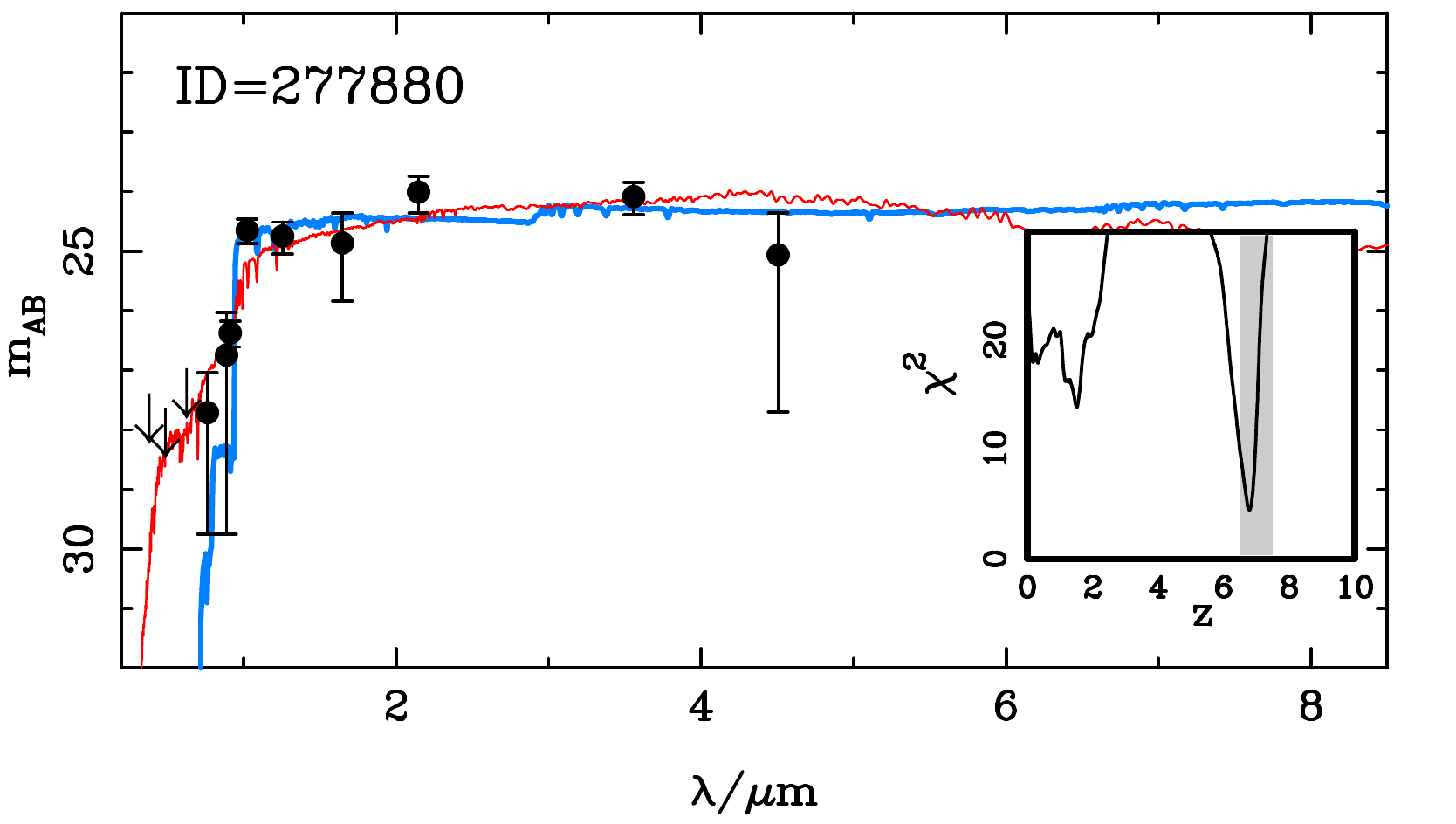}}\hspace{1cm}
\subfloat{\includegraphics[width=0.38\textwidth, trim = 1cm 8cm 10cm 4cm, clip = true]{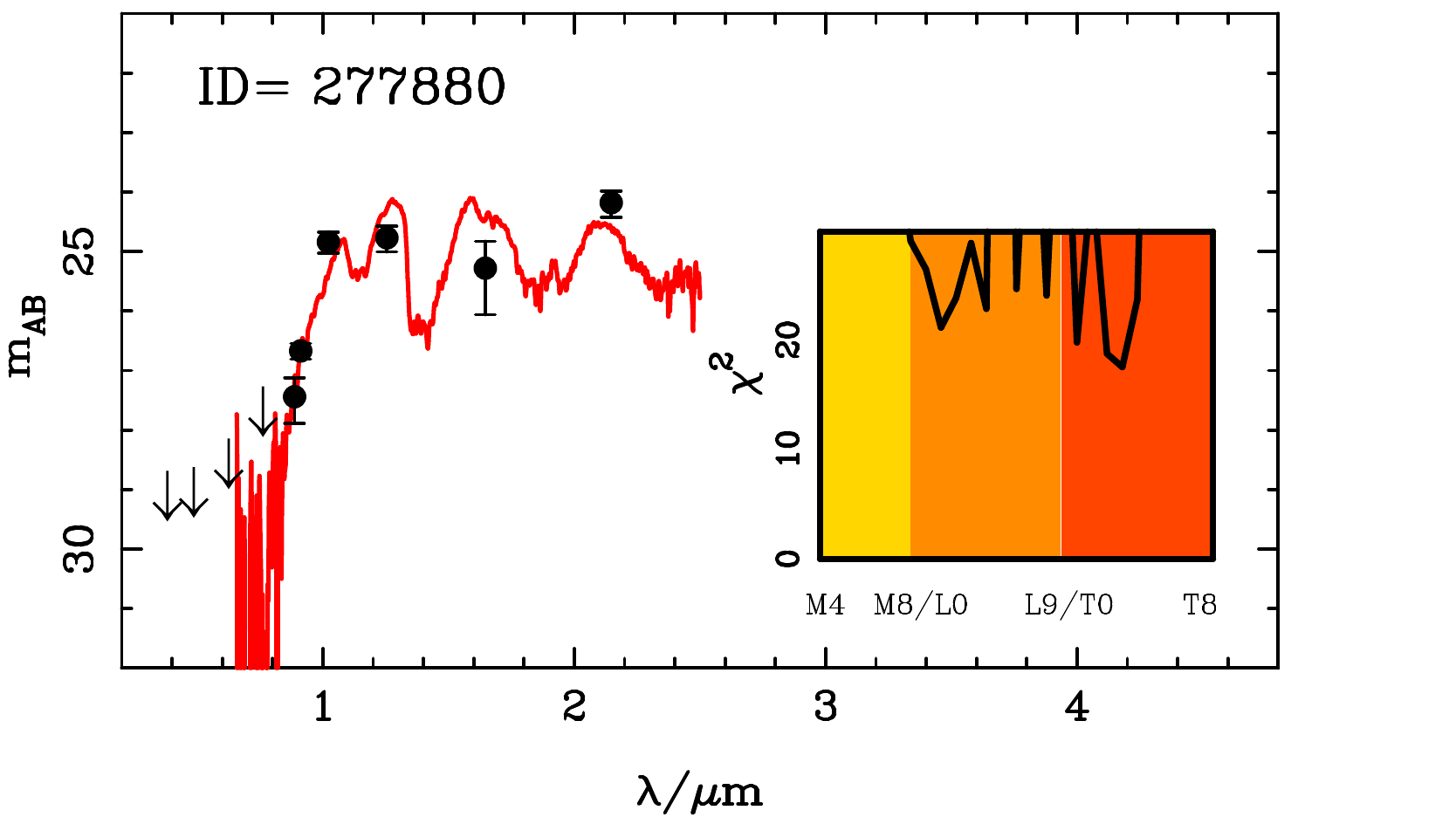}}\\

\caption{SED fits for each member of the final sample of ten high-redshift galaxies presented in this paper.  Where the candidate has a non-detection (below the $1\sigma$ level) in a given band, that point has been plotted with an arrow where the tip represents the $1\sigma$ limiting depth.  The best-fitting high-redshift 
galaxy template (without Ly$\alpha$) is shown in blue in the left hand panel, with the best-fitting alternative low-redshift template shown in red.  
The inset shows the redshift-$\chi^2$ distribution produced in the fitting process, where the redshift range $6.5 < z < 7.5$ targeted in this study is highlighted in grey.  The best-fitting low-redshift galaxy templates  
have a redshift around $z \simeq 1.2-1.6$, due to the fact that a 4000\AA/Balmer break in the model galaxy spectra can sometimes 
reproduce the observed spectral break at $\sim 1\mu$m if the data have insufficient dynamic range.
For each object the best-fitting stellar template is shown in the right-hand panel, where photometry measured in a 1.2\asec diameter aperture 
has been used.  The inset for the dwarf-star plots shows the $\chi^2$ distribution versus stellar type, with effective temperature decreasing 
from M4 through to T8.  The resulting best-fit photometric redshift, stellar type and derived parameters are detailed in Table~\ref{table:properties}.}
\label{fig:SEDfits}
\end{figure*}

\setcounter{figure}{4}  
\begin{figure*}
\subfloat{\includegraphics[width=0.38\textwidth, trim = 1cm 8cm 10cm 4cm, clip = true]{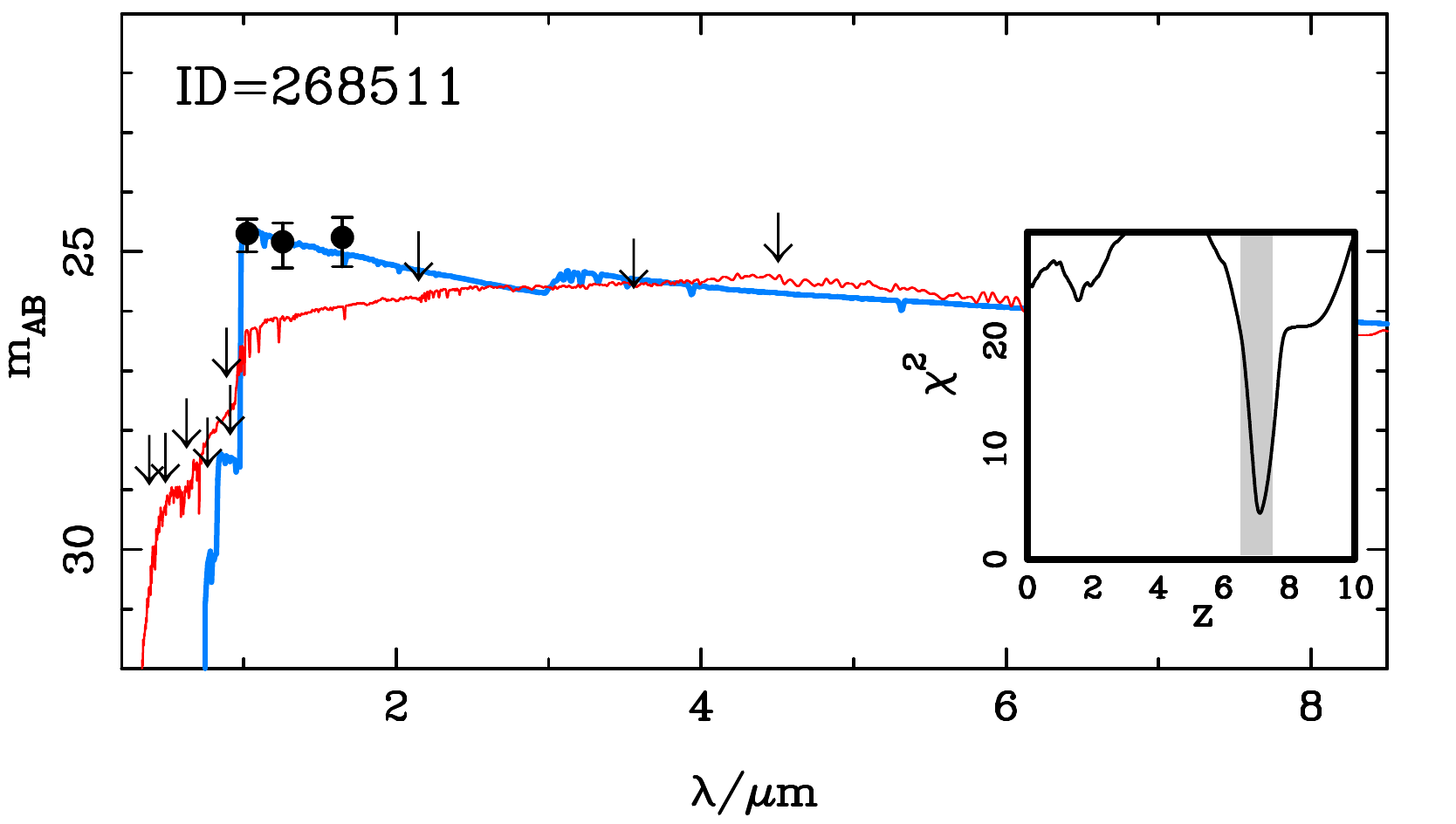}}\hspace{1cm}
\subfloat{\includegraphics[width=0.38\textwidth, trim = 1cm 8cm 10cm 4cm, clip = true]{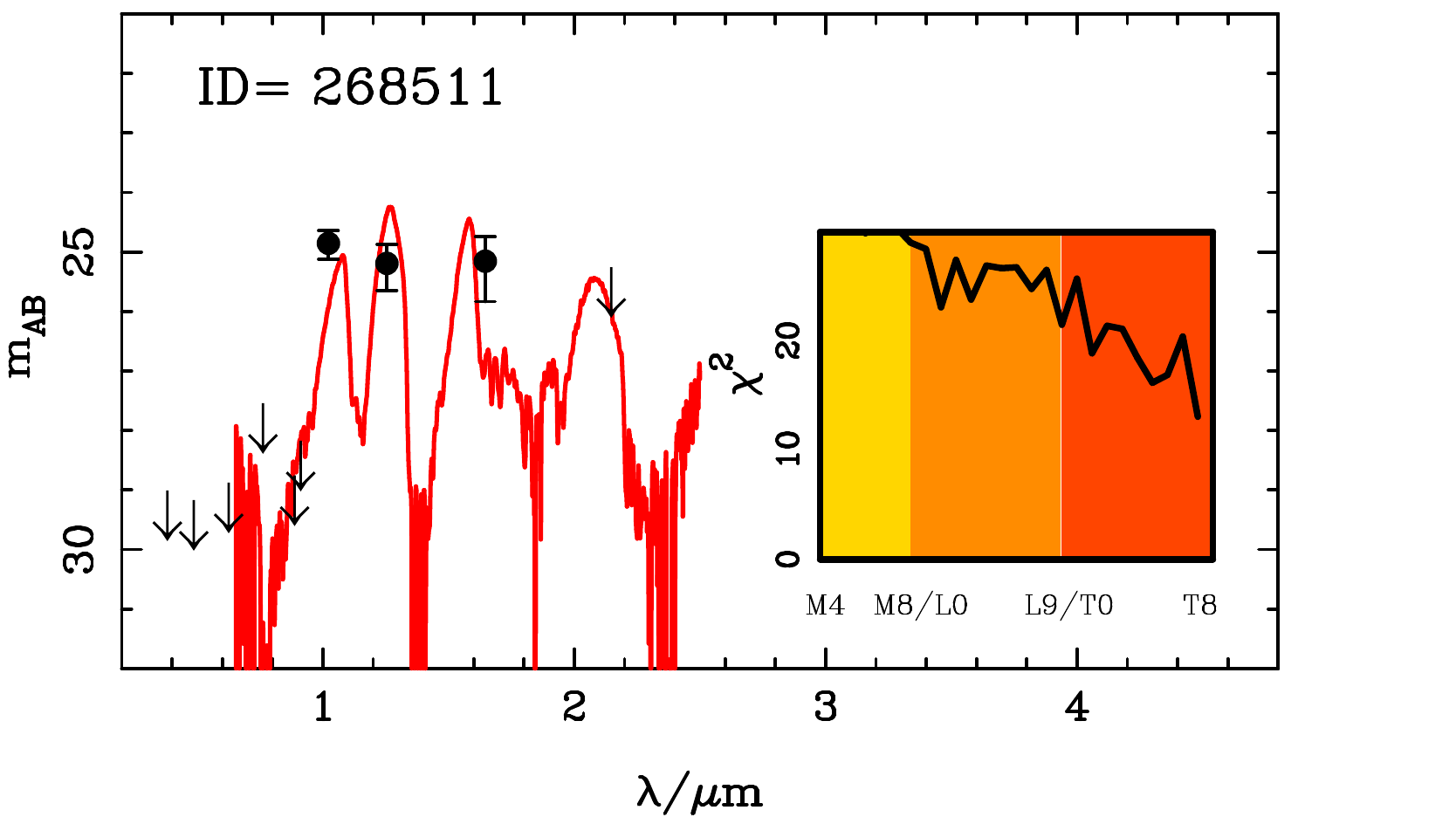}}\\
\vspace{0.5cm}
\subfloat{\includegraphics[width=0.38\textwidth, trim = 1cm 8cm 10cm 4cm, clip = true]{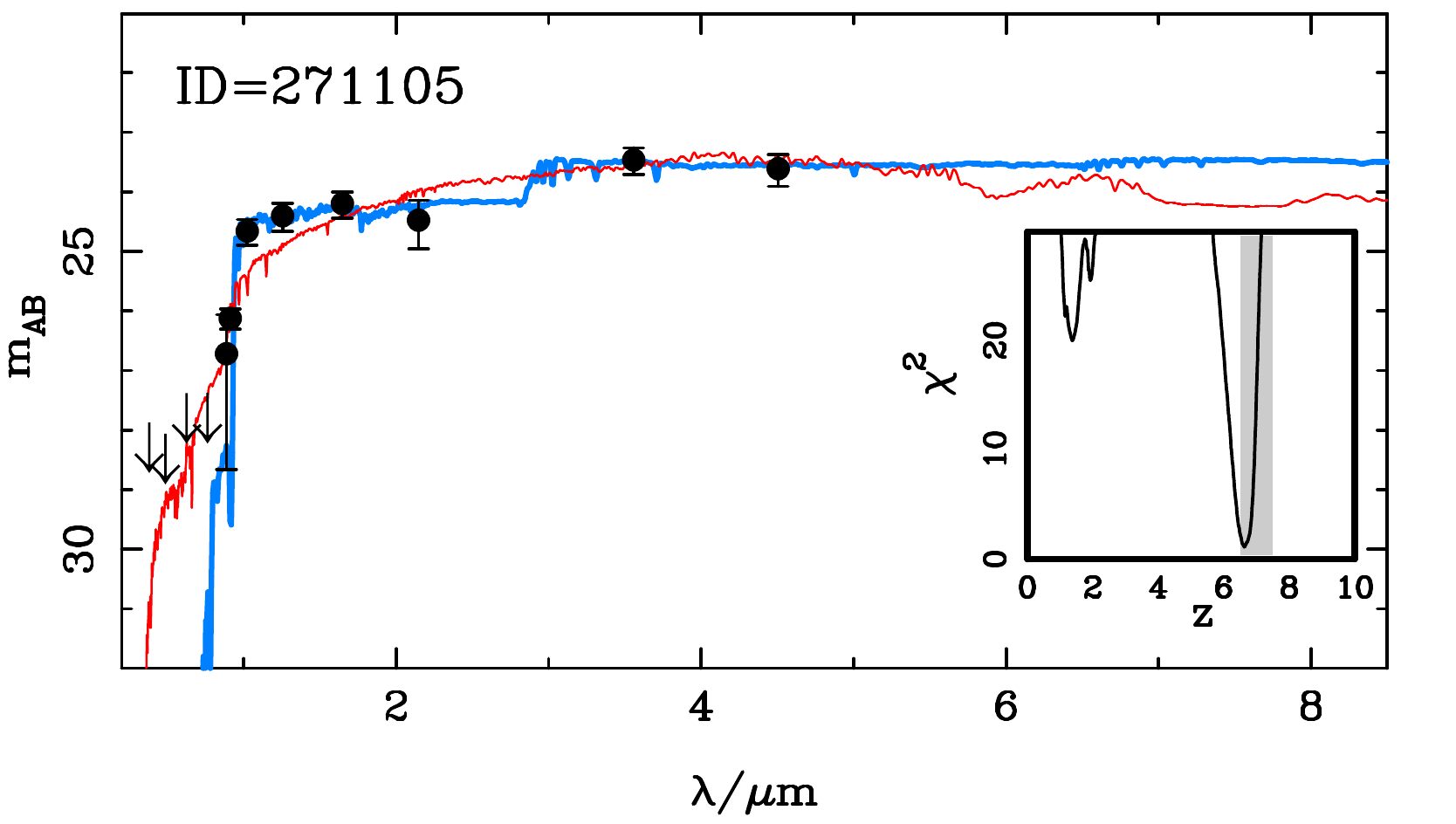}}\hspace{1cm}
\subfloat{\includegraphics[width=0.38\textwidth, trim = 1cm 8cm 10cm 4cm, clip = true]{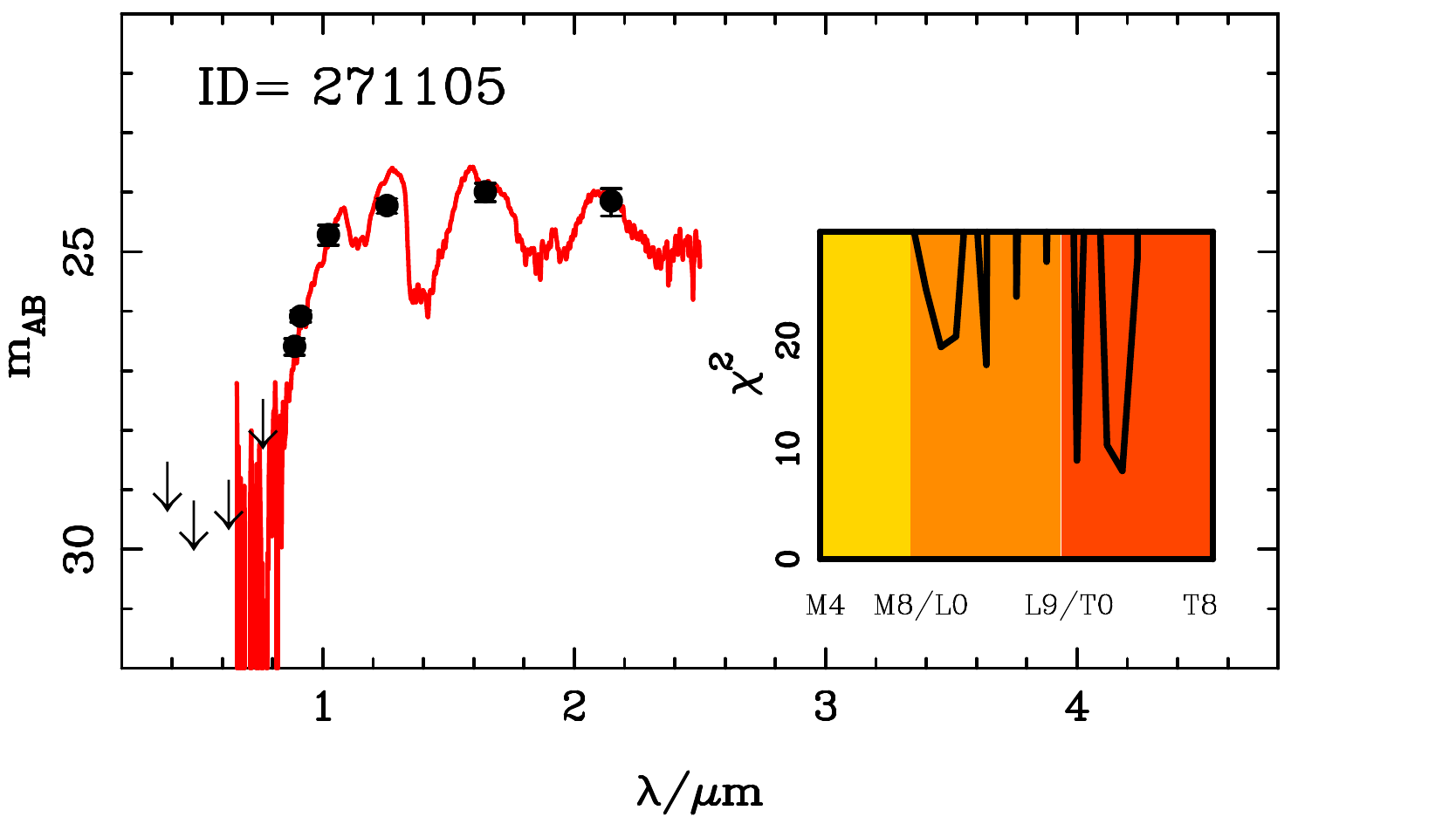}}\\
\vspace{0.5cm}
\subfloat{\includegraphics[width=0.38\textwidth, trim = 1cm 8cm 10cm 4cm, clip = true]{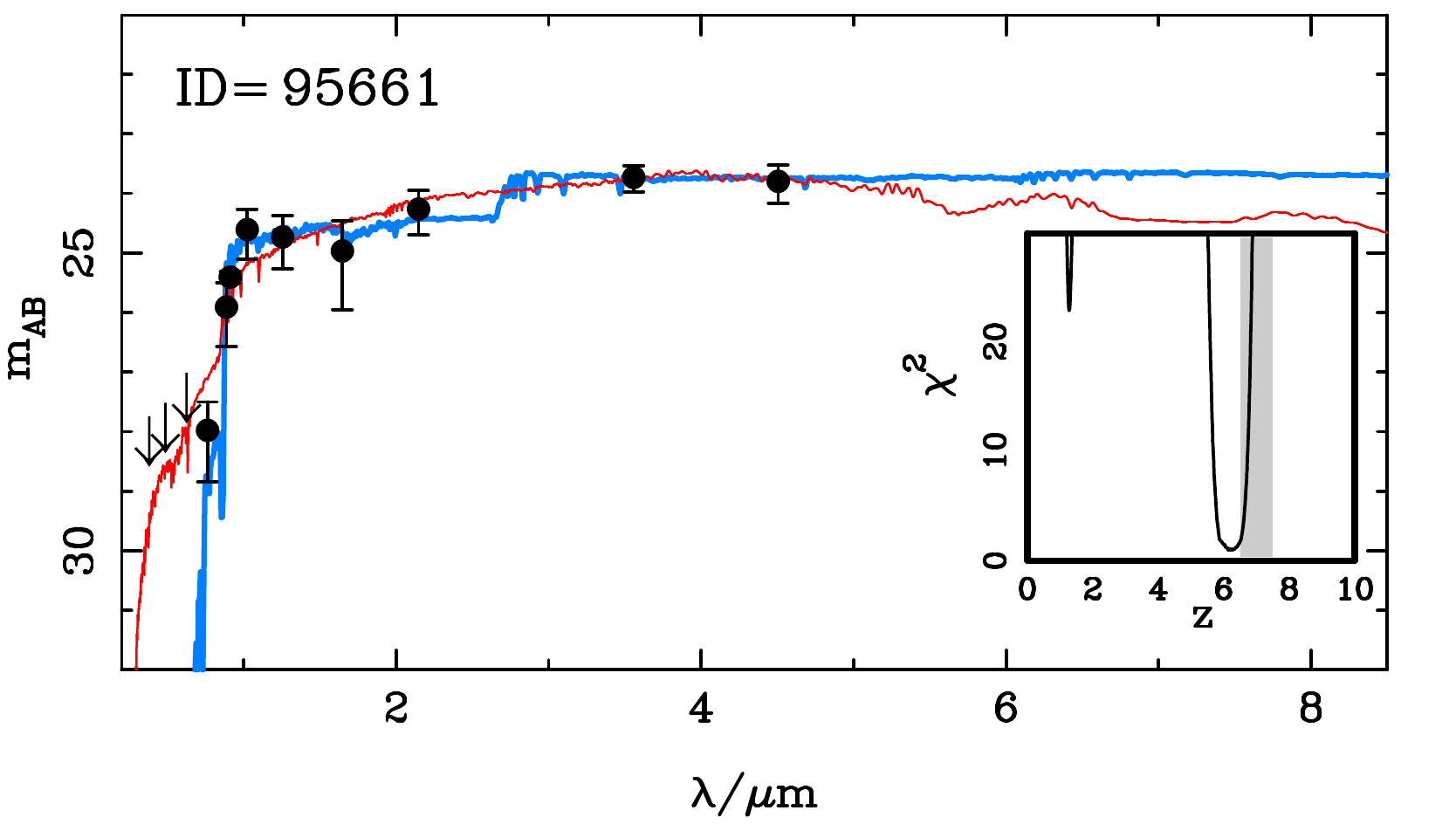}}\hspace{1cm}
\subfloat{\includegraphics[width=0.38\textwidth, trim = 1cm 8cm 10cm 4cm, clip = true]{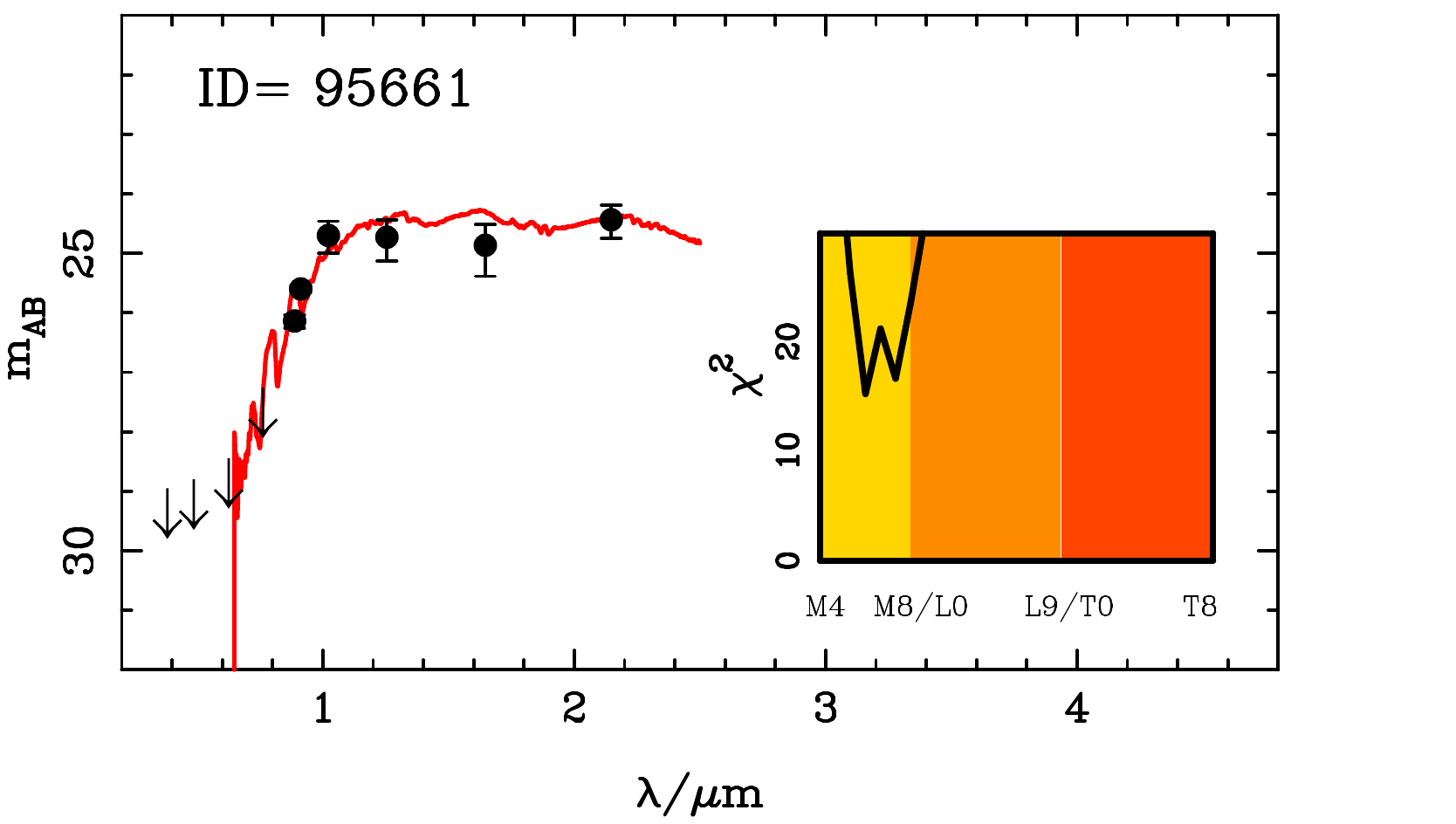}}\\
\vspace{0.5cm}
\subfloat{\includegraphics[width=0.38\textwidth, trim = 1cm 8cm 10cm 4cm, clip = true]{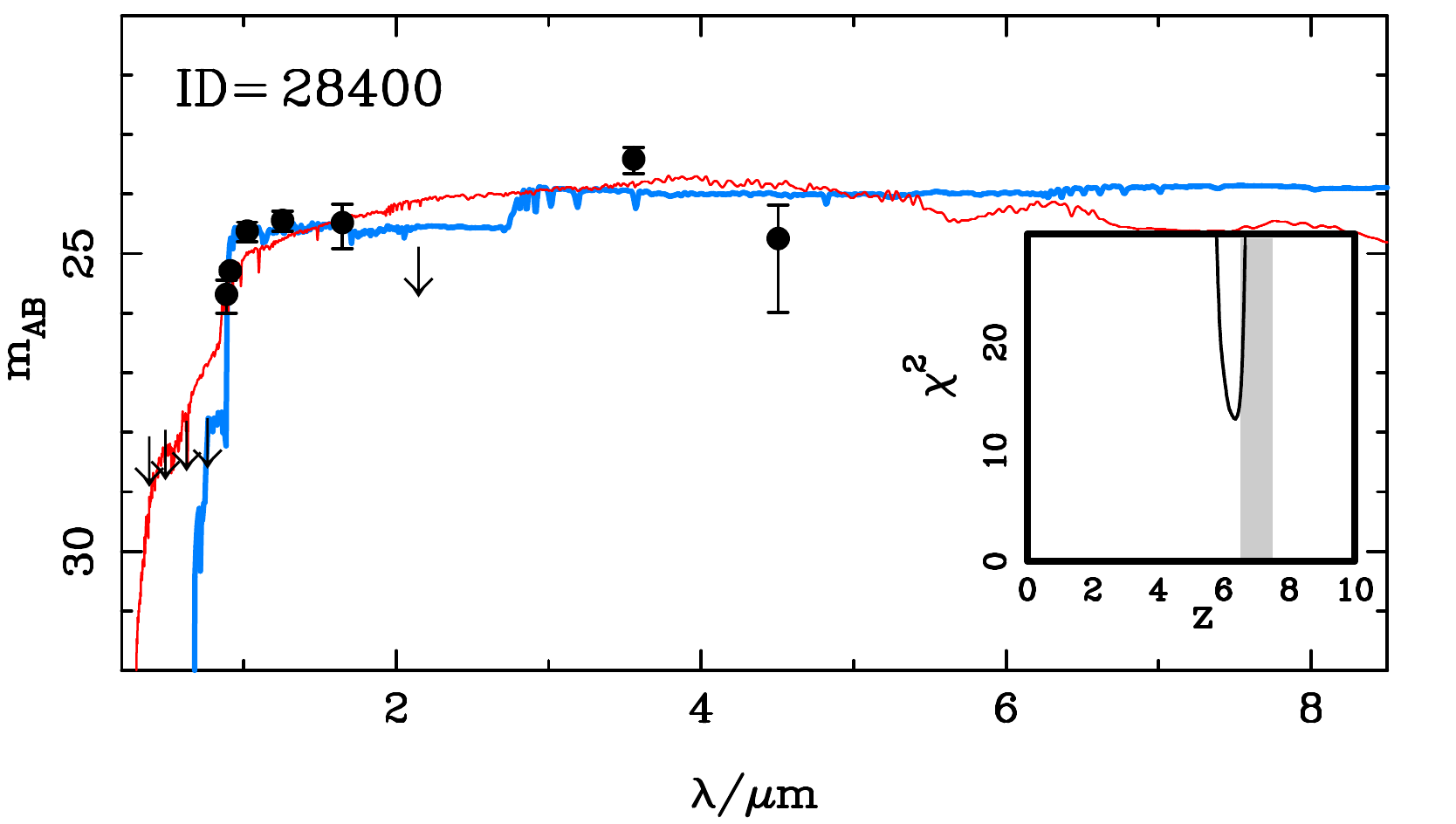}}\hspace{1cm}
\subfloat{\includegraphics[width=0.38\textwidth, trim = 1cm 8cm 10cm 4cm, clip = true]{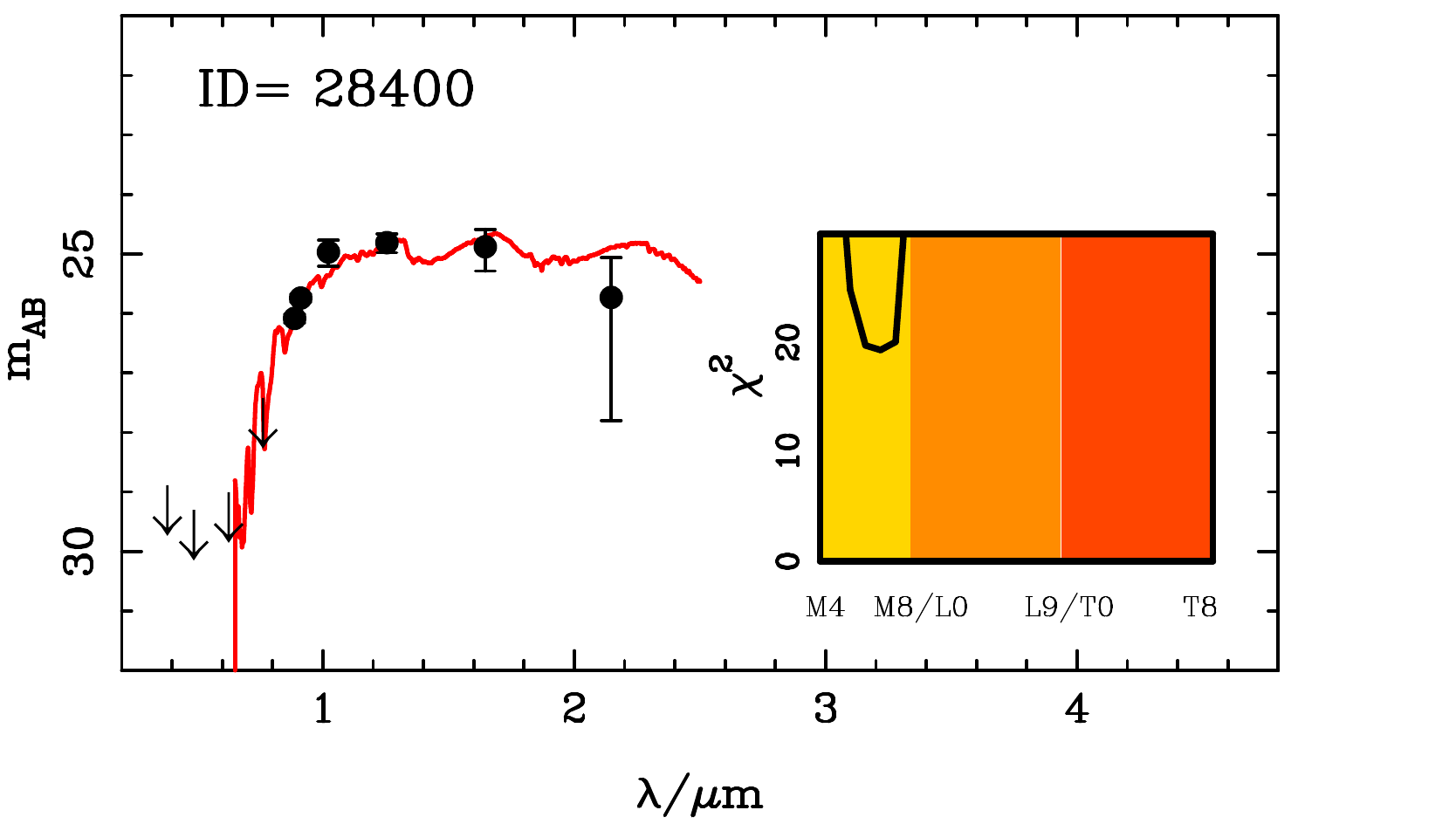}}\\
\vspace{0.5cm}
\subfloat{\includegraphics[width=0.38\textwidth, trim = 1cm 8cm 10cm 4cm, clip = true]{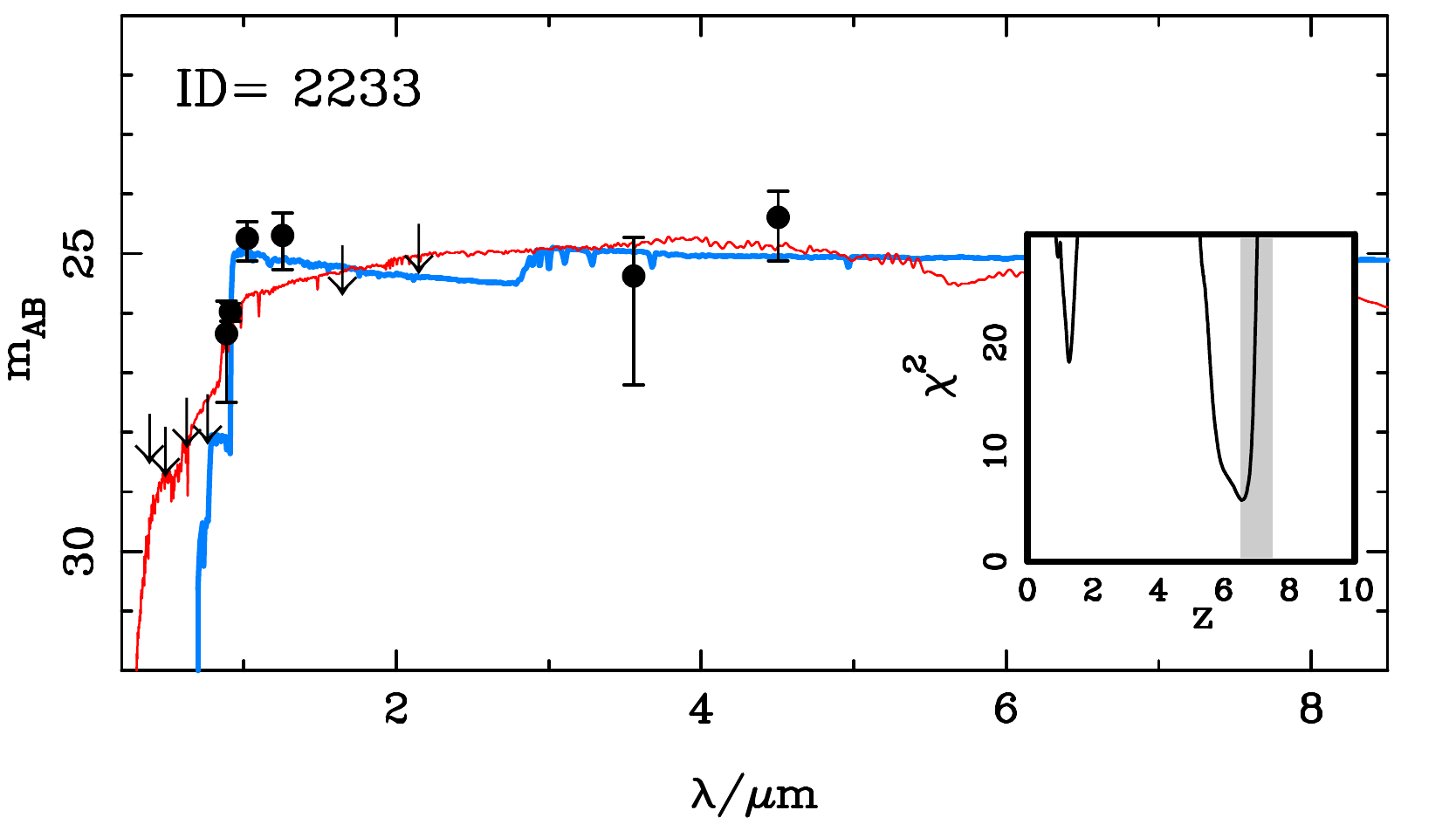}}\hspace{1cm}
\subfloat{\includegraphics[width=0.38\textwidth, trim = 1cm 8cm 10cm 4cm, clip = true]{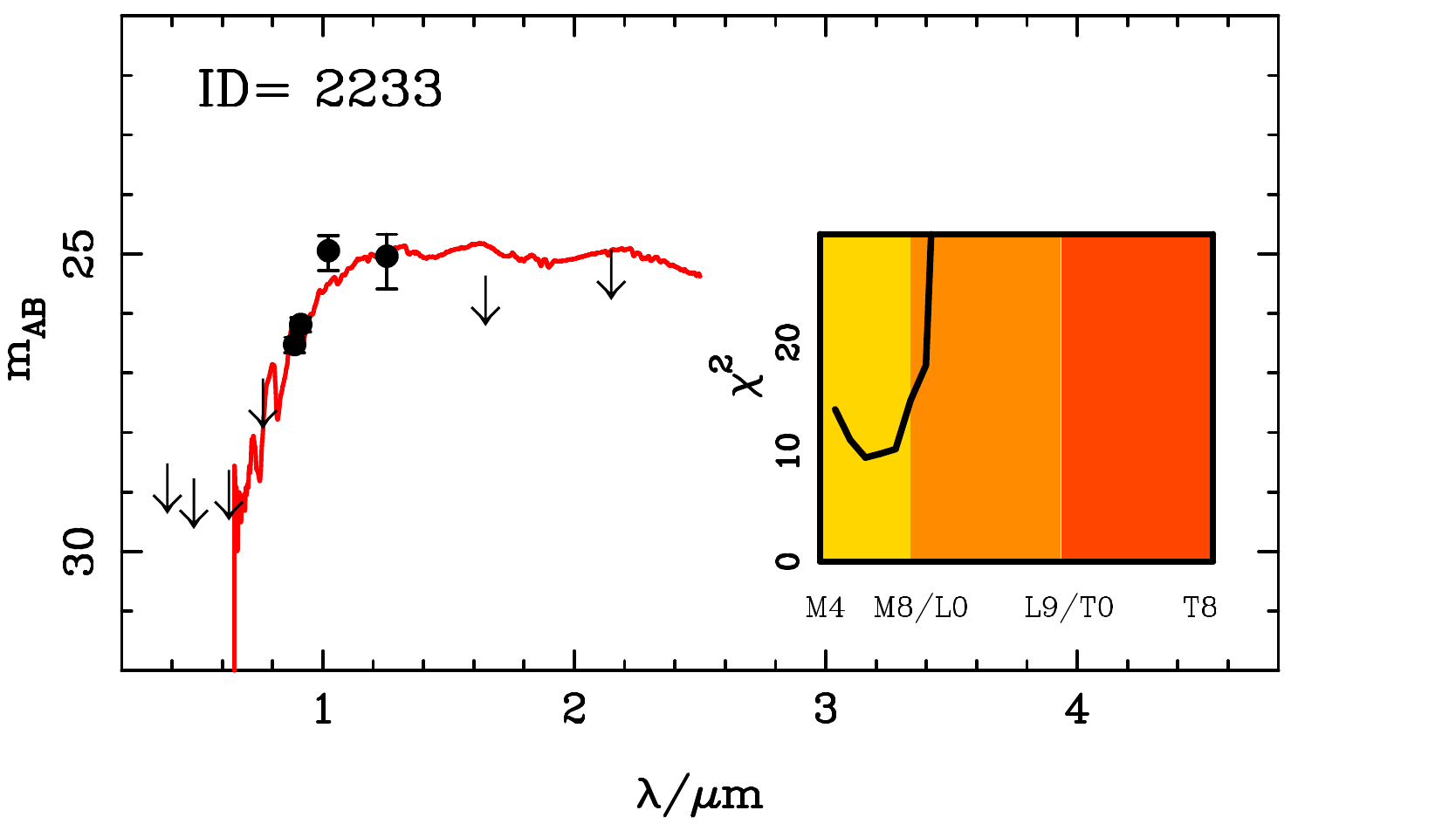}}\\

\caption{Continued.}
\end{figure*}

Finally, object 61432 has a flat near-infrared 
spectral slope and has a strong spectral break,  
with only a marginal detection in the $z'$-band.  
The somewhat red 3.6$\mu$m - 4.5$\mu$m colour might appear difficult to reconcile with a high-redshift SED as illustrated in Fig.~\ref{fig:Jch1ch2}, but 
the error bars are large and neither a late-type dwarf star nor a low-redshift galaxy provide remotely acceptable solutions.
As with the first object above (object 277912) allowance for a possible contribution from Lyman-$\alpha$ emission in the SED fitting does not change the photometric redshift,
with zero Lyman-$\alpha$ emission remaining the preferred option.

\begin{table*}
\caption{Results from SED fitting of the ten $z > 6.5$ galaxy candidates.  
The photometric redshift for the high-redshift solution is given, along with the raw $\chi_{\rm gal}^2$, for fitted galaxy templates 
with or without a contribution from \Lya line emission.   
The dust-extinction ($A_{V}$) and metallicity ($Z$)  values given in columns 4 and 5 are taken from the best-fitting 
model across all star-formation histories considered (but without \Lya).  For the fits involving \Lya we give simply 
the inferred photometric redshift and the rest-frame equivalent width ($EW_0$) of the \Lya line as selected by the best-fitting model. By 
definition, the $\chi_{\rm gal}^2$ values for the \Lya fits are either the same as, or lower than the $\chi_{\rm gal}^2$ values
given for the continuum-only fits as listed in column 3, while the photometric redshifts inferred from the \Lya fits are (inevitably) either the 
same as, or greater than those given in column 2.  Note that as a result of including the IRAC photometry in the final stages of SED fitting, the $\chi^2$ values here can exceed the $ \chi^2 < 11.3$ condition stated in Section~\ref{sect:refine}.  Columns 9 and 10 give the key information for 
the best-fitting dwarf-star SED fits (note that, because of the different number of degrees of freedom, the values 
of $\chi^2_{\rm stellar}$ cannot be directly compared with the corresponding values for the galaxy fits).  
Finally, in the last 5 columns we give derived physical parameters for these objects without \Lya in the fitting. For stellar mass ($M_*$) we simply quote the range of possible stellar masses as 
calculated from tau, constant or burst star-formation models, while the tabulated star formation rate (SFR) and specific SFR (sSFR) are 
calculated from the best-fitting constant star-formation model. The $A_V$ value for the best-fitting constant star-formation 
model is also presented, where the most extreme SFRs are inevitably coupled with the larger dust-reddening values 
(see the discussion in Section~\ref{sect:discussion}).  The final column gives the SFR estimated directly from the (observed) 
rest-frame UV luminosity (i.e. assuming zero dust extinction), 
using the~\citet*{Madau1998} prescription (dividing by a factor of 1.8 to convert from a Salpeter to a Chabrier IMF).}

\begin{tabular}{l | c l | r c l | c c  | c c c c c c c c}
\hline
 &   \multicolumn{4}{|c|}{No \Lya} &   \multicolumn{3}{|c|}{With \Lya}  & \multicolumn{2}{|c|}{Stars}   & &   \multicolumn{3}{|c|}{Constant SFH} &  \\ 
ID   &   $z_{\rm gal}$  &  $\chi^2_{\rm gal}$   &  $A_V$  &   $Z$ &   $z_{\rm gal}$  &  $\chi^2_{\rm gal}$  &   $EW_0$  &   Stellar &  $\chi^2_{\rm stellar}$   &  $M_{*}$  &  SFR   &  $A_V$   &   sSFR  & SFR$_{\rm UV}$ \\
   &    &   & mag  & ${\rm Z_{\sun}}$   & &  &   \AA   &   Type   &     &  $10^9{\rm M}_{\sun}$   &  ${\rm M}_{\sun}{\rm yr}^{-1}$   &   mag   &   $\rm Gyr^{-1}$   &   ${\rm M}_{\sun}{\rm yr}^{-1}$\\
   
\hline

     277912  &  $  6.97 ^{+ 0.06 }_{- 0.07 }$   &   1.1  &   0.7  &   1.0  &  6.97  &   1.1  &    0  &  T8  &  21.9  &   4.1 -  8.1  &   330  &   0.7  &  34.2  &    44  \\
      155880  &  $  6.78 ^{+ 0.06 }_{- 0.09 }$   &   3.7  &   0.5  &   0.2  &  6.98  &   2.8  &   40  &  T3  &  21.3  &   3.5 -  5.1  &   110  &   0.5  &  11.4  &    32  \\
      218467  &  $  7.04 ^{+ 0.10 }_{- 0.08 }$   &   6.1  &   0.0  &   1.0  &  7.20  &   4.9  &  110  &  T1  &  23.9  &   1.6 -  2.0  &    53  &   0.2  &   5.7  &    31  \\
       61432  &  $  7.07 ^{+ 0.14 }_{- 0.10 }$   &   2.8  &   0.2  &   0.2  &  7.07  &   2.8  &    0  &  T8  &  23.9  &   1.8 -  2.7  &    52  &   0.2  &   5.5  &    31  \\
       \hline
       
       Stack & $6.98^{+0.05}_{-0.05} $ &  2.6 & 0.3 & 1.0 & 7.12 & 1.5 & 40 & T3 & 37.9 & 3.0 - 4.0 & 98 & 0.4 & 32.7 & 32 \\
       
       \hline
      277880  &  $  6.77 ^{+ 0.07 }_{- 0.08 }$   &   4.5  &   0.8  &   1.0  &  6.77  &   4.5  &    0  &  T3  &  17.5  &   2.8 -  7.1  &   280  &   0.8  &  29.7  &    26  \\
      268511  &  $  7.09 ^{+ 0.12 }_{- 0.10 }$   &   4.2  &   0.0  &   0.2  &  7.22  &   3.9  &   50  &  T8  &  13.0  &   0.4 -  1.2  &    41  &   0.0  &   4.8  &    28  \\
      271105  &  $  6.62 ^{+ 0.13 }_{- 0.11 }$   &   1.1  &   0.0  &   1.0  &  6.97  &   0.8  &   70  &  T3  &   8.0  &   8.3 - 22.4  &   830  &   1.2  &  83.7  &    26  \\
      \hline
       95661  &  $  6.13 ^{+ 0.38 }_{- 0.27 }$   &   1.0  &   0.0  &   1.0  &  6.13  &   1.0  &    0  &  M6  &  15.2  &  13.2 - 24.5  &    69  &   0.5  &   6.6  &    25  \\
       28400  &  $  6.34 ^{+ 0.11 }_{- 0.19 }$   &  13.0  &   0.0  &   1.0  &  6.65  &  11.5  &   70  &  M7  &  19.3  &   8.7 - 15.1  &    39  &   0.2  &   3.8  &    26  \\
        2233  &  $  6.56 ^{+ 0.14 }_{- 0.23 }$   &   5.6  &   0.0  &   0.2  &  7.03  &   4.3  &  150  &  M6  &   9.5  &   2.3 -  5.9  &    15  &   0.0  &   1.6  &    25  \\

\hline
\end{tabular}
\label{table:properties}
\end{table*}

\subsection{Category 2 - Robust/Contaminant}

The three candidates that make up category 2 are all still consistent with being at high redshift, and 
the $z > 6.5$ galaxy solution is still formally preferred. However, 
with the current data we cannot exclude the possibility that these objects could be either at low redshift or galactic dwarf stars.

The best-fitting SED for object 277880 is a galaxy at $z_{\rm phot} = 6.77^{+0.07}_{-0.08}$. However, 
because of a strong $z'$-band detection in combination with a relatively red spectral-slope through the near-infrared bands, 
the low-redshift solution at $z = 2.0$ cannot be completely excluded.  
A stellar fit of type T3, although not as good as that of the 
high-redshift galaxy SED fit, cannot be completely excluded either.  
However, the flat $Y-J$ colour is hard to reproduce with either a 
low-redshift galaxy or dwarf-star fit, and so the $z > 6.5$ galaxy solution is still favoured for 277880.
Allowance for a possible contribution from Lyman-$\alpha$ emission in the SED fitting does not change the photometric redshift,
with zero Lyman-$\alpha$ emission remaining the preferred option.

For object 268511, the SED fitting again indicates that the high-redshift solution is the best fit, 
with the strong $z'-Y > 2.6$ break resulting in $z_{\rm phot} = 7.07^{+0.12}_{-0.10}$ and excluding 
all stellar fits except for the reddest T-dwarfs.  
Inspection of the postage-stamps for this object, however, shows that it is only clearly 
seen by eye in the $Y$-band where it is compact, and therefore it could be a transient object or a T8 star.
In this case, allowance for a possible contribution from Lyman-$\alpha$ emission in the SED fitting raises the photometric redshift 
to $z_{\rm phot} = 7.22$, with an implied Lyman-$\alpha$ rest-frame equivalent width of $EW_0 = 50$\,\AA.

Finally, 271105 has a best fitting high-redshift SED with $z_{\rm phot} = 6.62^{+0.13}_{-0.11}$, with a higher-redshift 
solution ($z_{\rm phot} = 6.97$) possible with the introduction of \Lya emission with $EW_0 = 70$\,\AA.  
A clear detection in the $z'$-band does, however, allow the stellar fit for a T3 dwarf to recreate the spectral 
break.  However, the 
red $J-$3.6$\mu$m colour of 271105 is inconsistent with a T-dwarf star (the object is well detected and unconfused in all near-infrared bands).  
Hence, although we cannot rule a stellar solution, it still seems highly likely that 271105 is indeed a high-redshift galaxy.

\subsection{Category 3 - Insecure}

The final three candidates all have good SED fits to a high-redshift galaxy template, 
with photometric redshifts $z < 7.0$ due to clear $z'$-band detections.  
However, for these objects an M-dwarf stellar fit cannot be completely excluded, and some of the photometric redshift estimates suggest 
a redshift just below $z = 6.5$. For both these reasons we classify these objects as insecure $z > 6.5 $ galaxies.

Object 95661 has a high-redshift galaxy fit with $z_{\rm phot} = 6.13^{+0.38}_{-0.27}$, and so still has some probability of being at $z > 6.5 $
(although here the introduction of \Lya emission does not increase the inferred redshift).  
A dwarf star of type M6 can also reproduce the flat near-infrared photometry as well as the spectral break, and so cannot be excluded.

Object 28400 has a high-redshift galaxy fit with $z_{\rm phot} = 6.34^{+0.11}_{-0.19}$ but in this 
case is actually better fitted with the inclusion of a \Lya emission line with $EW_0 = 70$\,\AA\,
increasing the inferred redshift above the $z = 6.5$ threshold to $z_{\rm phot} = 6.65$.  
The IRAC photometry shows a bright detection in the 3.6\,$\mu$m image but nothing in the 4.5\,$\mu$m data 
despite the similar depth, implying a break of at least 3.6$\mu$m - 4.5$\mu$m $ < -1.4$ 
(taking the 2$\sigma$ limit for 4.5\,$\mu$m).  A possible explanation would be an M-dwarf star 
which can produce this blue colour within the error bars as shown in Fig.~\ref{fig:Jch1ch2}.  One alternative explanation for the blue \choneminchtwo colour would be nebular emission.

Finally, object 2233 has a clear $z'$-band detection, but is weak in $Y$+\,$J$ and the other infrared bands.  
The high-redshift galaxy fit is comparable in quality to that of a M6 star and the weak near-infrared 
photometry allows the inclusion of a \Lya line of $EW_0 = 150$\,\AA\,which raises the photometric redshift to $z_{\rm phot} = 7.03$.

\newpage

\section{Comparison with Previous Studies}\label{sect:previous}

Several other ground-based studies have attempted to determine, or at least place a meaningful limit 
on the very bright end of the galaxy UV luminosity function at $z \sim 7$, using deep near-infrared photometry 
and colour cuts to attempt to exclude low-redshift galaxies and cool galactic dwarf-star contaminants.

\begin{figure}

\subfloat{\includegraphics[width=0.5\textwidth, trim = 1cm 8cm 10cm 4cm, clip = true]{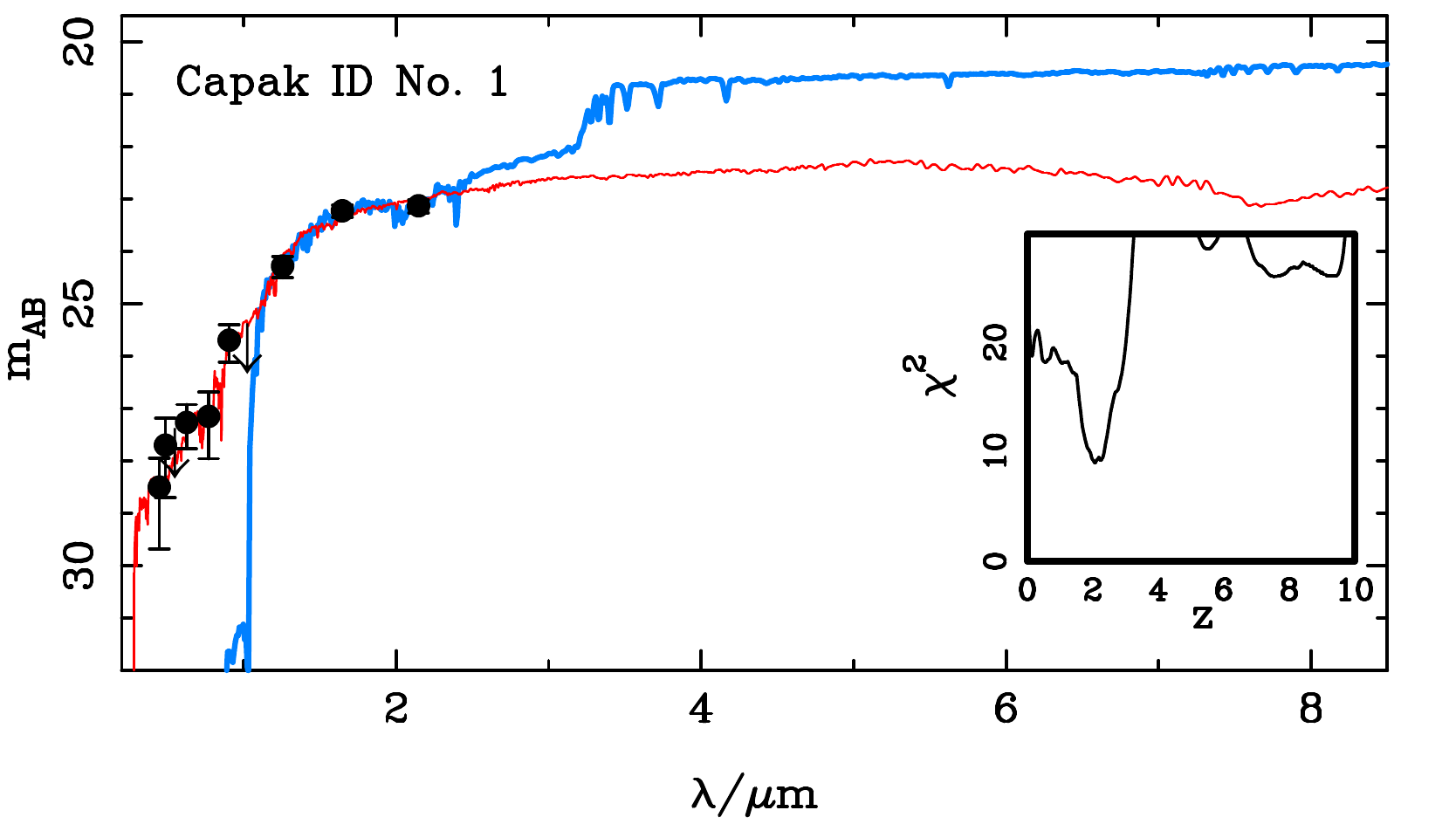}}\\
\subfloat{\includegraphics[width=0.5\textwidth, trim = 1cm 8cm 10cm 4cm, clip = true]{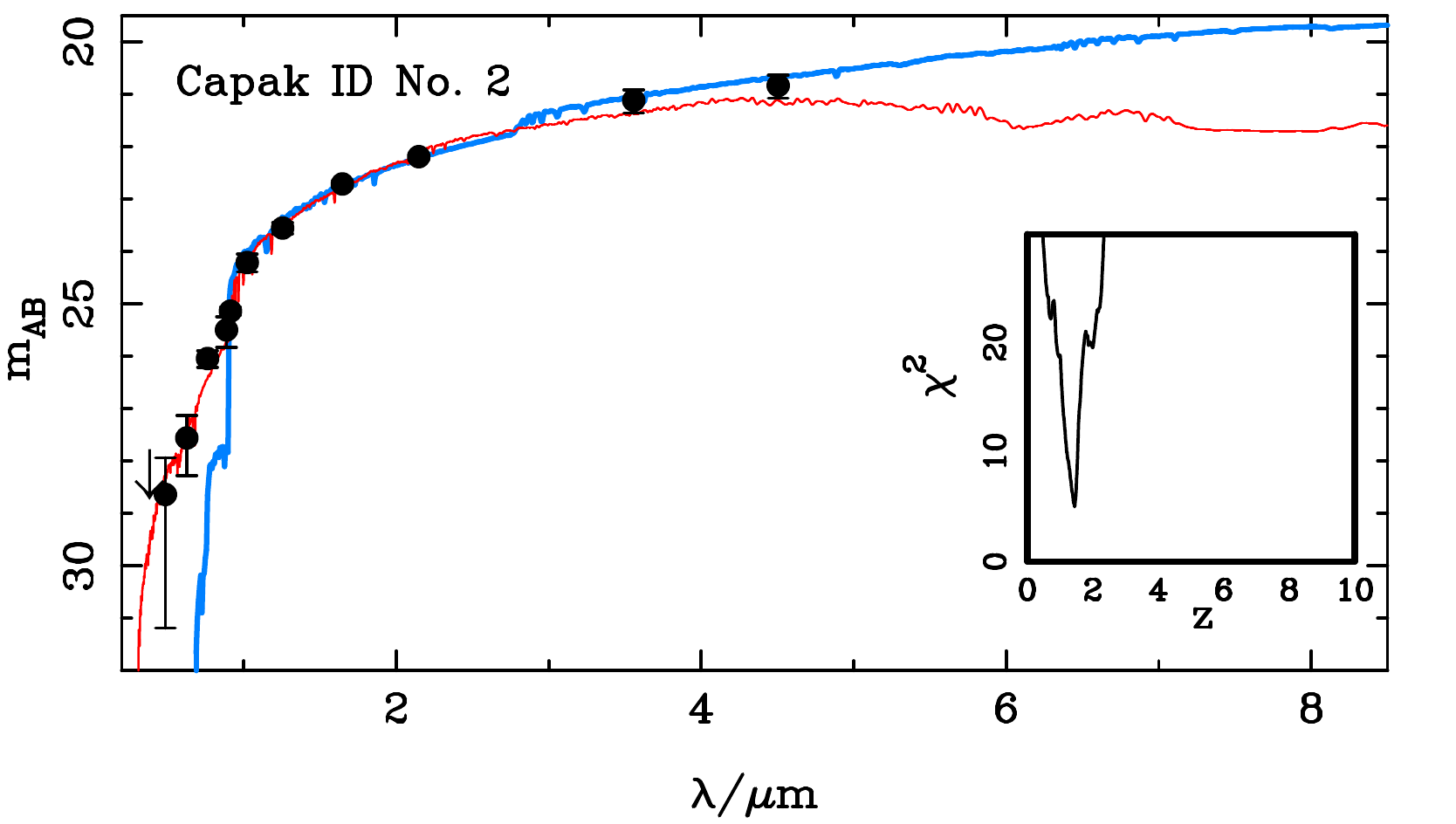}}\\
\subfloat{\includegraphics[width=0.5\textwidth, trim = 1cm 8cm 10cm 4cm, clip = true]{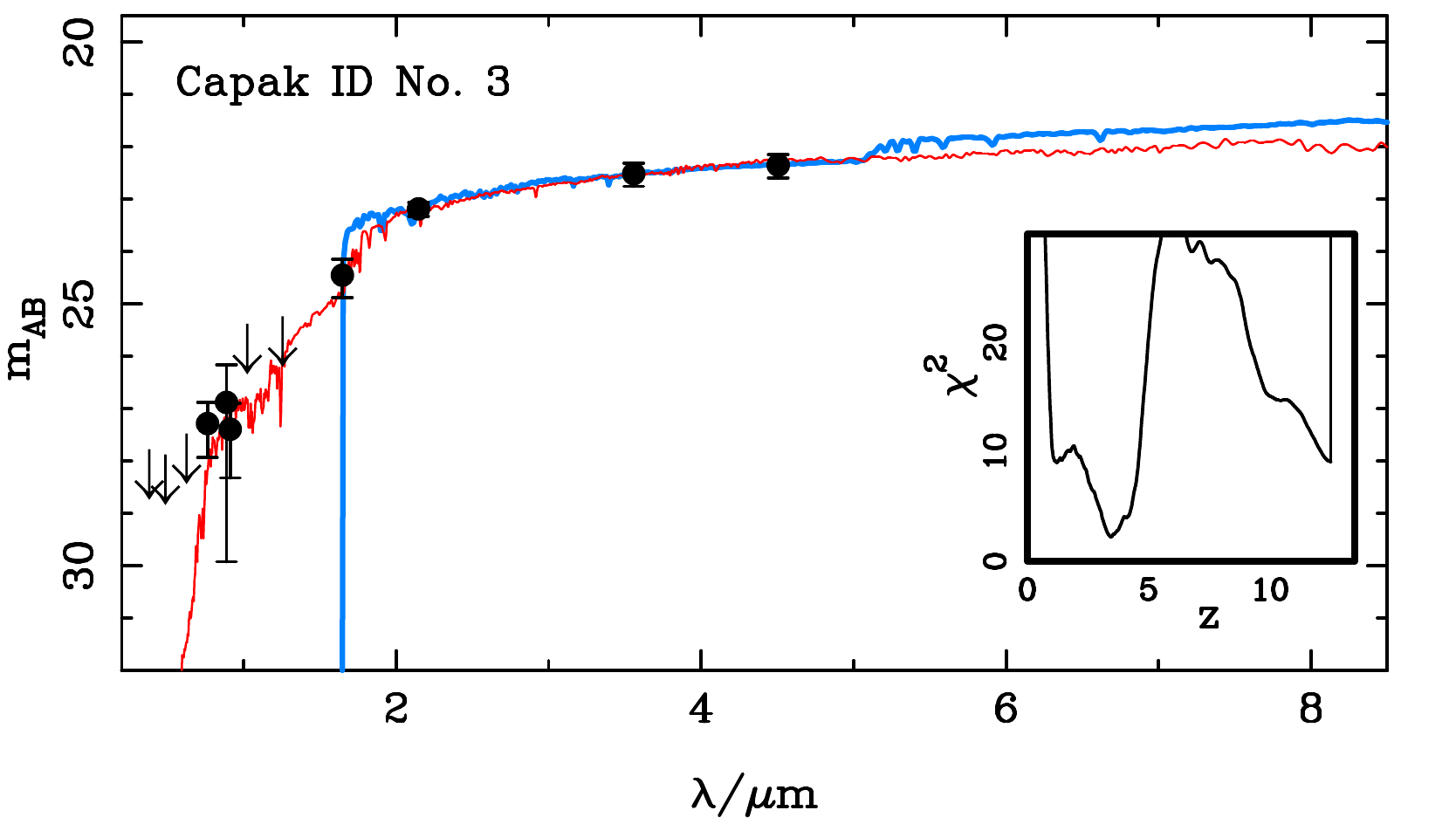}}

\caption{SED fits for the three $z > 7$ galaxy candidates from~\citet{Capak2011} using our revised photometry from the deeper UltraVISTA near-infrared 
imaging and the new $z'$-band Subaru imaging.  The UltraVISTA data extends to around one magnitude deeper in the $J$-band that the previous 
imaging available in the COSMOS field. In addition, it provides previously unavailable deep $Y$-band imaging,
invaluable for removing low-redshift galaxy contaminants and dwarf stars.  The blue line shows the best fitting high-redshift model (with $z > 4$) and the red line shows the best low-redshift fit, where the photometric redshift and $\chi^2$ values for these two solutions are Capak1: $z_{\rm phot} = 2.1 (7.6)$, $\chi^2 = 9.0 (26.0)$, Capak2: $z_{\rm phot} = 1.4 (6.9)$, $\chi^2 = 5.0 (42.2)$ and Capak3: $z_{\rm phot} = 3.4 (12.6)$, $\chi^2 = 2.2 (9.1)$. We have extended the redshift range of the Capak3 fit above to display the high-redshift model solution at $z \sim 13$, which although being a formally acceptable fit has an implied star-formation rate and stellar mass that make it unrealistic at this extreme redshift.
Hence, with our new photometry it can be seen that none of these objects has a robust high-redshift fit,
but rather all lie in the redshift range $z \simeq 1.5 - 3.5$.}
\label{fig:Capak2011fits}
\end{figure}

\subsection{\citet{Capak2011}}\label{sect:capak}

Of particular relevance to this paper is previous work within the COSMOS field undertaken by~\citet{Capak2011}.  They utilised the deepest multi-wavelength data over the full COSMOS 2\,deg$^2$ field available at the time, 
in particular optical data from Subaru including the $i^+$ and $z^{+}$-bands, near-infrared data in the 
$J$-band from UKIRT/WFCAM and $H$, $K_s$ imaging from CFHT/WIRCam~\citep{Bielby2011, McCracken2010}.  Galaxy candidates were selected in the $K_s$-band with 
a $5\sigma$ cut in the $J$ and $K_s$ bands of 23.7\,mag (3\asec diameter aperture) applied along with the colour conditions 
$z'-J \ge 1.5$, $K_s - 4.5\mu m > 0$ and $J-K_{s} > 0$.  They found three objects consistent with being at $z > 7$ as defined by 
their colour-colour criterion, and presented follow-up spectroscopy which, along with SED fitting, resulted in the rejection of one object (C2 below) 
as most likely lying at much lower redshift. Since the near-infrared data from UltraVISTA extends to around one magnitude deeper 
than that exploited by~\citet{Capak2011} in the $J$-band (with improved seeing) 
and also provides $Y$-band imaging that greatly improves the separation of contaminant populations (as described previously), 
we have extracted new photometry for all three of these objects and undertaken SED fitting 
in the same manner as described above for our own, new deeper sample of objects. 
Since none of the~\citet{Capak2011} candidates actually made it into the sample described above, 
it should not be a surprise that, as described below, we fail to confirm that they lie at $z > 6.5$.

The first object presented by~\citet{Capak2011}, hereafter C1, in fact lies just outside the area of overlapping deep data utilised 
in this paper and, as a consequence, the CFHTLS and new deeper Subaru $z'$-band imaging is unavailable.  
However, C1 is still within the 1.5\,deg$^2$ of the UltraVISTA imaging that extends beyond the central $\sim 1$\,deg$^2$ area, 
as seen in Fig.~\ref{fig:fieldmap}.  With the deeper near-infrared data, most importantly 
the $Y$-band, we obtained photometry for this object to check the consistency 
of the photometric redshift calculated in~\citet{Capak2011}.  Near-infrared photometry was measured within 
a 2\asec diameter aperture using {\sc sextractor} in dual-image mode with selection in the $K_s$-band, 
where the candidate was brightest.  
For the optical bands, cutouts from the COSMOS website in the $B_J$, $g^+$, $V_J$, $r^+$, $i^+$ 
and $z^+$-bands were extracted and the magnitudes (measured using {\sc GAIA}) and errors (taken from the 5$\sigma$ limits presented 
in~\citealt{Capak2011}) were included in the SED fitting analysis.  We find C1 to be undetected in $Y$, and to have $J = 24.4 \pm 0.2$, $H = 23.3 \pm 0.1$ and $K_s = 23.2 \pm 0.1$ (within 
a 2\asec aperture with a correction to 84\% enclosed flux). These values can be compared to the~\citet{Capak2011} magnitudes of 
$J = 23.21 \pm 0.05$, $H = 23.09 \pm 0.12$ and $K_s = 22.14 \pm 0.02$ within a 3\asec aperture.  
The large differences in photometry cannot be explained by aperture corrections, as the 3\asec aperture 
used by~\citet{Capak2011} was predicted to enclose 80\% of the flux in the PSF-matched $J, H, K_{s}$ imaging, 
implying the new magnitudes should be slightly brighter or comparable.  
An offset in photometry between the WIRCam and UltraVISTA data has been identified 
in the UltraVISTA data release documentation (see also~\citealt{McCracken2012}), 
implying that the UltraVISTA data is 0.15 mag fainter than the previous imaging, 
which is thought to be a problem with the COSMOS data rather than with VISTA; however this effect 
still cannot explain the change in the observed near-infrared colours. 
The IRAC imaging of C1 is highly-confused with a nearby object and 
so no attempt was made to measure this by either ourselves or by~\citet{Capak2011}.  
The large companion object is a low-redshift interloper at $z_{\rm phot} = 0.35$ from the 
COSMOS catalogue, which may have affected the accuracy of the 3\asec photometry.  
Our SED analysis of C1, with the fit shown in Fig.~\ref{fig:Capak2011fits}, yields no acceptable high-redshift solution, but rather reveals
this object to be a red galaxy at $z_{\rm phot} \simeq 2.1$.

The second object, C2, was already identified by~\citet{Capak2011} as a likely low-redshift interloper at $z_{\rm phot} = 1.59$, 
which we here confirm with our SED fit to the UltraVISTA data shown in the second panel of Fig.~\ref{fig:Capak2011fits}.  
This galaxy candidate was selected in our own preliminary catalogue, but excluded at a relatively early stage as a 
low-redshift galaxy because of the clear $i$-band detection.

Surprisingly, the third object, C3, was not selected in our analysis despite the claim by~\citet{Capak2011} that 
it has $J = 23.1 \pm 0.1$.  Visual inspection of the UltraVISTA $J$-band image of this object in fact reveals little, if any, evidence of
detectable flux, however the object is real as it is clearly visible in the $K_s$-band. Accordingly, 
photometry from a $K_s$-band  selected catalogue for the object was collected and the SED fit performed which, as shown in 
the third panel of Fig.~\ref{fig:Capak2011fits}, reveals the galaxy to lie at $z \simeq 3.5$.

This analysis therefore highlights the power of the new deeper and better-quality UltraVISTA data over the previously available near-infrared data
in the COSMOS field, and confirms that the new objects uncovered here from the $Y$+\,$J$ UltraVISTA imaging are the brightest credible $z > 6.5$ galaxy
candidates in this area of sky.

\subsection{\citet{Salvato2011}}

\begin{figure}

\includegraphics[width=0.5\textwidth, trim = 1cm 8cm 10cm 4cm, clip = true]{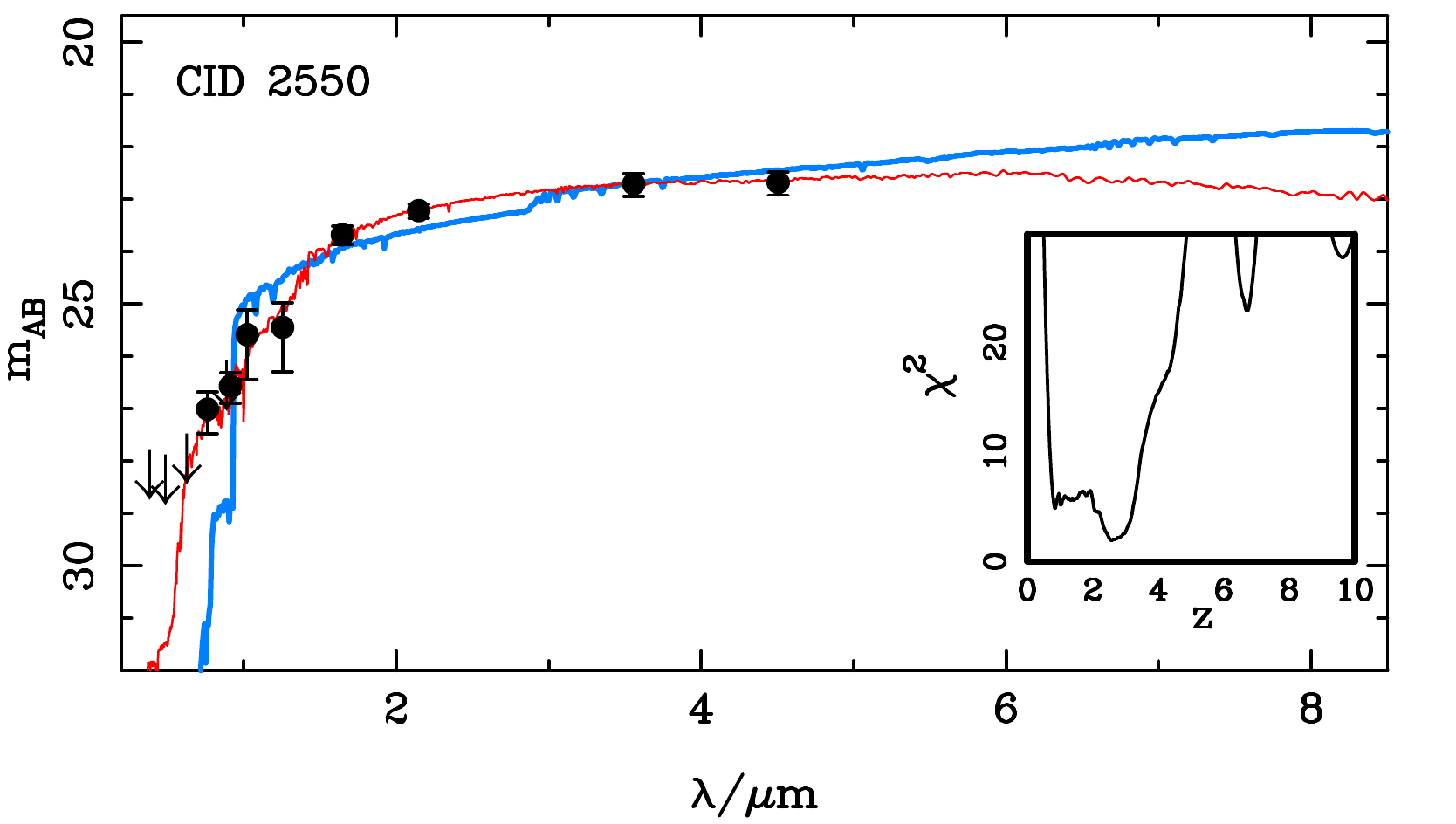}

\caption{SED fit for the candidate $z \sim 6.84$ X-ray-selected source from~\citet{Salvato2011} using our revised photometry.  The blue line shows the best fitting high-redshift model ($z_{\rm phot} = 6.7, \chi^2 = 22.9$) and the red line shows the best low-redshift fit ($z_{\rm phot} = 2.6, \chi^2 = 1.9$).  Due to the improved near-IR photometry, which shows the candidate to have a gradually rising SED rather than presenting a strong break, we conclude that the object is in-fact a lower redshift, dusty interloper.
}
\label{fig:salvato}
\end{figure}

Another particularly relevant high-redshift candidate within the field is the object CID-2550 identified by~\citet{Salvato2011}, that they claim could potentially be the highest redshift X-ray-selected source found to date.  By SED fitting to the optical and near-IR counterpart to a Chandra X-ray source (from the Chandra-COSMOS catalogue), they found a best fit photometric redshift of $z_{\rm phot} \sim 6.84$.
The previous magnitudes of this object CID-2550 are $i \approx 26.6,
z = 25.4, J = 23.6, H = 23.8, K = 23.0$ with the object becoming
brighter still in the Spitzer channels.  However, despite the extremely
bright $J$-band magnitude reported by~\citet{Salvato2011}, CID-2550 was not present in our
final sample.  CID-2550 is present in our original $Y+J$ selected
catalogue as you would expect, but was immediately excluded as it is unexpectedly below the
$5\sigma$-limit in the detection image (with $Y+J = 26.9$).  Inspection of the
UltraVISTA imaging and associated datasets shows the candidate to be
clearly detected in the $H$ and $K_s$-bands, with faint detections
in the $i$, $z'$, $Y$ and $J$-bands.  As the object is brightest in
the $K_s$-band we used $K_s$-band selected magnitudes and performed
SED fitting as for our own candidates.  Figure~\ref{fig:salvato}
shows our best fit to this object at $z_{\rm phot} = 2.6$, where the decrease in the photometric redshift compared to the previous determination by~\citet{Salvato2011} is a consequence of the reduced $J$-band magnitude and the inclusion of the $Y$-band, along with reduced errors throughout the near-IR that show the spectrum to be gradually rising rather than exhibiting a sudden break in the photometry.  Given that the candidate is an X-ray source, we conclude that it is most likely a dusty active galactic nucleus at a lower redshift of $z_{\rm phot} = 2.6$.

\subsection{\citet{Hsieh2012}}\label{sect:hsieh}

\citet{Hsieh2012} recently reported a search for $z_{850}$-dropout galaxies within the Extended Chandra Deep Field South (ECDFS), 
with objects detected in $J$-band data obtained as part of the Taiwan ECDFS Near-Infrared Survey (TENIS).  
Boxes in colour-colour space were used to select high-redshift galaxies, in particular the {\it Spitzer} 
IRAC channels were introduced in an attempt to eliminate low-redshift galaxy and dwarf-star contaminants.
One candidate high-redshift galaxy with $J = 25.12 \pm 0.23$ (total magnitude), TENIS-ZD1, was selected from the 0.25\,deg$^2$ surveyed, 
with a weighted photometric redshift of $z_{\rm phot} = 7.822^{+1.095}_{-0.725}$.  
The very large errors on the derived photometric redshift are a consequence of the proposed Lyman-break occurring 
between two widely-separated filters, the $z_{850}$- and $J$-band, due to the unfortunate lack of $Y$-band imaging.  
It is this lack of $Y$-band imaging which also makes it hard to exclude dwarf-star contaminants using the techniques discussed above (Fig.~\ref{fig:JKz}a),
leading \citet{Hsieh2012} to rely on the IRAC colours in an attempt to achieve this.

The infrared colours of dwarf star contaminants utilised by~\citet{Hsieh2012} were calculated from the AMES-dusty dwarf 
star models~\citep{Allard2001}. However, with the advent of recent cool dwarf-star photometry by 
\emph{Spitzer} from~\citet{Patten2006} and~\citet{DavyKirkpatrick2011} it is now clear that these models are inadequate. 
Specifically, while the largest observed colour over the spectral break of $z'-J \sim 3.5$ 
(see Fig.~\ref{fig:JKz}) agrees with the AMES-dusty model, where $z - J \sim 3$, the \choneminchthree colour, which is proposed by 
\citet{Hsieh2012} as the best way to discriminate between dwarf stars and galaxies, is underestimated by the model by over a magnitude.  
The results of~\citet{DavyKirkpatrick2011} in particular, who observed mainly T-dwarf stars, show that dwarf stars 
can be significantly redder than predicted in the AMES-dusty model, with \choneminchthree $ \simeq 0.5$ perfectly possible.  

Despite these concerns, the \choneminchthree colour of TENIS-ZD1 reported by 
\citet{Hsieh2012} is still redder than observed in the coolest T-dwarf stars 
to date, albeit now by only $\simeq 2\sigma$. However, it also exceeds the predictions of high-redshift galaxy templates
with reddening up to $A_V \simeq 1.0$. Unfortunately, because \citet{Hsieh2012} have neglected to provide the position of their 
putative high-redshift galaxy, we cannot check the IRAC photometry, or indeed the photometry and claimed non-detections
at any other wavelength, despite the fact that much of the key datasets are public. 
Our own SED fit to the photometry provided by \citet{Hsieh2012} supports the high-redshift solution (which they derived using the EAZY code; Brammer et al. 2008), but 
we note that the extreme redshift which results is largely a consequence of the arguably surprisingly bright magnitude 
reported at 5.8\,$\mu$m, while exclusion of the low-redshift alternative solution depends critically on the 
error ascribed to the single $J$-band data point (in the absence of any $Y$-band imaging of appropriate depth).
Given that we cannot make our own photometric measurements for this object, it is difficult to comment further on the validity or 
otherwise of this proposed $z = 7 - 8$ galaxy. What we can say is that, given our own results, it would not seem unreasonable 
to detect one galaxy at $z \simeq 7$ at $J \simeq 25$ in the ECDFS, but also that it is somewhat unexpected to see it detected in the 
two longest-wavelength IRAC bands, given that {\it none} of our own UltraVISTA candidate objects are detected in the S-COSMOS 
5.8\,$\mu$m or 8\,$\mu$m imaging. However, we note that the reported $5.8\,\mu$m and $8.0\,\mu$m magnitudes for TENIS-ZD1, of 
$m_{\rm AB} \sim 23.3$ and $23.6$ respectively, are deeper than the S-COSMOS limits.

\section{Discussion}\label{sect:discussion}
\subsection{Stacked photometry and physical properties}\label{sect:stack}

\begin{figure}
\includegraphics[width=0.5\textwidth, trim = 1cm 8cm 10cm 4cm, clip = true]{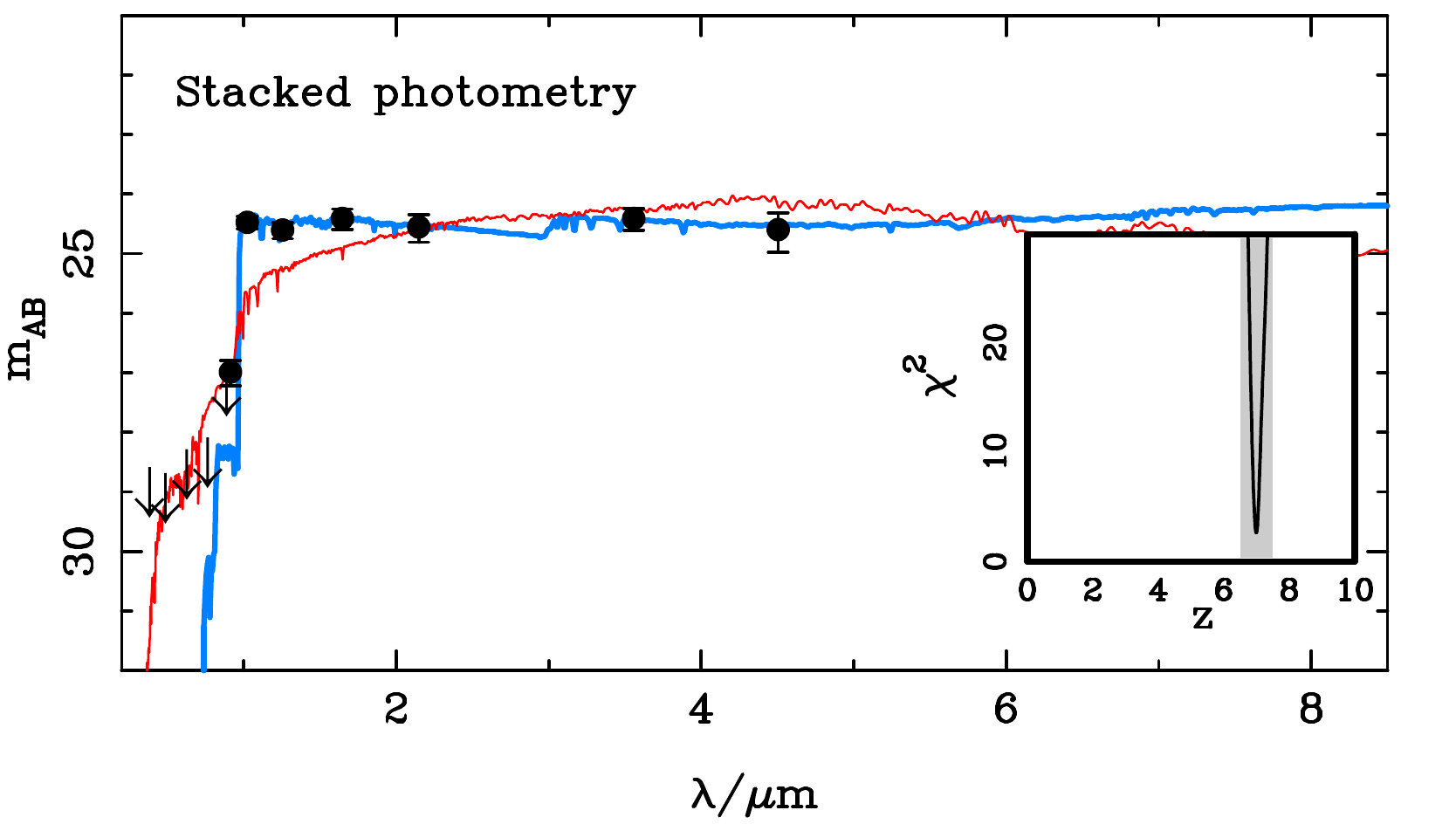}
\caption{The best SED fit (shown with a blue line) to the average photometry produced from a stack of our four most robust $z \simeq 7$ 
galaxy candidates as shown in Fig.~\ref{fig:stamps} and described in Section~\ref{sect:robust}.  The best-fitting photometric 
redshift for the stack is $z = 6.98 \pm 0.05$, rising to $z = 7.12$ with allowance for a contribution from Lyman-$\alpha$ emission (see
Table 2).  The best fitting low-redshift model ($z \simeq 1.5$) is shown as the red curve, which clearly cannot reproduce the large spectral break and flat near-IR colours of the stack photometry.}
\label{fig:stack}
\end{figure}

\subsubsection{Creation and analysis of the four-object stack}
An average stack was made from the postage-stamps of our four most robust $z > 6.5$ galaxy candidates as illustrated in Fig.~\ref{fig:stamps}. 
This was done both to create higher signal-to-noise photometry to better explore the implications of SED fitting, and 
also to double-check that no low-level $i$-band flux was present that might indicate contamination of 
our ``robust'' sample with objects at $z < 6.5$.  Photometric measurements were made in the same way as for the individual candidates, 
using {\sc SExtractor} in dual-image mode with selection in the $Y$+\,$J$ image.  
The resulting average photometry is included in Table~\ref{table:phot}, and the best-fitting SED is shown in Fig.~\ref{fig:stack}, 
where the variation in the rest-frame UV photometry observed in the four 
individual galaxies averages to produce a relatively flat UV slope, as discussed further below.  
The derived photometric redshift for the stack is $z = 6.98 \pm 0.05$, and the colour across the Lyman-break is $z'-Y = 2.5 \pm 0.2$, 
consistent with the marginal detection in the $z'$-band postage-stamp seen in Fig.~\ref{fig:stamps}, and yielding an extremely secure Lyman break.  
As detailed in Table 2, allowing the SED fitting to also include a contribution from Lyman-$\alpha$ emission causes the estimated redshift 
of the stack to rise to $z = 7.12$, with the best fit implying a rest-frame Lyman-$\alpha$ equivalent width of $EW_0 \simeq 40$\,\AA.

\subsubsection{Extinction and star-formation rates}
The best-fitting SED to the four-object stack is based on a tau-model with a characteristic star-formation timescale 
$\tau = 50\,{\rm Myr}$, with solar metallicity and moderate reddening of $A_V = 0.3$ mag. This is somewhat lower than the value  of $A_V = 0.75$ derived 
at $z \simeq 6$ by \citet{Willott2012} from a stack of 23 $z \simeq 6$ galaxies with $z' \simeq 25.0$.

However, we caution that 
dust reddening is, of course, degenerate with age/metallicity, and may also be exaggerated in single-component fits such as those utilised here, as there is evidence that a two-component
star-formation history may better reproduce the combination of fairly blue UV continuum (as displayed by this stack - see below) 
and (relatively) red UV-to-optical colour seen in the SEDs of 
many $z \simeq 7$ galaxies (Curtis-Lake et al., in preparation). The uncertainty in derived $A_V$ is important to bear in mind when considering the 
SFRs and sSFRs of these galaxies because, as can be seen from columns 12 and 13 in Table~\ref{table:properties}, 
the galaxies with the most extreme inferred SFRs all have correspondingly high values of $A_V$. It is for this reason that we have also judged it 
helpful, in the final column of Table 2, to provide an alternative estimate of SFR, based simply on the~\citet{Madau1998} conversion from UV flux density, 
assuming zero dust extinction. 

The SFR of the stack derived from the best-fitting SED assuming a constant SFH is 
98\,${\rm M}_{\sun}{\rm yr}^{-1}$ or, if no dust extinction is assumed (i.e. using the~\citealt{Madau1998} prescription), the ${\rm SFR} =  32\,{\rm M}_{\sun}{\rm yr}^{-1}$. 
From Table~\ref{table:properties} it can be seen that, with zero dust, the SFRs of the individual galaxies lie in the range $25 - 50\,{\rm M}_{\odot}{\rm yr}^{-1}$, but that adopting the best-fitting SED models can yield values 
as high as several hundred ${\rm M}_{\odot}{\rm yr}^{-1}$, as a result of best-fitting values of $A_V$ reaching as high as unity (although the single object with $A_V > 1$ is not from our `robust' sample of objects).

\subsubsection{UV slopes}
An estimate of the UV continuum spectral slope, $\beta$ ($f_{\lambda} \propto \lambda^\beta$), can be derived from the stacked near-infrared photometry in a number of ways.
Following~\citet{Bouwens2010} and~\citet{Dunlop2012a}, a simple estimate of $\beta$ can be derived from the $J-H$ colour (deliberately avoiding the $Y$-band 
which could be contaminated by Ly$\alpha$ emission). This yields a highly uncertain value of $\beta = -1.1 \pm 0.9$, but the availability of the $K_s$ photometry enables a more accurate estimate 
based on power-law fitting to the $J,H,K_s$ photometry which gives $\beta = -2.0 \pm 0.2$ (where the error is obtained from Monte Carlo simulations where the photometry is varied randomly according to the derived errors).
Including the $Y$-band photometry as well (i.e. fitting to $Y,J,H,K_s$) yields a slightly bluer value $\beta = -2.16 \pm 0.18$, but this could potentially be affected by Ly$\alpha$ emission and is, in any case, 
still clearly consistent with $\beta \simeq -2$. We thus adopt $\beta = -2.0 \pm 0.2$ as the best estimate of $\beta$ for the four-object stack, and note that this is perfectly 
consistent with the value deduced by~\citet{Dunlop2012a} and~\citet{McLure2011} for the most luminous of the $z \simeq 7$ galaxies uncovered by the 
HST WFC3/IR surveys to date (albeit these are $\gtrsim 1.5$ mag less luminous than 
those considered here). Our result is also consistent with the value obtained by~\citet{Finkelstein2011} who found $\beta = -2.04^{+0.17}_{-0.27}$ for a sample of lower-luminosity galaxies with $L > 0.75L^*$ at $z=7$, 
with a weak trend to redder colours with increasing luminosity. However, our result is significantly lower than the value of $\beta$ derived at $z \simeq 6$ by~\citet{Willott2012}, who inferred a significantly redder value of 
$\beta = -1.44 \pm 0.10$ from their stacked $Y,J,H,K_s$ photometry.

\subsubsection{Stellar masses and specific star-formation rates}
The stellar mass derived from the SED fitting of the stacked photometry is $M_* = 4 \times 10^9{\rm M}_{\sun}$, which can be compared with the value 
found by~\citet{Willott2012} of $M_*\simeq 10^{10}{\rm M}_{\sun}$ for bright galaxies at $z = 6$ (although the results are consistent within the errors);  
 we note that the galaxy detected by~\citet{Hsieh2012}, if really at $z_{phot} \sim 7.8$,  has a substantially 
larger inferred stellar mass of $M_* = 3.2 \times 10^{10}\,{\rm M}_{\sun}$. 

Based on the best-fitting model, the specific SFR (sSFR) of the stack is then $\simeq 30$\,Gyr$^{-1}$, falling to $\simeq$ 8\,Gyr$^{-1}$ if zero dust extinction is assumed. Clearly both 
these values (and indeed the values inferred from Table 2 for most of the individual objects) 
are higher than the average value of $\simeq 2-4$\,Gyr$^{-1}$ generally reported for studies of fainter Lyman-break galaxies 
at these redshifts (e.g.~\citealt{Gonzalez2010}), although they lie within 
the range of values found for individual objects by McLure et al. (2011). However it is probably premature to over-interpret 
these values as they could be biased high for potentially two reasons. First, if there really is a significant range 
in sSFR at a given stellar mass, then since the UltraVISTA data has only just reached the depth required to select genuine $z \simeq 7$ galaxies, it is 
likely that those galaxies detected on the basis of their rest-frame UV flux will be biased towards high values of sSFR. Second, even if the intrinsic 
range in sSFR is small, photometric errors will conspire to yield a systematic over-estimate of the rest-frame UV luminosities of our candidates, a point 
which is also important for the estimation of number density as a function of magnitude (as discussed further below).

\subsection{Luminosity Function}\label{sect:lf}

\begin{figure*}
\begin{center}
\subfloat{\includegraphics[width = 0.5\textwidth]{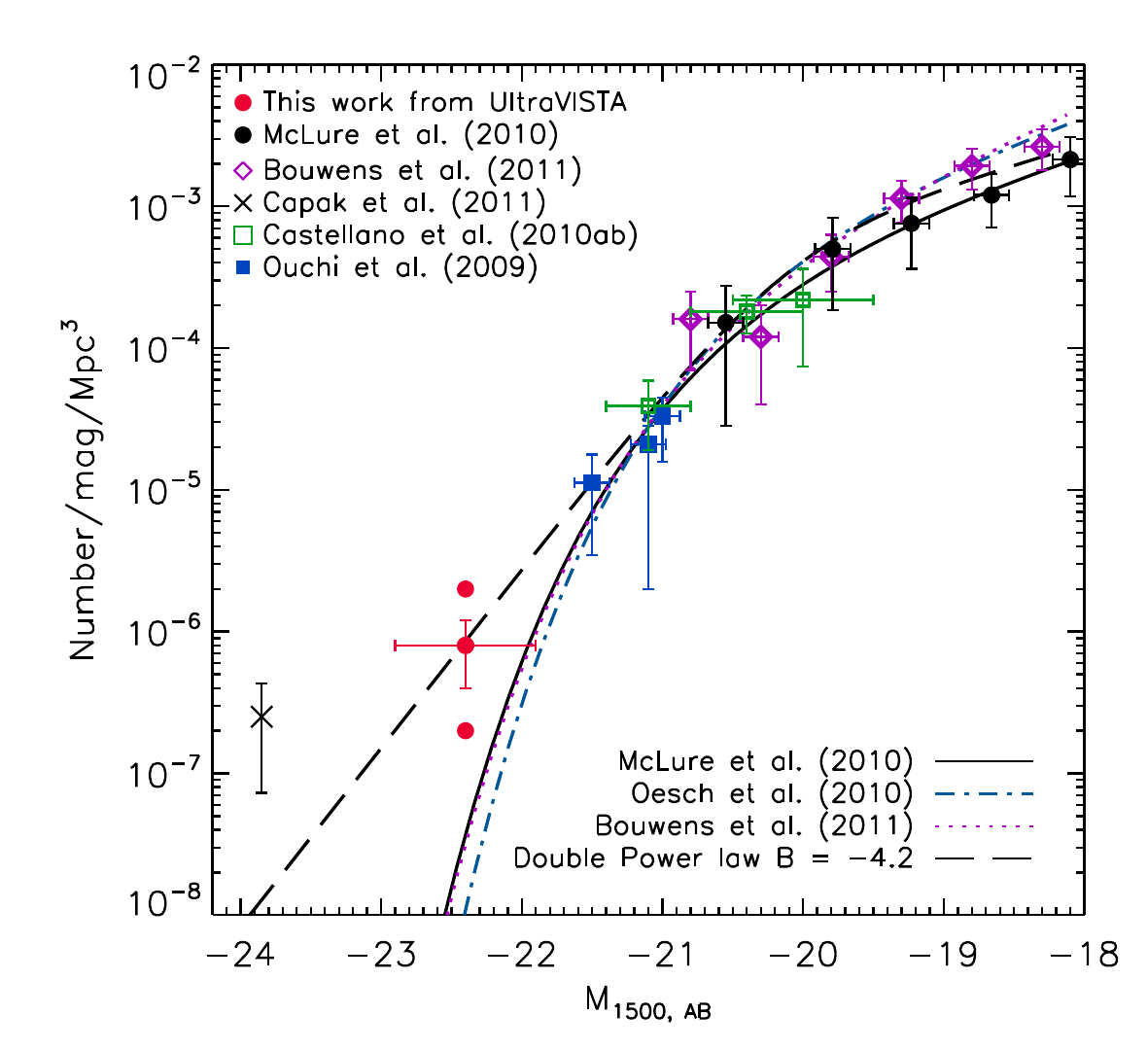}}
\subfloat{\includegraphics[width = 0.5\textwidth]{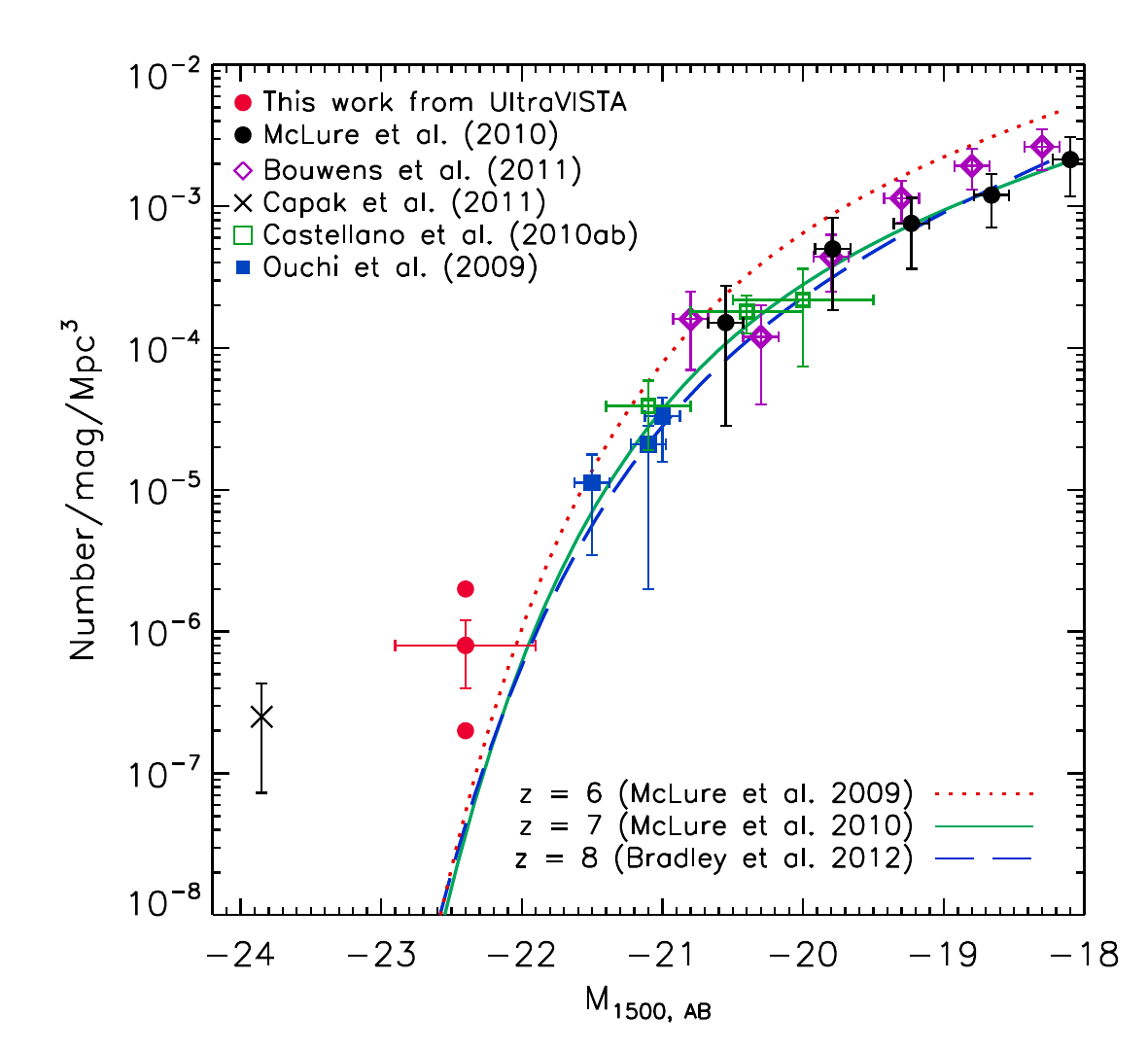}}
\caption{The left-hand panel shows the $z=7$ galaxy UV ($\simeq 1500$\,\AA) LF with data shown from~\citet{McLure2010},~\citet{Bouwens2011a},~\citet{Castellano2010a, Castellano2010b} 
and~\citet{Ouchi2009} overplotted on the best-fitting Schechter functions derived by \citet{Bouwens2011a},~\citet{Oesch2010} and~\citet{McLure2010}
(note the LFs derived by Ouchi et al. (2009) and McLure et al. (2010) are essentially identical).  
The estimated data-point from our new UltraVISTA study is shown in red, with Poissonian errors on the number density, with the upper, central and lower point given by ten, four or only one 
candidate in our sample being a confirmed as a $z > 6.5$ galaxy.  
A double power-law curve is also included for comparison, where the faint-end slope is matched to that of~\citet{McLure2011} and at the bright end we adopt
B$ = -4.2$ (see Section~\ref{sect:discussion} for parameterisation).  In the right-hand plot we show the same data, but this 
time compare with the the $z = 6,7,8$ LFs from~\citet{McLure2009},~\citet{McLure2010} and~\citet{Bradley2012} respectively (see text for further details and 
discussion).}
\label{fig:lf}
\end{center}
\end{figure*}

A full analysis of the implications of the new, luminous $z \simeq 7$ galaxies reported here for the form and evolution of the UV galaxy LF 
is deferred to a future paper (where we expect to be armed with a larger sample, and will have processed 
the required suite of detailed mock-data simulations). Nevertheless it is still instructive 
to briefly consider how our results compare with expectations based on existing determinations of the galaxy UV LF at these early times.  However, we caution that the LF parameters calculated below rely on an approximate estimate of the survey volume and should be considered a preliminary result.

In the left-hand panel of Fig. 8 we illustrate how our results measure up against currently-published determinations of the UV galaxy LF at $z \simeq 7$
as reported by \citet{Ouchi2009}, \citet{McLure2010}, \citet{Oesch2010} and \citet{Bouwens2011a}. As can be seen from this plot, despite
the continuing (important) debate over the faint-end slope (see Dunlop 2012), these independent determinations of the $z \simeq 7$ LF are in generally 
good agreement, especially between the break luminosity at $M^*_{1500} \simeq -20$ and $M^*_{1500} \simeq -21$, the brightest luminosity 
effectively sampled by the pre-existing ground-based data (in fact the McLure et al. (2010) and Ouchi et al. (2009) LFs are so similar that only one 
can be usefully plotted). This agreement also appears to extend out to the luminosity region probed here by UltraVISTA, at $M^*_{1500} < -22$, but 
it is important to stress that this is largely a result of the common assumption of a Schechter form, and that these curves are simply extrapolations
beyond the range of the pre-existing data.

To locate our new results on this plot we have adopted the  
$J$-band magnitude of the stack (i.e. the average $J$-band magnitude of our four most robust candidates), corrected to total magnitude, 
to calculate the rest-frame UV absolute magnitude at 1500\AA, $M_{1500} = -22.5$. We have then conducted a set of simulations and find that our selection
criteria, based on the $Y$+\,$J$ image, implies that incompleteness and flux-boosting virtually cancel each other in this magnitude regime, and that a 
relatively modest flux-boosting correction of only $\simeq 0.1$ mag is merited. This leads to a data-point in Fig. 8 plotted at $M_{1500} = -22.4$, at 
a comoving number density based on the discovery of four galaxies within our estimated comoving survey volume of 
$5 \times 10^{6} {\rm Mpc}^3$, based on our effective survey area of 0.9\,deg$^2$ (excluding the area unavailable for galaxy selection due to diffraction halos/spikes from bright stars) 
and assuming that our selection function is effectively unity over the redshift range $z \simeq 6.5 - 7.2$.  
The size of the bin in $M_{1500}$ was chosen to be 1\,mag to indicate the range of the observed $J$-band measurements for our four most secure galaxies (as seen in Table~\ref{table:phot}), 
while the error bar in density is based purely on the Poisson noise. Finally, to indicate possible extremes, we also show where this data-point would lie 
in the unlikely circumstance that all ten of the objects presented here are truly at $z > 6.5$, 
or alternatively, if only one high-redshift galaxy were to be confirmed.

It is clear from this plot that, while our results are much less extreme than would be implied from the results of Capak et al. (2011), nevertheless our discovery 
of four robust $z \simeq 7$ galaxies at $M_{1500} \simeq -22.4$ yields a somewhat higher number density than would be anticipated on the basis of simply
extrapolating the existing 
LFs. For example, the McLure et al. (2010) LF predicts only $\simeq 0.8$ galaxies with $J < 25$ in our survey volume. 

Given the small-number statistics, and the potential for cosmic variance (although with the low sample size available here, Poisson noise dominates that from cosmic variance), it would be premature to suggest that our results contradict the steep exponential fall-off 
at the bright end of the LF implied by the current Schechter-function fits to the $z \simeq 7$ LF. However, it also important to stress that, as is clear from Fig. 8, 
our results are certainly not in conflict with the existing {\it data} at this redshift, and would in fact lie naturally on a power-law extrapolation 
from the data at $M^*_{1500} \simeq -21$. To illustrate this we have plotted a simple double power-law LF through the data, with the functional form:

\begin{equation}
   \phi (M) = \frac{\phi^*}{10^{0.4 (A + 1)(M - M^*)} + 10^{0.4(B + 1)(M-M^*)}}.
 \end{equation}
 
\noindent
We set $\phi^*$, $M^*$ and the faint end slope, $A$ ($\equiv \alpha$), to the values determined in the Schechter-function fit by~\citet{McLure2010}, and have simply set the value of the bright-end 
slope, $B$, to make the function pass through our data-point ($B = -4.2$). Even without a detailed analysis it is obvious that this functional form provides a good 
description of all the available data, indicating that it may be premature to conclude that the LF at $z \simeq 7$ is best described by a Schechter function.

Finally, in the second panel of Fig. 8, we show the same $z \simeq 7$ data points, but here we plot the best-fitting galaxy UV LFs at $z \simeq 6$ (McLure et al. 2009; Willott et al 2012),
$z \simeq 7$ (McLure et al. 2010) and $z \simeq 8$ (Bradley et al. 2012). This figure demonstrates that there is in fact currently very little (if any) significant evidence for 
evolution in the {\it bright} end of the galaxy LF in the redshift range $z \simeq 6 - 8$. We note that the only other work directly probing 
the galaxy LF at $M_{1500} < -22$ at these epochs is the study at $ z \simeq 6$ by \citet{Willott2012}, who exploited $\simeq 4$\,deg$^2$ of ground-based CFHT imaging 
to search for $z \simeq 6$ galaxies brighter than $M_{1350} \simeq -21.5$. \citet{Willott2012} found 40 galaxies, many with $M_{1350} < -22$, but 
relatively few with $M_{1350} < -22.5$. Thus, even at $z \simeq 6$ the basic statistics at $M_{UV} < -22.5$ remain poor, but nevertheless 
a maximum likelihood fit to the $z \simeq 6$ LF 
led~\citet{Willott2012} to conclude in favour of the steep exponential cutoff implied by the Schechter-function fit of McLure et al. (2009). 
However, the near-infrared data analysed by Willott et al. (2012) are somewhat inhomogeneous, of inadequate depth to detect the $z \simeq 6$ galaxies
in $J$-band in two of the four survey fields, and lack the $Y$-band data which, ideally, would enable determination of $M_{1500}$ free from 
corrections due to IGM extinction (which can impact on the $z'$-band at these redshifts). Moreover, in the sample reported by Willott et al. 
(2012) the COSMOS field alone contains two $z \simeq 6$ galaxies with $J$-band magnitudes brighter than the most luminous galaxy candidate 
(as based on the $z$-band mag) reported from the full 4\,deg$^2$ survey. We conclude that the form of the bright-end of the LF is still open to debate at both $z \simeq 6$ and $z \simeq 7$.
At $z = 8$,~\citet{Bradley2012} have also claimed to find no evidence for an excess of sources at the bright-end. However, 
due to the (relatively) small effective field size of the current BoRG survey, 
the galaxies detected are at $L < 2L^*$ (the galaxies presented here are $\simeq 9L^*$) and so the constraints on the very bright-end of the $z \simeq 8$ LF
remain weak. The final UltraVISTA imaging is designed to 
improve our knowledge of the bright end of the galaxy UV LF at $z \simeq 8$ as well as at $z \simeq 7$.

\section{Conclusion}
We have exploited the new, deep, near-infrared UltraVISTA imaging of the COSMOS field, in tandem with deep optical and mid-infrared
imaging, to conduct a new search for luminous galaxies at redshifts $z \simeq 7$. The unique multi-wavelength dataset provided by VISTA, CFHT, Subaru, HST and {\it Spitzer} 
over a common area of $\simeq 1$\,deg$^2$ has allowed us to select galaxy candidates at redshifts $z > 6.5$ by searching first for
UltraVISTA $Y$+\,$J$-detected ($<25$\,mag) objects which are {\it undetected} in the CFHT and HST optical data.

This sample was then refined using a photometric redshift fitting code, enabling the rejection of lower-redshift galaxy
contaminants, and cool galactic M, L, T dwarf stars; brown-dwarf contamination is a much more serious problem for wide-area ground-based surveys than for
deeper/narrower HST WFC3/IR surveys, and
so we have taken great care to utilise the full multi-wavelength dataset (fitting the latest stellar templates), including IRAC colours, to minimise dwarf-star contamination.

The final result of this process is a small sample of (at most) 10
credible galaxy candidates at $z > 6.5$ (from over 200,000 galaxies detected in the year-one UltraVISTA data). The first four of these objects appear to be
robust galaxies at $z > 6.5$, and fitting to their stacked spectral energy distribution yields $z_{phot} = 6.98 \pm 0.05$ with a stellar mass $M_*\simeq 5\times10^9\,{\rm M_{\odot}}$
and rest-frame UV spectral slope $\beta \simeq -2.0 \pm 0.2$ (where $f_{\lambda} \propto \lambda^{\beta}$).
The next three are also good candidates for $z > 6.5$ galaxies,
but the possibility that they are low redshift interlopers or dwarf stars cannot be excluded.
Our final subset of three additional candidates is afflicted not only by potential
dwarf-star contamination, but also contains objects likely to lie at redshifts just below $z = 6.5$.

We have also been able to demonstrate that the three even-brighter $z \gtrsim 7$ galaxy candidates reported in the COSMOS field by~\citet{Capak2011} are in fact all lower-redshift galaxies at $z \simeq 1.5-3.5$. Consequently
the new $z \simeq 7$ galaxies reported here are the first credible $z \simeq 7$ Lyman-break galaxies discovered in the COSMOS field and, as the most UV-luminous discovered to date at these redshifts, are
prime targets for deep follow-up spectroscopy.
We have investigated the physical properties of these galaxies as inferred from the broad-band photometry, and have discussed the uncertainties in, and implications of their estimated 
star-formation rates and stellar masses. Finally, we have considered only briefly the 
implications of the inferred number density of these ``bright'' galaxies for the form of the galaxy luminosity function at these early epochs, 
deferring a full re-analysis of the $z \simeq 7$ UV luminosity function to a future paper (when we expect to be armed with a larger sample of objects).

\section*{Acknowledgements}
RAAB and JSD acknowledge the support of the European Research Council via
the award of an Advanced Grant. JSD and RJM acknowledge the support 
of the Royal Society via a Wolfson Research Merit Award and a University 
Research Fellowship respectively.  RJM acknowledges the support of the Leverhulme Trust via the award of a Philip Leverhulme research prize.
HM was supported by ANR grant ANR-07-BLAN-0228. BMJ and JPUF acknowledge support from the ERC-StG grant EGGS-278202.  
The Dark Cosmology Centre is funded by the Danish National Research Foundation.

This work is based on data products from observations made with ESO Telescopes at the La Silla Paranal Observatories under ESO programme ID 179.A-2005 and on data products produced by TERAPIX and the Cambridge Astronomy survey Unit on behalf of the UltraVISTA consortium. This study was based in part on observations obtained with MegaPrime/MegaCam, 
a joint project of CFHT and CEA/DAPNIA, at the Canada-France-Hawaii Telescope (CFHT) which is operated by the National Research Council (NRC) of Canada, the Institut National des Science de l'Univers of the Centre National de la Recherche Scientifique (CNRS) of France, and the University of Hawaii. This work is based in part on data products produced at TERAPIX and the Canadian Astronomy Data Centre as part of the Canada-France-Hawaii Telescope Legacy Survey, a collaborative project of NRC and CNRS. This work is based in part on observations made with the NASA/ESA {\it Hubble Space Telescope}, which is operated by the Association 
of Universities for Research in Astronomy, Inc, under NASA contract NAS5-26555.
This work is based in part on observations made with the {\it Spitzer Space Telescope}, which is operated by the Jet Propulsion Laboratory, 
California Institute of Technology under NASA contract 1407. We thank the staff of the Subaru telescope for their assistance with the $z'$-band imaging utilised here.

This research has benefitted from the SpeX Prism Spectral Libraries, maintained by Adam Burgasser at http://pono.ucsd.edu/$\sim$adam/browndwarfs/spexprism.

\label{lastpage}

\bibliographystyle{mn2e}
\bibliography{library_abbrv_revised}

\begin{thebibliography}{59}
\expandafter\ifx\csname natexlab\endcsname\relax\def\natexlab#1{#1}\fi

\bibitem[{Allard {et~al}\mbox{.}(2001)Allard, Hauschildt, Alexander, Tamanai,
  \& Schweitzer}]{Allard2001}
Allard F., Hauschildt P.~H., Alexander D.~R., Tamanai A., Schweitzer A., 2001,
  ApJ, 556, 357

\bibitem[{Bertin \& Arnouts(1996)}]{Bertin1996}
Bertin E., Arnouts S., 1996, A\&AS, 117, 393

\bibitem[{Bertin {et~al}\mbox{.}(2002)Bertin, Mellier, Radovich, Missonnier,
  Didelon, \& Morin}]{Bertin2002}
Bertin E., Mellier Y., Radovich M., Missonnier G., Didelon P., Morin B., 2002,
  ASP Conference Proceedings, 281

\bibitem[{Bielby {et~al}\mbox{.}(2011)Bielby, Hudelot, McCracken, Ilbert,
  Daddi, {Le F\`{e}vre}, Gonzalez-Perez, Kneib, Marmo, Mellier, Salvato,
  Sanders, \& Willott}]{Bielby2011}
Bielby R. {et~al.}, 2011, A\&A, submitted (arXiv:1111.6997)

\bibitem[{Bouwens {et~al}\mbox{.}(2007)Bouwens, Illingworth, Franx, \&
  Ford}]{Bouwens2007}
Bouwens R.~J., Illingworth G.~D., Franx M., Ford H., 2007, ApJ, 670, 928

\bibitem[{Bouwens {et~al}\mbox{.}(2011)Bouwens, Illingworth, Oesch, Labb\'{e},
  Trenti, van Dokkum, Franx, Stiavelli, Carollo, Magee, \&
  Gonzalez}]{Bouwens2011a}
Bouwens R.~J. {et~al.}, 2011, ApJ, 737, 90

\bibitem[{Bouwens {et~al}\mbox{.}(2010)Bouwens, Illingworth, Oesch, Trenti,
  Stiavelli, Carollo, Franx, van Dokkum, Labb\'{e}, \& Magee}]{Bouwens2010}
Bouwens R.~J. {et~al.}, 2010, ApJ, 708, L69

\bibitem[{Bradley {et~al}\mbox{.}(2012)Bradley, Trenti, Oesch, Stiavelli, Treu,
  Bouwens, Shull, Holwerda, \& Pirzkal}]{Bradley2012}
Bradley L.~D. {et~al.}, 2012, ApJ, submitted (arXiv:1204.3641)

\bibitem[{Bruzual \& Charlot(2003)}]{Bruzual2003}
Bruzual G., Charlot S., 2003, MNRAS, 344, 1000

\bibitem[{Burningham {et~al}\mbox{.}(2010)Burningham, Pinfield, Lucas, Leggett,
  Deacon, Tamura, Tinney, Lodieu, Zhang, Huelamo, Jones, Murray, Mortlock,
  Patel, {Barrado y Navascu\'{e}s}, {Zapatero Osorio}, Ishii, Kuzuhara, \&
  Smart}]{Burningham2010}
Burningham B. {et~al.}, 2010, MNRAS, 406, 1885

\bibitem[{Calzetti {et~al}\mbox{.}(2000)Calzetti, Armus, Bohlin, Kinney,
  Koornneef, \& Storchi-Bergmann}]{Calzetti2000}
Calzetti D., Armus L., Bohlin R.~C., Kinney A.~L., Koornneef J.,
  Storchi-Bergmann T., 2000, ApJ, 533, 682

\bibitem[{Capak {et~al}\mbox{.}(2007)Capak, Aussel, Ajiki, McCracken, Mobasher,
  Scoville, Shopbell, Taniguchi, Thompson, Tribiano, Sasaki, Blain, Brusa,
  Carilli, Comastri, Carollo, Cassata, Colbert, Ellis, Elvis, Giavalisco,
  Green, Guzzo, Hasinger, Ilbert, Impey, Jahnke, Kartaltepe, Kneib, Koda,
  Koekemoer, Komiyama, Leauthaud, Lefevre, Lilly, Liu, Massey, Miyazaki,
  Murayama, Nagao, Peacock, Pickles, Porciani, Renzini, Rhodes, Rich, Salvato,
  Sanders, Scarlata, Schiminovich, Schinnerer, Scodeggio, Sheth, Shioya, Tasca,
  Taylor, Yan, \& Zamorani}]{Capak2007}
Capak P. {et~al.}, 2007, ApJS, 172, 99

\bibitem[{Capak {et~al}\mbox{.}(2011)Capak, Mobasher, Scoville, McCracken,
  Ilbert, Salvato, Men\'{e}ndez-Delmestre, Aussel, Carilli, Civano, Elvis,
  Giavalisco, Jullo, Kartaltepe, Leauthaud, Koekemoer, Kneib, LeFloch, Sanders,
  Schinnerer, Shioya, Shopbell, Tanaguchi, Thompson, \& Willott}]{Capak2011}
Capak P. {et~al.}, 2011, ApJ, 730, 68

\bibitem[{Castellano {et~al}\mbox{.}(2010{\natexlab{a}})Castellano, Fontana,
  Boutsia, Grazian, Pentericci, Bouwens, Dickinson, Giavalisco, Santini,
  Cristiani, Fiore, Gallozzi, Giallongo, Maiolino, Mannucci, Menci, Moorwood,
  Nonino, Paris, Renzini, Rosati, Salimbeni, Testa, \&
  Vanzella}]{Castellano2010a}
Castellano M. {et~al.}, 2010{\natexlab{a}}, A\&A, 511, A20

\bibitem[{Castellano {et~al}\mbox{.}(2010{\natexlab{b}})Castellano, Fontana,
  Paris, Grazian, Pentericci, Boutsia, Santini, Testa, Dickinson, Giavalisco,
  Bouwens, Cuby, Mannucci, Cl\'{e}ment, Cristiani, Fiore, Gallozzi, Giallongo,
  Maiolino, Menci, Moorwood, Nonino, Renzini, Rosati, Salimbeni, \&
  Vanzella}]{Castellano2010b}
Castellano M. {et~al.}, 2010{\natexlab{b}}, A\&A, 524, A28

\bibitem[{Chabrier(2003)}]{Chabrier2003}
Chabrier G., 2003, PASP, 115, 763

\bibitem[{Curtis-Lake {et~al}\mbox{.}(2012)Curtis-Lake, McLure, Pearce, Dunlop,
  Cirasuolo, Stark, Almaini, Bradshaw, Chuter, Foucaud, \&
  Hartley}]{CurtisLake2012}
Curtis-Lake E. {et~al.}, 2012, MNRAS, 422, 1425

\bibitem[{Dalton {et~al}\mbox{.}(2006)Dalton, Caldwell, Ward, Whalley,
  Woodhouse, Edeson, Clark, Beard, Gallie, Todd, Strachan, Bexawada,
  Sutherland, \& Emerson}]{Dalton2006}
Dalton G.~B. {et~al.}, 2006, Proc. SPIE, 6269, 30

\bibitem[{Dunlop(2012)}]{Dunlopbook2012}
Dunlop J.~S., 2012, arXiv:1205.1543

\bibitem[{Dunlop {et~al}\mbox{.}(2012)Dunlop, McLure, Robertson, Ellis, Stark,
  Cirasuolo, \& de~Ravel}]{Dunlop2012a}
Dunlop J.~S., McLure R.~J., Robertson B.~E., Ellis R.~S., Stark D.~P.,
  Cirasuolo M., de~Ravel L., 2012, MNRAS, 420, 901

\bibitem[{Emerson \& Sutherland(2010)}]{Emerson2010}
Emerson J.~P., Sutherland W.~J., 2010, Proc. SPIE, 7733, 4

\bibitem[{Findlay {et~al}\mbox{.}(2012)Findlay, Sutherland, Venemans,
  Reyl\'{e}, Robin, Bonfield, Bruce, \& Jarvis}]{Findlay2012}
Findlay J.~R., Sutherland W.~J., Venemans B.~P., Reyl\'{e} C., Robin A.~C.,
  Bonfield D.~G., Bruce V.~A., Jarvis M.~J., 2012, MNRAS, 419, 3354

\bibitem[{Finkelstein {et~al}\mbox{.}(2010)Finkelstein, Papovich, Giavalisco,
  Reddy, Ferguson, Koekemoer, \& Dickinson}]{Finkelstein2010}
Finkelstein S.~L., Papovich C., Giavalisco M., Reddy N.~A., Ferguson H.~C.,
  Koekemoer A.~M., Dickinson M., 2010, ApJ, 719, 1250

\bibitem[{Finkelstein {et~al}\mbox{.}(2011)Finkelstein, Papovich, Salmon,
  Finlator, Dickinson, Ferguson, Giavalisco, Koekemoer, Reddy, Bassett,
  Conselice, Dunlop, Faber, Grogin, Hathi, Kocevski, Lai, Lee, McLure,
  Mobasher, \& Newman}]{Finkelstein2011}
Finkelstein S.~L. {et~al.}, 2011, ApJ, in press (arXiv:1110.3785)

\bibitem[{Finlator {et~al}\mbox{.}(2011)Finlator, Oppenheimer, \&
  Dav\'{e}}]{Finlator2011}
Finlator K., Oppenheimer B.~D., Dav\'{e} R., 2011, MNRAS, 410, 1703

\bibitem[{Gonz\'{a}lez {et~al}\mbox{.}(2010)Gonz\'{a}lez, Labb\'{e}, Bouwens,
  Illingworth, Franx, Kriek, \& Brammer}]{Gonzalez2010}
Gonz\'{a}lez V., Labb\'{e} I., Bouwens R.~J., Illingworth G., Franx M., Kriek
  M., Brammer G.~B., 2010, ApJ, 713, 115

\bibitem[{Grogin {et~al}\mbox{.}(2011)Grogin, Kocevski, Faber, Ferguson,
  Koekemoer, Riess, Acquaviva, Alexander, Almaini, Ashby, Barden, Bell,
  Bournaud, Brown, Caputi, Casertano, Cassata, Castellano, Challis, Chary,
  Cheung, Cirasuolo, Conselice, Cooray, Croton, Daddi, Dahlen, Dav\'{e},
  de~Mello, Dekel, Dickinson, Dolch, Donley, Dunlop, Dutton, Elbaz, Fazio,
  Filippenko, Finkelstein, Fontana, Gardner, Garnavich, Gawiser, Giavalisco,
  Grazian, Guo, Hathi, H\"{a}ussler, Hopkins, Huang, Huang, Jha, Kartaltepe,
  Kirshner, Koo, Lai, Lee, Li, Lotz, Lucas, Madau, McCarthy, McGrath, McIntosh,
  McLure, Mobasher, Moustakas, Mozena, Nandra, Newman, Niemi, Noeske, Papovich,
  Pentericci, Pope, Primack, Rajan, Ravindranath, Reddy, Renzini, Rix, Robaina,
  Rodney, Rosario, Rosati, Salimbeni, Scarlata, Siana, Simard, Smidt,
  Somerville, Spinrad, Straughn, Strolger, Telford, Teplitz, Trump, van~der
  Wel, Villforth, Wechsler, Weiner, Wiklind, Wild, Wilson, Wuyts, Yan, \&
  Yun}]{Grogin2011}
Grogin N.~A. {et~al.}, 2011, ApJS, 197, 35

\bibitem[{Hsieh {et~al}\mbox{.}(2012)Hsieh, Wang, Yan, Lin, Karoji, Lim, Ho, \&
  Tsai}]{Hsieh2012}
Hsieh B., Wang W., Yan H., Lin L., Karoji H., Lim J., Ho P. T.~P., Tsai C.,
  2012, ApJ, 749, 88

\bibitem[{Ilbert {et~al}\mbox{.}(2008)Ilbert, Salvato, Capak, {Le Floc'h},
  Aussel, McCracken, Arnouts, Mobasher, Sanders, Scoville, \&
  Taniguchi}]{Ilbert2008}
Ilbert O. {et~al.}, 2008, ASP Conf. Ser., 399

\bibitem[{Jaacks {et~al}\mbox{.}(2012)Jaacks, Choi, Nagamine, Thompson, \&
  Varghese}]{Jaacks2012}
Jaacks J., Choi J.-H., Nagamine K., Thompson R., Varghese S., 2012, MNRAS, 420,
  1606

\bibitem[{Kirkpatrick {et~al}\mbox{.}(2011)Kirkpatrick, Cushing, Gelino,
  Griffith, Skrutskie, Marsh, Wright, Mainzer, Eisenhardt, McLean, Thompson,
  Bauer, Benford, Bridge, Lake, Petty, Stanford, Tsai, Bailey, Beichman, Bloom,
  Bochanski, Burgasser, Capak, Cruz, Hinz, Kartaltepe, Knox, Manohar, Masters,
  Morales-Calder\'{o}n, Prato, Rodigas, Salvato, Schurr, Scoville, Simcoe,
  Stapelfeldt, Stern, Stock, \& Vacca}]{DavyKirkpatrick2011}
Kirkpatrick J.~D. {et~al.}, 2011, ApJS, 197, 19

\bibitem[{Knapp {et~al}\mbox{.}(2004)Knapp, Leggett, Fan, Marley, Geballe,
  Golimowski, Finkbeiner, Gunn, Hennawi, Ivezi\'{c}, Lupton, Schlegel, Strauss,
  Tsvetanov, Chiu, Hoversten, Glazebrook, Zheng, Hendrickson, Williams, Uomoto,
  Vrba, Henden, Luginbuhl, Guetter, Munn, Canzian, Schneider, \&
  Brinkmann}]{Knapp2004}
Knapp G.~R. {et~al.}, 2004, AJ, 127, 3553

\bibitem[{Koekemoer {et~al}\mbox{.}(2007)Koekemoer, Aussel, Calzetti, Capak,
  Giavalisco, Kneib, Leauthaud, {Le Fevre}, McCracken, Massey, Mobasher,
  Rhodes, Scoville, \& Shopbell}]{Koekemoer2007}
Koekemoer A.~M. {et~al.}, 2007, ApJS, 172, 196

\bibitem[{Koekemoer {et~al}\mbox{.}(2011)Koekemoer, Faber, Ferguson, Grogin,
  Kocevski, Koo, Lai, Lotz, Lucas, McGrath, Ogaz, Rajan, Riess, Rodney,
  Strolger, Casertano, Castellano, Dahlen, Dickinson, Dolch, Fontana,
  Giavalisco, Grazian, Guo, Hathi, Huang, van~der Wel, Yan, Acquaviva,
  Alexander, Almaini, Ashby, Barden, Bell, Bournaud, Brown, Caputi, Cassata,
  Challis, Chary, Cheung, Cirasuolo, Conselice, Cooray, Croton, Daddi,
  Dav\'{e}, de~Mello, de~Ravel, Dekel, Donley, Dunlop, Dutton, Elbaz, Fazio,
  Filippenko, Finkelstein, Frazer, Gardner, Garnavich, Gawiser, Gruetzbauch,
  Hartley, H\"{a}ussler, Herrington, Hopkins, Huang, Jha, Johnson, Kartaltepe,
  Khostovan, Kirshner, Lani, Lee, Li, Madau, McCarthy, McIntosh, McLure,
  McPartland, Mobasher, Moreira, Mortlock, Moustakas, Mozena, Nandra, Newman,
  Nielsen, Niemi, Noeske, Papovich, Pentericci, Pope, Primack, Ravindranath,
  Reddy, Renzini, Rix, Robaina, Rosario, Rosati, Salimbeni, Scarlata, Siana,
  Simard, Smidt, Snyder, Somerville, Spinrad, Straughn, Telford, Teplitz,
  Trump, Vargas, Villforth, Wagner, Wandro, Wechsler, Weiner, Wiklind, Wild,
  Wilson, Wuyts, \& Yun}]{Koekemoer2011}
Koekemoer A.~M. {et~al.}, 2011, ApJS, 197, 36

\bibitem[{Lawrence {et~al}\mbox{.}(2007)Lawrence, Warren, Almaini, Edge,
  Hambly, Jameson, Lucas, Casali, Adamson, Dye, Emerson, Foucaud, Hewett,
  Hirst, Hodgkin, Irwin, Lodieu, McMahon, Simpson, Smail, Mortlock, \&
  Folger}]{Lawrence2007}
Lawrence A. {et~al.}, 2007, MNRAS, 379, 1599

\bibitem[{Madau(1995)}]{Madau1995}
Madau P., 1995, ApJ, 441, 18

\bibitem[{Madau {et~al}\mbox{.}(1998)Madau, Pozzetti, \& Dickinson}]{Madau1998}
Madau P., Pozzetti L., Dickinson M., 1998, ApJ, 498, 106

\bibitem[{Massey {et~al}\mbox{.}(2010)Massey, Stoughton, Leauthaud, Rhodes,
  Koekemoer, Ellis, \& Shaghoulian}]{Massey2010}
Massey R., Stoughton C., Leauthaud A., Rhodes J., Koekemoer A., Ellis R.,
  Shaghoulian E., 2010, MNRAS, 401, 371

\bibitem[{McCracken {et~al}\mbox{.}(2010)McCracken, Capak, Salvato, Aussel,
  Thompson, Daddi, Sanders, Kneib, Willott, Mancini, Renzini, Cook, {Le
  F\`{e}vre}, Ilbert, Kartaltepe, Koekemoer, Mellier, Murayama, Scoville,
  Shioya, \& Tanaguchi}]{McCracken2010}
McCracken H.~J. {et~al.}, 2010, ApJ, 708, 202

\bibitem[{McCracken {et~al}\mbox{.}(2012)McCracken, Milvang-Jensen, Dunlop,
  Franx, Fynbo, {Le F\`{e}vre}, Holt, Caputi, Goranova, Buitrago, Emerson,
  Freudling, Hudelot, L\'{o}pez-Sanjuan, Magnard, Mellier, M\o~ller, Nilsson,
  Sutherland, Tasca, \& Zabl}]{McCracken2012}
McCracken H.~J. {et~al.}, 2012, A\&A, in press (arXiv:1204.6586)

\bibitem[{McLure {et~al}\mbox{.}(2009)McLure, Cirasuolo, Dunlop, Foucaud, \&
  Almaini}]{McLure2009}
McLure R.~J., Cirasuolo M., Dunlop J.~S., Foucaud S., Almaini O., 2009, MNRAS,
  395, 2196

\bibitem[{McLure {et~al}\mbox{.}(2006)McLure, Cirasuolo, Dunlop, Sekiguchi,
  Almaini, Foucaud, Simpson, Watson, Hirst, Page, \& Smail}]{McLure2006}
McLure R.~J. {et~al.}, 2006, MNRAS, 372, 357

\bibitem[{McLure {et~al}\mbox{.}(2010)McLure, Dunlop, Cirasuolo, Koekemoer,
  Sabbi, Stark, Targett, \& Ellis}]{McLure2010}
McLure R.~J., Dunlop J.~S., Cirasuolo M., Koekemoer A.~M., Sabbi E., Stark
  D.~P., Targett T.~A., Ellis R.~S., 2010, MNRAS, 403, 960

\bibitem[{McLure {et~al}\mbox{.}(2011)McLure, Dunlop, de~Ravel, Cirasuolo,
  Ellis, Schenker, Robertson, Koekemoer, Stark, \& Bowler}]{McLure2011}
McLure R.~J. {et~al.}, 2011, MNRAS, 418, 2074

\bibitem[{Oesch {et~al}\mbox{.}(2010)Oesch, Bouwens, Illingworth, Carollo,
  Franx, Labb\'{e}, Magee, Stiavelli, Trenti, \& van Dokkum}]{Oesch2010}
Oesch P.~A. {et~al.}, 2010, ApJ, 709, L16

\bibitem[{Oesch {et~al}\mbox{.}(2012)Oesch, Bouwens, Illingworth, Gonzalez,
  Trenti, van Dokkum, Franx, Labbe, Carollo, \& Magee}]{Oesch2012}
Oesch P.~A. {et~al.}, 2012, ApJ, submitted (arXiv:1201.0755)

\bibitem[{Oke \& Gunn(1983)}]{Oke1983}
Oke J.~B., Gunn J.~E., 1983, ApJ, 266, 713

\bibitem[{Ouchi {et~al}\mbox{.}(2009)Ouchi, Mobasher, Shimasaku, Ferguson,
  Fall, Ono, Kashikawa, Morokuma, Nakajima, Okamura, Dickinson, Giavalisco, \&
  Ohta}]{Ouchi2009}
Ouchi M. {et~al.}, 2009, ApJ, 706, 1136

\bibitem[{Patel(2010)}]{Patel2010}
Patel M., 2010, PhD thesis, Imperial College London

\bibitem[{Patten {et~al}\mbox{.}(2006)Patten, Stauffer, Burrows, Marengo, Hora,
  Luhman, Sonnett, Henry, Raghavan, Megeath, Liebert, \& Fazio}]{Patten2006}
Patten B.~M. {et~al.}, 2006, ApJ, 651, 502

\bibitem[{Robertson {et~al}\mbox{.}(2010)Robertson, Ellis, Dunlop, McLure, \&
  Stark}]{Robertson2010}
Robertson B.~E., Ellis R.~S., Dunlop J.~S., McLure R.~J., Stark D.~P., 2010,
  Nat, 468, 49

\bibitem[{Salvato {et~al}\mbox{.}(2011)Salvato, Ilbert, Hasinger, Rau, Civano,
  Zamorani, Brusa, Elvis, Vignali, Aussel, Comastri, Fiore, {Le Floc'h},
  Mainieri, Bardelli, Bolzonella, Bongiorno, Capak, Caputi, Cappelluti,
  Carollo, Contini, Garilli, Iovino, Fotopoulou, Fruscione, Gilli, Halliday,
  Kneib, Kakazu, Kartaltepe, Koekemoer, Kovac, Ideue, Ikeda, Impey, {Le Fevre},
  Lamareille, Lanzuisi, {Le Borgne}, {Le Brun}, Lilly, Maier, Manohar, Masters,
  McCracken, Messias, Mignoli, Mobasher, Nagao, Pello, Puccetti, Perez-Montero,
  Renzini, Sargent, Sanders, Scodeggio, Scoville, Shopbell, Silvermann,
  Taniguchi, Tasca, Tresse, Trump, \& Zucca}]{Salvato2011}
Salvato M. {et~al.}, 2011, ApJ, 742, 61

\bibitem[{Sanders {et~al}\mbox{.}(2007)Sanders, Salvato, Aussel, Ilbert,
  Scoville, Surace, Frayer, Sheth, Helou, Brooke, Bhattacharya, Yan,
  Kartaltepe, Barnes, Blain, Calzetti, Capak, Carilli, Carollo, Comastri,
  Daddi, Ellis, Elvis, Fall, Franceschini, Giavalisco, Hasinger, Impey,
  Koekemoer, {Le Fevre}, Lilly, Liu, McCracken, Mobasher, Renzini, Rich,
  Schinnerer, Shopbell, Taniguchi, Thompson, Urry, \& Williams}]{Sanders2007}
Sanders D.~B. {et~al.}, 2007, ApJS, 172, 86

\bibitem[{Scoville {et~al}\mbox{.}(2007{\natexlab{a}})Scoville, Abraham,
  Aussel, Barnes, Benson, Blain, Calzetti, Comastri, Capak, Carilli, Carlstrom,
  Carollo, Colbert, Daddi, Ellis, Elvis, Ewald, Fall, Franceschini, Giavalisco,
  Green, Griffiths, Guzzo, Hasinger, Impey, Kneib, Koda, Koekemoer, Lefevre,
  Lilly, Liu, McCracken, Massey, Mellier, Miyazaki, Mobasher, Mould, Norman,
  Refregier, Renzini, Rhodes, Rich, Sanders, Schiminovich, Schinnerer,
  Scodeggio, Sheth, Shopbell, Taniguchi, Tyson, Urry, {Van Waerbeke},
  Vettolani, White, \& Yan}]{Scoville2007a}
Scoville N. {et~al.}, 2007{\natexlab{a}}, ApJS, 172, 38

\bibitem[{Scoville {et~al}\mbox{.}(2007{\natexlab{b}})Scoville, Aussel, Brusa,
  Capak, Carollo, Elvis, Giavalisco, Guzzo, Hasinger, Impey, Kneib, LeFevre,
  Lilly, Mobasher, Renzini, Rich, Sanders, Schinnerer, Schminovich, Shopbell,
  Taniguchi, \& Tyson}]{Scoville2007}
Scoville N. {et~al.}, 2007{\natexlab{b}}, ApJS, 172, 1

\bibitem[{Stanway {et~al}\mbox{.}(2008)Stanway, Bremer, Squitieri, Douglas, \&
  Lehnert}]{Stanway2008}
Stanway E.~R., Bremer M.~N., Squitieri V., Douglas L.~S., Lehnert M.~D., 2008,
  MNRAS, 386, 370

\bibitem[{Trenti {et~al}\mbox{.}(2011)Trenti, Bradley, Stiavelli, Oesch, Treu,
  Bouwens, Shull, MacKenty, Carollo, \& Illingworth}]{Trenti2011}
Trenti M. {et~al.}, 2011, ApJ, 727, L39

\bibitem[{Trenti {et~al}\mbox{.}(2010)Trenti, Stiavelli, Bouwens, Oesch, Shull,
  Illingworth, Bradley, \& Carollo}]{Trenti2010}
Trenti M., Stiavelli M., Bouwens R.~J., Oesch P., Shull J.~M., Illingworth
  G.~D., Bradley L.~D., Carollo C.~M., 2010, ApJ, 714, L202

\bibitem[{Willott {et~al}\mbox{.}(2012)Willott, McLure, Hibon, Bielby,
  McCracken, Kneib, Ilbert, Bonfield, Bruce, \& Jarvis}]{Willott2012}
Willott C.~J. {et~al.}, 2012, AJ, submitted (arXiv:1202.5330)

\end{thebibliography}

\end{document}